\documentclass[sn-mathphys,Numbered]{sn-jnl}

\usepackage{graphicx}%
\usepackage{multirow}%
\usepackage{amsmath,amssymb,amsfonts}%
\usepackage{amsthm}%
\usepackage{mathrsfs}%
\usepackage[title]{appendix}%
\usepackage{xcolor}%
\usepackage{textcomp}%
\usepackage{manyfoot}%
\usepackage{booktabs}%
\usepackage{algorithm}%
\usepackage{algorithmicx}%
\usepackage{algpseudocode}%
\usepackage{listings}%
\usepackage{stackrel}

\newcommand{\updatered}[1]{#1}

\newcommand{\euu}[0]{\epsilon_{\uparrow\uparrow}}
\newcommand{\eud}[0]{\epsilon_{\uparrow\downarrow}}
\newcommand{\edu}[0]{\epsilon_{\downarrow\uparrow}}
\newcommand{\edd}[0]{\epsilon_{\downarrow\downarrow}}
\newcommand{\jcomm}[3]{\left\{#1,#2,#3\right\}_{\circ}}

\theoremstyle{thmstyleone}%
\theoremstyle{thmstyletwo}%

\theoremstyle{thmstylethree}%

\begin{document}

\title{Dixon-Rosenfeld Lines and the Standard Model}


\author*[1]{\fnm{David} \sur{Chester}}
\email{DavidC@QuantumGravityResearch.org}

\author*[2]{\fnm{Alessio} \sur{Marrani}}
\email{alessio.marrani@um.es}

\author*[3]{\fnm{Daniele} \sur{Corradetti}}
\email{a55944@ualg.pt}

\author*[1]{\fnm{Raymond} \sur{Aschheim}}
\email{Raymond@QuantumGravityResearch.org}

\author*[1]{\fnm{Klee} \sur{Irwin}}
\email{Klee@QuantumGravityResearch.org}

\affil*[1]{\centering\orgname{Quantum Gravity Research}, \\
\orgaddress{\street{Topanga Canyon Rd 101 S.}, \city{Topanga}, \state{CA} \postcode{90290}, \country{USA}}}

\affil*[2]{\centering\orgdiv{Instituto de F\'\i sica Teorica, Departamento de F\'\i sica},\\
\orgname{Universidad de Murcia}, \orgaddress{Campus de Espinardo, E-30100, Spain}}

\affil*[3]{\centering\orgname{Universidade do Algarve}, \orgdiv{Departamento de Matem\'{a}tica},\\
\orgaddress{Campus de Gambelas, 8005-139 Faro, Portugal}}

\abstract{
We present three new coset manifolds named Dixon-Rosenfeld lines that are similar to Rosenfeld projective lines except over the Dixon algebra $\mathbb{C}\otimes\mathbb{H}\otimes\mathbb{O}$.
Three different Lie groups are found as isometry groups of these coset
\updatered{manifolds using Tits' formula. We demonstrate how Standard Model interactions with the Dixon algebra in recent work from Furey and Hughes can be uplifted to tensor products of division algebras and Jordan algebras for a single generation of fermions.}
The Freudenthal-Tits construction clarifies how the three Dixon-Rosenfeld
projective lines are contained within $\mathbb{C}\otimes\mathbb{H}\otimes J_{2}(\mathbb{O})$,
$\mathbb{O}\otimes J_{2}(\mathbb{C}\otimes\mathbb{H})$, and $\mathbb{C}\otimes\mathbb{O}\otimes J_{2}(\mathbb{H})$.
}




\date{\today}

\maketitle

\tableofcontents

\section{Introduction and Motivation}

We focus on the definition of three coset manifolds of dimension 64
that we call \emph{Dixon-Rosenfeld lines}. Each contains an
isometry group whose Lie algebra is obtained from Tits' magic formula.
These three constructions are obtained similarly to how projective lines are obtained over
$\mathbb{R},\mathbb{C},\mathbb{H}$ and $\mathbb{O}$; therefore,
they can be thought of as ``generalized'' projective lines over the
Dixon algebra $\mathbb{T}\equiv\mathbb{C}\otimes\mathbb{H}\otimes\mathbb{O}$
in the sense presented by Rosenfeld in \cite{Rosenfeld-1993,Rosenfeld Group2-1,Rosenf98}.

The division algebras have been used for a wide variety of applications in physics \cite{Conway 1937,Gunaydin 1973,Manogue Dray,Huerta 1,Todorov 2018,Masi Nature}.
In 1973, G\"{u}rsey and G\"{u}naydin discussed the relationship of
octonions \updatered{and split octonions} to QCD \cite{Gunaydin 1973,Gunaydin 1976}.
Later, Dixon introduced the algebra $\mathbb{T}\equiv\mathbb{C}\otimes\mathbb{H}\otimes\mathbb{O}$
for a single generation of fermions in the Standard Model \cite{Di83,Di84,Di85,Di90,Di94}.
This line of investigation was revived when Furey further explored
the Standard Model with the Dixon algebra \cite{Furey2012,Furey2014,Furey2015,Furey2015B,Furey2018A,Furey2018,Furey2018C}
and Castro introduced gravitational models involving the Dixon algebra
\cite{Castro2019,Castro2019b,Castro2021}\footnote{\updatered{While G\"{u}rsey and G\"{u}naydin explored the use of split octonions $\mathbb{O}_s$ for quarks, Furey clarified how $\mathbb{C}\otimes \mathbb{O}$ leads to quarks and leptons \cite{Furey2012}, which contains both $\mathbb{O}$ and $\mathbb{O}_s$ as subalgebras. QCD gauges the compact $SU(3)$, which is a maximal subgroup of the automorphism group
over the octonions $\mbox{Aut}(\mathbb{O})=G_{2}$.}}. Recently, Furey and Hughes
focused on Weyl spinors for one generation of the Standard Model fermions
with $\mathbb{T}$ \cite{FureyHughes,FureyHughesB}.

\bigskip

Our work on Dixon-Rosenfeld lines defines three homogeneous spaces
that locally embed a representation of $\mathbb{T}$ to encode one
generation of fermions in the Standard Model. Section \eqref{ThreeLines}
shows that three coset manifolds of real dimension 64 are possible,
giving three non-simple Lie algebras as isometry groups that are obtained
from Tits formula. Section \eqref{OctoRosenfeld} analyzes the relationship
between the new Dixon-Rosenfeld lines with the Rosenfeld lines. Section
\eqref{CHphysics} uplifts scalar, spinor, vector, and 2-form representations
of the Lorentz group representations with $\mathbb{C}\otimes\mathbb{H}$
from Furey \cite{Furey2012} to $\mathbb{C}\otimes J_{2}(\mathbb{O})$.
Section \eqref{COphysics} uplifts the Standard Model fermionic charge
sector described by Furey with $\mathbb{C}\otimes\mathbb{O}$ \cite{Furey2015}
to $\mathbb{C}\otimes J_{2}(\mathbb{O})$. Section \eqref{CHOphysics}
uplifts recent work by Furey and Hughes for encoding Standard Model
interactions with $\mathbb{C}\otimes\mathbb{H}\otimes\mathbb{O}$
\cite{FureyHughes} to the three different realizations of the Dixon-Rosenfeld
lines via $\mathbb{C}\otimes\mathbb{H}\otimes J_{2}(\mathbb{O})$,
$\mathbb{O}\otimes J_{2}(\mathbb{C}\otimes\mathbb{H})$, and $\mathbb{C}\otimes\mathbb{O}\otimes J_{2}(\mathbb{H})$.
Section \eqref{Conclusions} concludes with a summary of our work and outlines prospects for future work.

\subsection{Tensor products on unital composition algebras}

An\emph{ algebra} is a vector space $X$ with a bilinear multiplication.
Different properties of the multiplication give rise to numerous kind
of algebras. Indeed, for what it will be used in the following sections,
an algebra $X$ is said to be \emph{commutative} if $xy=yx$ for every
$x,y\in X$, is \emph{associative} if satisfies $x\left(yz\right)=\left(xy\right)z$,
is \emph{alternative} if $x(yx)=(xy)x$, \emph{flexible}
if $x(yy)=(xy)y$ and, finally, \emph{power-associative}
if $x(xx)=(xx)x$ and $(xx)(xx)=((xx)x)x$. 
It is worth noting that the last four proprieties are progressive
and proper refinements of associativity, i.e.
\[
\text{associative}\Rightarrow\text{alternative}\Rightarrow\text{flexible}\Rightarrow\text{power\text{-}associative}.
\]

Every algebra has a zero element $0\in X$, since $X$ has to be a
group in respect to the sum, but if it also does not have zero divisors,
then $X$ is called a \emph{division} algebra, i.e.~if $xy=0$ then
or $x=0$ or $y=0$. While the zero element is always present in any
algebra, if it exists an element $1\in X$ such that $1x=x1=x$ for
all $x\in X$ then the algebra is \emph{unital}. Finally, if we can
define over $X$ an involution, called \emph{conjugation}, and a quadratic
form $N$, called \emph{norm}, such that
\begin{align}
N\left(x\right) & =x\overline{x},\\
N\left(xy\right) & =N\left(x\right)N\left(y\right),
\end{align}
with $x,y\in X$ and $\overline{x}$ as the conjugate of $x$, then
the algebra is called a \emph{composition} algebra.

A well-known theorem due to Hurwitz \cite{Hurwitz} states that $\mathbb{R}$,
$\mathbb{C}$, $\mathbb{H}$ and $\mathbb{O}$ are the only four normed
division algebras that are also unital and composition \cite{ElDuque Comp,Baez}.
More specifically, $\mathbb{R}$ is also totally ordered, commutative
and associative, $\mathbb{C}$ is just commutative and associative,
$\mathbb{H}$ is only associative and, finally, $\mathbb{O}$ is only
alternative, as summarized in Table \eqref{tab:div}.

\begin{table}
\begin{tabular}{|c|c|c|c|c|c|c|}
\hline
Algebra  & Ord.  & Comm.  & Ass.  & Alter.  & Flex.  & Pow.~Ass.\tabularnewline
\hline
\hline
$\mathbb{R}$  & Yes  & Yes  & Yes  & Yes  & Yes  & Yes\tabularnewline
\hline
$\mathbb{C}$  & No  & Yes  & Yes  & Yes  & Yes  & Yes\tabularnewline
\hline
$\mathbb{H}$  & No  & No  & Yes  & Yes  & Yes  & Yes\tabularnewline
\hline
$\mathbb{O}$  & No  & No  & No  & Yes  & Yes  & Yes\tabularnewline
\hline
\end{tabular}
\caption{Ordinality, commutativity, associativity, alternativity, flexibility, and power associativity are summarized for the division algebras.}
\label{tab:div}
\end{table}

Since all four normed division algebras are vector spaces over the
field of reals $\mathbb{R}$ we are able to define a tensor product
$\mathbb{A}\otimes\mathbb{B}$ of two normed division algebras, with
a bilinear product defined by
\begin{equation}
\left(a\otimes b\right)\left(c\otimes d\right)=ac\otimes bd,
\end{equation}
where $a,c\in\mathbb{A}$ and $b,d\in\mathbb{B}$. The resulting tensor
products are well known tensor algebras called $\mathbb{C}\otimes\mathbb{C}$
\emph{Bicomplex}, $\mathbb{C}\otimes\mathbb{H}$ \emph{Biquaternions},
$\mathbb{H}\otimes\mathbb{H}$ \emph{Quaterquaternions}, $\mathbb{C}\otimes\mathbb{O}$
\emph{Bioctonions}, $\mathbb{H}\otimes\mathbb{O}$ \emph{Quateroctonions}
and $\mathbb{O}\otimes\mathbb{O}$ \emph{Octooctonions}. By the definition
of the product, it is clear that all algebras involving the Octonions
are not associative. Moreover, while Bioctonions $\mathbb{C}\otimes\mathbb{O}$
is an alternative algebra, Quateroctonions $\mathbb{H}\otimes\mathbb{O}$
and Octooctonions $\mathbb{O}\otimes\mathbb{O}$ are not alternative
nor power-associative. Every alternative algebra tensor a commutative
algebra yields again to an alternative algebra, so that with few additional
efforts we can easily find all properties for triple tensor products
listed in Table \eqref{tab:Commutativity,-associativity,-al}.

\begin{table}
\begin{tabular}{|c|c|c|c|c|c|}
\hline
Algebra  & Comm.  & Ass.  & Alter.  & Flex.  & Pow.~Ass.\tabularnewline
\hline
\hline
$\mathbb{C}\otimes\mathbb{C}$  & Yes  & Yes  & Yes  & Yes  & Yes\tabularnewline
\hline
$\mathbb{C}\otimes\mathbb{H}$  & No  & Yes  & Yes  & Yes  & Yes\tabularnewline
\hline
$\mathbb{H}\otimes\mathbb{H}$  & No  & Yes  & Yes  & Yes  & Yes\tabularnewline
\hline
$\mathbb{C}\otimes\mathbb{O}$  & No  & No  & Yes  & Yes  & Yes\tabularnewline
\hline
$\mathbb{H}\otimes\mathbb{O}$  & No  & No  & No  & No  & No\tabularnewline
\hline
$\mathbb{O}\otimes\mathbb{O}$  & No  & No  & No  & No  & No\tabularnewline
\hline
\hline
$\mathbb{C\otimes\mathbb{C}}\otimes\mathbb{C}$  & Yes  & Yes  & Yes  & Yes  & Yes\tabularnewline
\hline
$\mathbb{C\otimes\mathbb{C}}\otimes\mathbb{H}$  & Yes  & Yes  & Yes  & Yes  & Yes\tabularnewline
\hline
$\mathbb{C\otimes\mathbb{H}}\otimes\mathbb{H}$  & No  & Yes  & Yes  & Yes  & Yes\tabularnewline
\hline
$\mathbb{H\otimes\mathbb{H}}\otimes\mathbb{H}$  & No  & Yes  & Yes  & Yes  & Yes\tabularnewline
\hline
$\mathbb{C\otimes\mathbb{C}}\otimes\mathbb{O}$  & No  & No  & Yes  & Yes  & Yes\tabularnewline
\hline
$\mathbb{C\otimes\mathbb{H}}\otimes\mathbb{O}$  & No  & No  & No  & No  & No\tabularnewline
\hline
\end{tabular}
\caption{\label{tab:Commutativity,-associativity,-al}Commutativity, associativity,
alternativity, flexibility and power associativity of two and three tensor products
of normed division algebras $\mathbb{R}$, $\mathbb{C}$, $\mathbb{H}$
and $\mathbb{O}$ are shown. The split version of the algebras obeys the same
property of the division version.}
\end{table}

\subsection{The Dixon algebra}

\label{DixonAlgebra}

The \emph{Dixon Algebra} $\mathbb{T}$ is the $\mathbb{R}$-linear
tensor product of the four normed division algebras, i.e.~$\mathbb{R}\otimes\mathbb{C}\otimes\mathbb{H}\otimes\mathbb{O}$
or equivalently $\mathbb{C}\otimes\mathbb{H}\otimes\mathbb{O}$, with
linear product defined by
\begin{equation}
\left(z\otimes q\otimes w\right)\left(z'\otimes q'\otimes w'\right)=zz'\otimes qq'\otimes ww',
\end{equation}
with $z,z'\in\mathbb{C}$, $q,q'\in\mathbb{H}$ and $w,w'\in\mathbb{O}$.
From the previous formula it is evident that $\mathbb{T}$ is unital
with unit element $\text{\textbf{1}}=1\otimes1\otimes1.$ As a real
vector space, the Dixon Algebra has an $\mathbb{R}^{64}$ decomposition
for which every element $t$ is of the form
\begin{align}
t & =\stackrel[\alpha=0]{63}{\sum}t^{\alpha}\,\,\,z\otimes q\otimes w,\label{eq:decomposizione 64-1}
\end{align}
where $t^{\alpha}\in\mathbb{R}$, and $z,q,w$ are elements of a basis
for $\mathbb{C}$, $\mathbb{H}$, $\mathbb{O}$ respectively, i.e.
$z\in\left\{ 1,I\right\} $, $q\in\left\{ 1,i,j,k\right\} $ and $w\in\left\{ 1,e_{1},...,e_{7}\right\} $
with
\begin{align}
I^{2} & =i^{2}=j^{2}=k^{2}=e_{\alpha}^{2}=-1,\\
\left[I\,,i\right] & =\left[I\,,j\right]=\left[I\,,k\right]=\left[I\,,e_{\alpha}\right]=0,\\
\left[e_{\alpha},i\right] & =\left[e_{\alpha},j\right]=\left[e_{\alpha},k\right]=0,
\end{align}
and the other rules of multiplication given in Fig.~\eqref{fig:Multiplication-rule-of-1}.

It is straightforward to see that every element in the set
\begin{equation}
D=\left\{ \left(I\,q\pm1\right),\left(I\,e_{\alpha}\pm1\right),\,\,\left(qe_{\alpha}\pm1\right):q\in\left\{ i,j,k\right\} \right\} ,
\end{equation}
is a \emph{zero divisor }and therefore $\mathbb{T}$ is not a division
algebra. Moreover, the Dixon algebra is not commutative, neither associative,
nor alternative or flexible and, finally, not even power-associative,
i.e.~in general $x\left(xx\right)\neq\left(xx\right)x$ .
\begin{figure}
\begin{centering}
\includegraphics[scale=0.06]{DixonMultiplicationRules2}
\par\end{centering}
\caption{\label{fig:Multiplication-rule-of-1}Multiplication rule of Octonions
$\mathbb{O}$ (\emph{right}), Quaternions $\mathbb{H}$ (\emph{middle})
and Complex $\mathbb{C}$ (\emph{left}).}
\end{figure}

Nevertheless, it is possible to define a quadratic norm $N$ over
$\mathbb{T}$, starting from the decomposition in Eq.~\eqref{eq:decomposizione 64-1},
i.e.
\begin{equation}
N\left(t\right)=\stackrel[\alpha=0]{63}{\sum}\left(t^{\alpha}\right)^{2},
\end{equation}
with an associated \emph{polar form} $\left\langle \cdot,\cdot\right\rangle $
given by the symmetric bilinear form
\begin{equation}
2\left\langle t_{1},t_{2}\right\rangle =N\left(t_{1}+t_{2}\right)-N\left(t_{1}\right)-N\left(t_{2}\right).
\end{equation}

\section{Dixon-Rosenfeld lines}

\label{ThreeLines}

The geometrical motivation for defining Dixon-Rosenfeld lines as coset
manifolds relies on the study of the octonionic planes explored by
Tits, Freudenthal and Rosenfeld in a series of seminal works \cite{Tits,Freud 1965,Rosenfeld-1993,Vinberg}
that led to a geometric interpretation of Lie algebras and to the
construction of the Tits-Freudenthal Magic Square. While Freudenthal
interpreted the entries of the Magic Square as different forms of automorphisms
of the projective plane such as isometries, collineations, homography
etc., Rosenfeld thought of every row of the magic square as the Lie algebra
of the isometry groups of a ``generalized'' projective plane over
a tensorial product of Hurwitz algebras \cite{Rosenfeld-1993} (see
also \cite{MCCAI Magic Rosenfeld} for a recent systematic review).
In fact, tensor products over Hurwitz algebras are not division algebras, which
therefore do not allow the definition of a projective plane in a strict
sense. Nevertheless, later works of Atsuyama proved the insight of
Rosenfeld to be correct and that it is possible to use these
algebras to define projective planes in a ``wider sense'' \cite{Atsuyama,Atiyah Bernd,Lands-Maniv}.
A similar analysis was then carried out for generalized projective
lines making use the Tits-Freudenthal Magic Square of order two instead
of three, thus relating the resulting Lie algebras with isometries
of generalized projective lines, instead of planes (see \cite{MCCAI Magic Rosenfeld},
for more details).

\subsection{Dixon lines as coset manifold }

\emph{Coset manifolds} arise from coset spaces over a Lie group $G$
given by an equivalence relation of the type
\begin{equation}
g\sim g'\Longleftrightarrow gh=g',
\end{equation}
where $g,g'\in G$ and $h\in H$ and $H$ is a closed subgroup of
$G$. In this case, the coset space $G/H$, obtained from the equivalence
classes $gH$, inherits a manifold structure from $G$ and is therefore
a manifold of dimension
\begin{equation}
\text{dim}\left(G/H\right)=\text{dim}\left(G\right)-\text{dim}\left(H\right).
\end{equation}
Moreover, $G/H$ can be endowed with invariant metrics such that all
elements of the original group $G$ are isometries of the constructed
metric \cite{Fre and Fedotov,MCCAI Magic Rosenfeld}. More specifically,
the structure constants of the Lie algebra $\mathfrak{g}$ of the
Lie group $G$ define completely the metric and therefore all the
metric-dependent tensors, such as the curvature tensor, the Ricci
tensor, etc. Finally, the coset space $G/H$ is a
homogeneous manifold by construction, i.e.~the group $G$ acts transitively, and its
\emph{isotropy subgroup} is precisely $H$, i.e.~the group $H$ is
such that for any given point $p$ in the manifold $hp=p$. Therefore,
for our purposes in the definition of the Dixon-Rosenfeld lines, it
will be sufficient to define the isometry group and the isotropy group
of the coset manifold to have them completely defined in its topological
and metrical descriptions.

\subsection{Tits' magic formula}

We now proceed defining three Dixon projective lines as three different
coset spaces of real dimension 64 obtained from three isometry algebras $%
\mathfrak{a}_{I}$, $\mathfrak{a}_{II}$ and $\mathfrak{a}_{III}$ making the
use of Tits' magic formula \cite{Tits} for $n=2$, i.e.
\updatered{
\begin{equation}
\mathcal{L}_{2}\left( \mathbb{A},\mathbb{B}\right) =\mathfrak{der}\left(
\mathbb{A}\right) \oplus \mathfrak{der}\left( J_{2}\left( \mathbb{B}\right)
\right) \oplus \left( \mathbb{A}^{\prime }\otimes J_{2}^{\prime }\left(
\mathbb{B}\right) \right) ,  \label{Tits Formula}
\end{equation}
}
where $\mathbb{A},\mathbb{B}$ are alternative algebras and $J_{2}\left(
\mathbb{B}\right) $ is a Jordan algebra over Hermitian two by two matrices
\footnote{\updatered{
After \cite{BS}, when $\mathbb{A}=\mathbb{O}$ (see case $\mathbf{II}$
below), the formula (\ref{Tits Formula}) has $\mathfrak{der}\left( \mathbb{A}%
\right) $ replaced by $\mathfrak{so}(\mathbb{A}^{\prime })$.}}.
Brackets on $\mathcal{L}_{2}\left( \mathbb{A},\mathbb{B}\right) $ can be
defined following notation in \cite[sec.~3]{BS} for which, given the an
algebra $\mathbb{A}$, we define
\begin{equation}
X^{\prime }=X-\frac{1}{2}\text{Tr}\left( X\right) \text{\textbf{1}},
\end{equation}%
as the projection of an element of the algebra in the subspace orthogonal to
the identity denoted as $\text{\textbf{1}}$. We then define $J_{2}^{\prime
}\left( \mathbb{B}\right) $ the algebra obtained by such elements with the
product $\bullet $ given by the projection back on the subspace orthogonal
to the identity of the Jordan product, i.e.
\begin{equation}
X^{\prime }\bullet Y^{\prime }=X^{\prime }\cdot Y^{\prime }-2\left\langle
X^{\prime },Y^{\prime }\right\rangle \text{\textbf{1}},
\end{equation}%
where as usual, we intended $X\cdot Y=XY+YX$ and $\left\langle
X,Y\right\rangle =\frac{1}{2}\text{Tr}\left( X\cdot Y\right) $ for every $X,Y%
\in J_{2}\left( \mathbb{B}\right) $.
\updatered{
With this notation, the vector space (\ref{Tits Formula}) is endowed with
the following brackets \cite{BS}
}

\begin{enumerate}
\item The usual brackets on the Lie subalgebra
\updatered{
$\mathfrak{der}\left( \mathbb{A}\right) \oplus \mathfrak{der}\left(
J_{2}\left( \mathbb{B}\right) \right) $.
}

\item When
\updatered{
$a\in \mathfrak{der}\left( \mathbb{A}\right) \oplus \mathfrak{der}\left(
J_{2}\left( \mathbb{B}\right) \right) $
} and $A\in \mathbb{A}^{\prime }\otimes J_{2}^{\prime }\left( \mathbb{B}%
\right) $ then
\begin{equation}
\left[ a,A\right] =a\left( A\right) .  \label{eq:der action}
\end{equation}

\item When
\updatered{
$a\otimes A,b\otimes B \in \mathbb{A}^{\prime }\otimes
J_{2}^{\prime }\left(\mathbb{B}\right)$
}
then
\begin{equation}
\left[a\otimes A,b\otimes B\right]=\frac{1}{2}\left\langle A,B\right\rangle
D_{a,b}-\left\langle a,b\right\rangle \left[L_{A},L_{B}\right]+\frac{1}{2}%
\left[a,b\right]\otimes\left(A\bullet B\right),
\end{equation}
where $L_{x}$ and $R_{x}$ are the left and right action on the algebra and $%
D_{x,y}$ is given by 
\begin{equation}
D_{x,y}=\left[L_{x},L_{y}\right]+\left[L_{x},R_{y}\right]+\left[R_{x},R_{y}%
\right].  \label{eq:der bracket}
\end{equation}
\end{enumerate}

\updatered{
Applying now formula (\ref{eq:der bracket}) to the Jordan algebra with left
and right Jordan product, the left and right products are the same and
\begin{equation}
D_{X,Y}=3 \left[L_{X},R_{Y}\right].  \label{eq:der jordan bracket}
\end{equation}
and
\begin{equation}
D_{X,Y}(Z)=3 \left[X, Z, Y\right] = 3[[X,Y],Z] .
\label{eq:der jordan associator}
\end{equation}
}

%

\subsection{Three isometry groups}

Tits' formula is the most general formula compared to those of Vinberg \cite%
{Vinberg}, Atsuyama \cite{Atsuyama1}, Santander and Herranz \cite{Santander}%
, Barton and Sudbery \cite{BS}, and Elduque \cite{EldMS2} since it does not
require the use of two composition algebras, but only the use of an
alternative algebra and a Jordan algebra obtained from another alternative
algebra.

\updatered{
Next, we consider all tensor products of the form $\mathbb{A}\otimes J_{2}(%
\mathbb{B})$ with $\mathbb{A}$ and $\mathbb{B}$ alternative such that $%
\mathbb{A}\otimes \mathbb{B}$ corresponds to the Dixon algebra $\mathbb{C}%
\otimes \mathbb{H}\otimes \mathbb{O}$.
}
Since $\mathbb{H}\otimes \mathbb{O}$ is not alternative, the possible
candidates can be \emph{a priori} only related with the following four
different $\mathbb{A}$ and $\mathbb{B}$, i.e.
\begin{align}
I:\mathbb{A}& =\left( \mathbb{C}\otimes \mathbb{H}\right) ,\quad \mathbb{B}=%
\mathbb{O}, \\
II:\mathbb{A}& =\mathbb{O},\quad \mathbb{B}=\left( \mathbb{C}\otimes \mathbb{%
H}\right) , \\
III:\mathbb{A}& =\left( \mathbb{C}\otimes \mathbb{O}\right) ,\quad \mathbb{B}%
=\mathbb{H},
\end{align}%
and, finally, $\mathbb{A}=\mathbb{H},\mathbb{B}=\left( \mathbb{C}\otimes
\mathbb{O}\right) .$ However the latter case, i.e.~$\mathbb{A}=\mathbb{H},%
\mathbb{B}=\left( \mathbb{C}\otimes \mathbb{O}\right) $, would need the
existence of a Jordan algebra $J_{2}\left( \mathbb{C}\otimes \mathbb{O}%
\right) $ over bioctonions $\mathbb{C}\otimes \mathbb{O}$, which is not
possible.\footnote{
\updatered{
We excluded on purpose the cases where the involution on Hermitian matrices
does not involve whole blocks of imaginary units in the algebra, giving rise
to $\mathbb{C}\otimes J_{2}(\mathbb{O})$ and $\mathbb{O}\otimes J_{2}(%
\mathbb{C})$, for which the analogy with Rosenfeld construction does not
hold. Alternatively, one might consider an involution given by the composition of
the conjugation of $\mathbb{C}$ and the conjugation of $\mathbb{O}$, or also
an involution flipping all imaginary units of $\mathbb{C}\otimes \mathbb{O}$%
; while the former yields to non-real diagonal elements of the corresponding
Hermitian $2\times 2$ matrices over $\mathbb{C}\otimes \mathbb{O}$, the
latter has real diagonal elements. However, we checked that in both cases
the resulting rank-2 Hermitian matrices do not form a Jordan algebra
satisfying the Jordan identity.
}
}

\begin{table}
\begin{tabular}{|c|c|c|c|}
\hline
 & $\mathbb{T}P_{I}^{1}$  & $\mathbb{T}P_{II}^{1}$  & $\mathbb{T}P_{III}^{1}$\tabularnewline
\hline
\hline
$\mathfrak{isom}$  & $\mathfrak{so}_{9}\oplus \mathfrak{su}_{2}\oplus \left( \mathbf{9},2\cdot\mathbf{3}+\mathbf{1}\right)$  & $\mathfrak{so}_{7}\oplus \mathfrak{so}_{6}\oplus 3\cdot \left( \mathbf{7},\mathbf{4}\right) \oplus \left( \mathbf{1},\mathbf{1}\right)$  & $\mathfrak{g}_{2}\oplus \mathfrak{so}_{5}\oplus \left( 2\cdot \mathbf{7}+\mathbf{1},\mathbf{5}\right)$ \tabularnewline
\hline
$\mathfrak{isot}$  & $\mathfrak{so}_{8}\oplus \mathfrak{su}_{2}\oplus \left( \mathbf{1},2\cdot \mathbf{3}+\mathbf{1}\right)$ & $\mathfrak{so}_{7}\oplus \mathfrak{su}_{2}\oplus \left( \mathbf{7}, mathbf{3+1}\right) \oplus \left( \mathbf{1},\mathbf{5}\right)$ & $\mathfrak{g}_{2}\oplus \mathfrak{su}_{2}\oplus \left( 2\cdot \mathbf{7}+\mathbf{1},\mathbf{1}\right) \oplus \left( \mathbf{1},\mathbf{3}\right)$ \tabularnewline
\hline
\end{tabular}
\caption{\updatered{Isometry and isotropy Lie algebras of the three Dixon-Rosenfeld lines. For $\mathbb{T}P_{II}^{1}$, the \textquotedblleft minimal\textquotedblright\ enhancement (\ref{pre}) is considered.}}
\label{TABLE:3Lines isometry-isotropy}
\end{table}

We are therefore left with only three different possibilities, i.e.
\begin{equation}
\begin{array}{c}
\mathfrak{a}_{I}=\mathcal{L}_{2}\left(\mathbb{\mathbb{C}\otimes\mathbb{H}},%
\mathbb{O}\right), \\
\mathfrak{a}_{II}=\mathcal{L}_{2}\left(\mathbb{\mathbb{O}},\mathbb{C}\otimes%
\mathbb{H}\right), \\
\mathfrak{a}_{III}=\mathcal{L}_{2}\left(\mathbb{\mathbb{C}\otimes\mathbb{O}},%
\mathbb{H}\right).%
\end{array}
\label{eq:Three cases}
\end{equation}

\updatered{
We will now discuss how, due to the three possible cases in Eq.~%
\eqref{eq:Three cases}, there exist three \textquotedblleft homogeneous
realizations\textquotedblright\ of the Dixon-Rosenfeld projective line $%
\mathbb{T}P^{1}$, which will be distinguished by the subscript $I$, $II$ and
$III$, respectively.

We start and observe that
\begin{equation}
\mathfrak{der}\left( \mathbb{C}\otimes \mathbb{H}\right) \simeq \mathfrak{der%
}\left( \mathbb{C}\right) \oplus \mathfrak{der}\left( \mathbb{H}\right)
\simeq \mathfrak{der}\left( \mathbb{H}\right) \simeq \mathfrak{su}_{2},
\label{tthis}
\end{equation}%
such that%
\begin{equation}
\mathbb{C}\otimes \mathbb{H}\simeq \left( 2\cdot \mathbf{1}\right) \otimes
\left( \mathbf{1}+\mathbf{3}\right) =2\cdot \left( \mathbf{1}\oplus \mathbf{3%
}\right) ~\text{of~}\mathfrak{su}_{2},  \label{res2}
\end{equation}%
implying that the imaginary biquaternions are%
\begin{equation}
\left( \mathbb{C}\otimes \mathbb{H}\right) ^{\prime }\simeq \mathbf{1}\oplus
2\cdot \mathbf{3}~\text{of~}\mathfrak{su}_{2}.
\end{equation}%
This can be understood by observing that%
\begin{equation}
\left.
\begin{array}{l}
\mathbb{C\simeq }\left\{ 1,I\right\}  \\
\\
\mathbb{H\simeq }\left\{ 1,i,j,k\right\}
\end{array}%
\right\} \Rightarrow \mathbb{C}\otimes \mathbb{H}\simeq \left\{ \underset{%
\mathbf{1}\oplus \mathbf{1}}{\underbrace{1,I}},\underset{\mathbf{3}}{%
\underbrace{i,j,k}},\underset{\mathbf{3}}{\underbrace{Ii,Ij,Ik}}\right\} .
\label{res2-bis}
\end{equation}%
Note that $\mathfrak{der}\left( \mathbb{C}\otimes \mathbb{H}\right) $ is
next-to-maximal into $\mathfrak{der}\left( \mathbb{O}\right) \simeq
\mathfrak{g}_{2}$, because it can be obtained by a chain of two maximal (and
symmetric) embeddings,%
\begin{eqnarray}
\mathfrak{g}_{2} &\supset &\mathfrak{su}_{2}\oplus \mathfrak{su}_{2}\supset
\mathfrak{su}_{2,d},  \label{jh} \\
\mathbf{7} &=&(\mathbf{1},\mathbf{3})+(\mathbf{2},\mathbf{2})=\mathbf{3}+%
\mathbf{3}+\mathbf{1},  \notag \\
\mathbf{14} &=&(\mathbf{3},\mathbf{1})+(\mathbf{1},\mathbf{3})+(\mathbf{4},%
\mathbf{2})=3\cdot \mathbf{3}+\mathbf{5},  \notag
\end{eqnarray}%
or equivalently by a chain of two maximal (one non-symmetric and one
symmetric) embeddings,%
\begin{eqnarray}
\mathfrak{g}_{2} &\supset &\mathfrak{su}_{3}\supset \mathfrak{su}_{2,P},
\label{jhh} \\
\mathbf{7} &=&\mathbf{3}+\overline{\mathbf{3}}+\mathbf{1}=\mathbf{3}+\mathbf{%
3}+\mathbf{1},  \notag \\
\mathbf{14} &=&\mathbf{3}+\overline{\mathbf{3}}+\mathbf{8}=3\cdot \mathbf{3}+%
\mathbf{5}.  \notag
\end{eqnarray}

In all cases, the Dixon algebra $\mathbb{T}$ will have the same covariant
realization in terms of%
\begin{equation}
\mathfrak{der}\left( \mathbb{T}\right) \simeq \mathfrak{der}\left( \mathbb{C}%
\otimes \mathbb{H}\right) \oplus \mathfrak{der}\left( \mathbb{O}\right)
\simeq \mathfrak{der}\left( \mathbb{H}\right) \oplus \mathfrak{der}\left(
\mathbb{O}\right) \simeq \mathfrak{su}_{2}\oplus \mathfrak{g}_{2},\label%
{der(T)}
\end{equation}%
i.e.%
\begin{equation}
\mathbb{T\simeq }T\left( \mathbb{T}P_{I}^{1}\right) \mathbb{\simeq }T\left(
\mathbb{T}P_{II}^{1}\right) \mathbb{\simeq }T\left( \mathbb{T}%
P_{III}^{1}\right) \simeq 2\cdot \left( \mathbf{3}+\mathbf{1},\mathbf{7}+%
\mathbf{1}\right) ~\text{of~}\mathfrak{su}_{2}\oplus \mathfrak{g}_{2},
\label{TI}
\end{equation}%
which can enjoy the following enhancements of (manifest) covariance,%
\begin{eqnarray}
\mathbb{T} &\simeq &2\cdot \left( \mathbf{1}+\mathbf{3},\mathbf{7}+\mathbf{1}%
\right) ~\text{of~}\mathfrak{su}_{2}\oplus \mathfrak{so}_{7}\label{pre-TI-2}
\\
&\simeq &2\cdot \left( \mathbf{1}+\mathbf{3},\mathbf{8}_{v}\right) ~\text{of~%
}\mathfrak{su}_{2}\oplus \mathfrak{so}_{8}.  \label{TI-2}
\end{eqnarray}

\begin{description}
\item[$\mathbf{I}.$] In the case $\mathbb{A}=\mathbb{C}\otimes \mathbb{H}$
and $\mathbb{B}=\mathbb{O}$, Tits' formula (\ref{Tits Formula}) yields (cf.(%
\ref{tthis}))%
\begin{eqnarray}
\mathfrak{a}_{I} &=&\mathcal{L}_{2}\left( \mathbb{C}\otimes \mathbb{H},%
\mathbb{O}\right) =\mathfrak{isom}\left( \mathbb{T}P_{I}^{1}\right)   \notag
\\
&:=&\mathfrak{der}\left( \mathbb{C}\otimes \mathbb{H}\right) \oplus
\mathfrak{der}\left( J_{2}(\mathbb{O})\right) \oplus \left( \mathbb{C}%
\otimes \mathbb{H}\right) ^{\prime }\otimes J_{2}^{\prime }(\mathbb{O})
\notag \\
&=&\mathfrak{su}_{2}\oplus \mathfrak{so}_{9}\oplus \left( \mathbf{1}+2\cdot
\mathbf{3},\mathbf{9}\right) ,  \label{isom1}
\end{eqnarray}%
because%
\begin{eqnarray}
\mathfrak{der}\left( J_{2}(\mathbb{O})\right)  &=&\mathfrak{so}_{9}, \\
J_{2}^{\prime }(\mathbb{O}) &\simeq &\mathbf{9},
\end{eqnarray}%
The Lie algebra $\mathfrak{isom}\left( \mathbb{T}P_{I}^{1}\right) $ has
therefore dimension $3+36+63=102$.

\item[$\mathbf{II}.$] In the case $\mathbb{A}=\mathbb{O}$ and $\mathbb{B}=%
\mathbb{\mathbb{C}\otimes \mathbb{H}}$, after the treatment given in Sec. 8
of \cite{BS}, Tits' formula (\ref{Tits Formula}) gets $\mathfrak{der}\left(
\mathbb{O}\right) $ replaced by $\mathfrak{so}\left( \mathbb{O}^{\prime
}\right) $, and thus one obtains
\begin{eqnarray}
\mathfrak{a}_{II} &=&\mathcal{L}_{2}\left( \mathbb{O},\mathbb{C}\otimes
\mathbb{H}\right) =\mathfrak{isom}\left( \mathbb{T}P_{II}^{1}\right)   \notag
\\
&:=&\mathfrak{so}\left( \mathbb{O}^{\prime }\right) \oplus \mathfrak{der}%
\left( J_{2}(\mathbb{C}\otimes \mathbb{H})\right) \oplus \mathbb{O}^{\prime
}\otimes J_{2}^{\prime }(\mathbb{C}\otimes \mathbb{H}).\label{a_II}
\end{eqnarray}%
$J_{2}(\mathbb{C}\otimes \mathbb{H})$ is a rank-2 Jordan algebra, defined as
the algebra of $2\times 2$ matrices over $\mathbb{C}\otimes \mathbb{H}$ (cf.
(\ref{res2}) and (\ref{res2-bis})) which are Hermitian with respect to the
involution $\imath $ given by the composition of the conjugation of $\mathbb{%
C}$ and of the conjugation of $\mathbb{H}$:%
\begin{equation}
\imath :\left\{
\begin{array}{lll}
I & \rightarrow  & -I; \\
i,j,k & \rightarrow  & -i,-j,-k; \\
Ii,Ij,Ik & \rightarrow  & Ii,Ij,Ik.%
\end{array}%
\right.
\end{equation}%
Interestingly, this implies that the diagonal elements of the matrices of $%
J_{2}(\mathbb{C}\otimes \mathbb{H})$ are non-real, being of the form $%
d=d_{1}+Iid_{2}+Ijd_{3}+Ikd_{4}$, with $d_{1},d_{2},d_{3},d_{4}\in \mathbb{R}
$, and\footnote{%
Despite this, such diagonal elements do not belong to $\mathbb{H}_{s}$,
because the composite imaginary units $Ii,Ij,Ik$ are not a basis for $%
\mathbb{H}_{s}^{\prime }$.} $\left( Ii\right) ^{2}=\left( Ij\right)
^{2}=\left( Ik\right) ^{2}=1$. Thus, with respect to $\mathfrak{der}\left(
\mathbb{C}\otimes \mathbb{H}\right) \simeq \mathfrak{su}_{2}$ (\ref{tthis}),
$J_{2}(\mathbb{C}\otimes \mathbb{H})$ fit into the following representations:%
\begin{equation}
J_{2}(\mathbb{C}\otimes \mathbb{H})\simeq \left(
\begin{array}{ccc}
\mathbf{1}\oplus \mathbf{3} &  & 2\cdot \left( \mathbf{1}\oplus \mathbf{3}%
\right)  \\
&  &  \\
\ast  &  & \mathbf{1}\oplus \mathbf{3}%
\end{array}%
\right) \simeq 4\cdot \left( \mathbf{1}\oplus \mathbf{3}\right) ~\text{of~}%
\mathfrak{su}_{2},\label{traceful}
\end{equation}%
yielding for the traceless part that%
\begin{equation}
J_{2}^{\prime }(\mathbb{C}\otimes \mathbb{H})\simeq 3\cdot \left( \mathbf{1}%
\oplus \mathbf{3}\right) ~\text{of~}\mathfrak{su}_{2}.\label{traceless}
\end{equation}%
On the other hand, as proved in Appendix \ref{App-der(J2(biquat))}, it holds that%
\begin{equation}
\mathfrak{der}\left( J_{2}(\mathbb{C}\otimes \mathbb{H})\right) \simeq
\mathfrak{so}_{6},\label{ecco}
\end{equation}%
and thus (\ref{traceful}) and (\ref{traceless}) respectively enjoy the
following enhancements\footnote{%
(\ref{traceful}) resp. (\ref{traceless}) can be obtained from (\ref%
{traceful-2}) resp. (\ref{traceless-2}) by observing that $\mathfrak{der}%
\left( \mathbb{C}\otimes \mathbb{H}\right) \simeq \mathfrak{su}_{2}$ (\ref%
{tthis}) is a diagonal subalgebra of $\mathfrak{der}\left( J_{2}(\mathbb{C}%
\otimes \mathbb{H})\right) \simeq \mathfrak{so}_{6}$ (\ref{ecco}) obtained
by the following chain of maximal and symmetric embeddings: $\mathfrak{so}%
_{6}\simeq \mathfrak{su}_{4}\rightarrow \mathfrak{so}_{4}\simeq \mathfrak{su}%
_{2}\oplus \mathfrak{su}_{2}\overset{d}{\rightarrow }\mathfrak{su}_{2}$.}:%
\begin{eqnarray}
J_{2}(\mathbb{C}\otimes \mathbb{H}) &\simeq &\left(
\begin{array}{ccc}
\mathbf{4} &  & \mathbf{4}\oplus \mathbf{4} \\
&  &  \\
\ast  &  & \mathbf{4}%
\end{array}%
\right) \simeq 4\cdot \mathbf{4}~\text{of~}\mathfrak{so}_{6};\label%
{traceful-2} \\
J_{2}^{\prime }(\mathbb{C}\otimes \mathbb{H}) &\simeq &3\cdot \mathbf{4}~%
\text{of~}\mathfrak{so}_{6}.\label{traceless-2}
\end{eqnarray}%
Therefore, since%
\begin{eqnarray}
\mathfrak{so}\left( \mathbb{O}^{\prime }\right)  &=&\mathfrak{so}_{7}=%
\mathfrak{g}_{2}\oplus \mathbf{7};  \label{ttthis} \\
\mathbb{O}^{\prime } &\simeq &\mathbf{7}~\text{of~}\mathfrak{so}_{7}=\mathbf{%
7}~\text{of~}\mathfrak{g}_{2},
\end{eqnarray}%
formula (\ref{a_II}) can be made explicit as follows:%
\begin{eqnarray}
\mathfrak{a}_{II} &=&\mathcal{L}_{2}\left( \mathbb{O},\mathbb{C}\otimes
\mathbb{H}\right) =\mathfrak{isom}\left( \mathbb{T}P_{II}^{1}\right) =%
\mathfrak{so}_{7}\oplus \mathfrak{so}_{6}\oplus 3\cdot \left( \mathbf{7},%
\mathbf{4}\right) \label{a_II-2} \\
&=&\mathfrak{g}_{2}\oplus \mathfrak{so}_{6}\oplus 3\cdot \left( \mathbf{7},%
\mathbf{4}\right) \oplus \left( \mathbf{7},\mathbf{1}\right)   \notag \\
&=&\mathfrak{g}_{2}\oplus \mathfrak{so}_{4}\oplus 3\cdot \left( \mathbf{7},%
\mathbf{2,2}\right) \oplus \left( \mathbf{7},\mathbf{1,1}\right) \oplus
\left( \mathbf{1},\mathbf{3},\mathbf{3}\right)   \notag \\
&=&\mathfrak{g}_{2}\oplus \mathfrak{su}_{2}\oplus 3\cdot \left( \mathbf{7},%
\mathbf{3}\right) \oplus 4\cdot \left( \mathbf{7},\mathbf{1}\right) \oplus
\left( \mathbf{1},\mathbf{5}\right) \oplus 2\cdot \left( \mathbf{1},\mathbf{3%
}\right) \oplus \left( \mathbf{1},\mathbf{1}\right) .\label{questa}
\end{eqnarray}%
The last line (\ref{questa}) has a manifest $\left( \mathfrak{g}_{2}\oplus
\mathfrak{su}_{2}\right) $-covariance, which is the natural one for the
Dixon algebra $\mathbb{T}$ (cf. (\ref{TI})), giving%
\begin{equation}
\mathbb{T\nsubseteq }\mathfrak{a}_{II},
\end{equation}%
because there is only one singlet $\left( \mathbf{1},\mathbf{1}\right) $ in (\ref{questa}). See Appendix \ref{aIInotFT} for a more exhaustive treatment of all $\mathfrak{su}_2$'s inside $\mathfrak{so}_6$.

However, it is anticipated that $\mathbb{T}\in \mathfrak{a}_{II}$. Therefore, there are two possibilities to resolve this issue. First, it was asserted above that $\mathbb{H}$ corresponds to $\mathbf{1}\oplus \mathbf{3}$ of $\mathfrak{su}_2$, which led to $\mathbb{T}\in\mathfrak{a}_I$. However, $\mathbb{C}\otimes \mathbb{H}$ is known to allow for three different representations of $\mathfrak{sl}_{2,\mathbb{C}}$ \cite{Furey2012}. If the spinor representations were chosen instead, then $\mathbb{T}\in \mathfrak{a}_{II}.$ Second, one may claim that the $2\times 2$ Freudenthal-Tits formula does not apply to the case where $\mathbb{A}=\mathbb{O}$ and $\mathbb{B} = \mathbb{C}\otimes\mathbb{H}$. Freudenthal and Tits' formula was designed for $3\times 3$, but the $2\times 2$ case already has a precedent of the formula depending on the algebras chosen, as $\mathbb{A}=\mathbb{O}$ leads to a difference from $\mathbb{A}=\mathbb{C}$ or $\mathbb{H}$ in the $2\times 2$ case. In this work, we merely claim that a non-simple Lie algebra $\mathfrak{a}_{II}$ exists, but we do not fully determine its precise structure.

Assuming that the Freudenthal-Tits formula does not apply to 
The \textquotedblleft minimal\textquotedblright\
enhancement of $\mathfrak{a}_{II}$ such that it contains $\mathbb{T}$ with ${\bf 1}\oplus {\bf 3}$ of $\mathfrak{su}_2$ 
amounts to adding a $\left( \mathfrak{g}_{2}\oplus \mathfrak{su}_{2}\right) $%
-singlet generator:%
\begin{eqnarray}
\mathfrak{a}_{II} &\longrightarrow &\mathfrak{a}_{II,\text{enh.}}:=\mathfrak{%
a}_{II}\oplus \left( \mathbf{1},\mathbf{1}\right) \label{pre} \\
&=&\mathfrak{g}_{2}\oplus \mathfrak{su}_{2}\oplus 3\cdot \left( \mathbf{7},%
\mathbf{3}\right) \oplus 4\cdot \left( \mathbf{7},\mathbf{1}\right) \oplus
\left( \mathbf{1},\mathbf{5}\right) \oplus 2\cdot \left( \mathbf{1},\mathbf{3%
}\right) \oplus 2\cdot \left( \mathbf{1},\mathbf{1}\right)   \notag \\
&=&\mathfrak{g}_{2}\oplus \mathfrak{su}_{2}\oplus \left( \mathbf{7},\mathbf{3%
}\right) \oplus 2\cdot \left( \mathbf{7},\mathbf{1}\right) \oplus \left(
\mathbf{1},\mathbf{5}\right) \oplus \mathbb{T}.\label{a_II-enh}
\end{eqnarray}%
Thanks to (\ref{pre-TI-2}), the last line (\ref{a_II-enh}) enjoys the
further symmetry enhancement%
\begin{equation}
\mathfrak{a}_{II,\text{enh.}}=\mathfrak{so}_{7}\oplus \mathfrak{su}%
_{2}\oplus \left( \mathbf{7},\mathbf{3}\right) \oplus \left( \mathbf{7},%
\mathbf{1}\right) \oplus \left( \mathbf{1},\mathbf{5}\right) \oplus \mathbb{T%
}. \label{TTitsEnh}
\end{equation}%
Note however that a further symmetry enhancement to $\mathfrak{so}_{8}\oplus
\mathfrak{su}_{2}$ is not possible without breaking $\mathbb{T}$ itself%
\footnote{%
A further $\left( \mathbf{1},\mathbf{3}\right) $-generator would be needed
in the r.h.s. of (\ref{pre}).}. If the Lie algebra $\mathfrak{a}_{II}$ should be enhanced, then $\mathfrak{a}_{II,\text{enh.}%
}\equiv \mathfrak{isom}\left( \mathbb{T}P_{II}^{1}\right) _{\text{enh.}}$
given by (\ref{a_II-enh}) has dimension $21+15+3\cdot 28+1=121$. Alternatively, it is possible that $\mathfrak{a}_{II}$ is 120-dimensional such that $\mathbb{H}\in \mathbb{T}$ contains spinor representations, such as ${\bf 2}\oplus {\bf2}$ of $\mathfrak{der}(\mathbb{H}) = \mathfrak{su}_2$. 

\item[$\mathbf{III}.$] Finally, in the case $\mathbb{A}=\mathbb{C}\otimes
\mathbb{O}$ (non-associative) and $\mathbb{B}=\mathbb{H}$, Tits' formula (%
\ref{Tits Formula}) yields
\begin{eqnarray}
\mathfrak{a}_{III} &=&\mathcal{L}_{2}\left( \mathbb{C}\otimes \mathbb{O},%
\mathbb{H}\right) =\mathfrak{isom}\left( \mathbb{T}P_{III}^{1}\right)
\notag \\
&:=&\mathfrak{der}\left( \mathbb{C}\otimes \mathbb{O}\right) \oplus
\mathfrak{der}\left( J_{2}(\mathbb{H})\right) \oplus \left( \mathbb{C}%
\otimes \mathbb{O}\right) ^{\prime }\otimes J_{2}^{\prime }(\mathbb{H})
\notag \\
&=&\mathfrak{g}_{2}\oplus \mathfrak{so}_{5}\oplus \left( 2\cdot \mathbf{7}+%
\mathbf{1},\mathbf{5}\right) ,  \label{isom3}
\end{eqnarray}%
because%
\begin{eqnarray}
\mathfrak{der}\left( J_{2}(\mathbb{H})\right)  &\simeq &\mathfrak{so}_{5},
\notag \\
J_{2}^{\prime }(\mathbb{H}) &\simeq &\mathbf{5},
\end{eqnarray}%
and%
\begin{eqnarray}
\mathfrak{der}\left( \mathbb{C}\otimes \mathbb{O}\right)  &\simeq &\mathfrak{%
der}\left( \mathbb{O}\right) \simeq \mathfrak{g}_{2}, \\
\left( \mathbb{C}\otimes \mathbb{O}\right) ^{\prime } &\simeq &2\cdot
\mathbf{7}+\mathbf{1}.
\end{eqnarray}%
The Lie algebra $\mathfrak{isom}\left( \mathbb{T}P_{III}^{1}\right) $ has
dimension $14+10+75=99$.
\end{description}

}

\subsection{Three Dixon lines}

A Dixon-Rosenfeld projective line $\mathbb{T}P^{1}$ can be realized as an
homogeneous space of dimension dim$_{\mathbb{R}}\mathbb{T}P^{1}=\text{dim}_{%
\mathbb{R}}\mathbb{T}=64$, whose
\updatered{
corresponding Lie algebra generators
}
$\mathfrak{Lie}\left( \mathbb{T}P^{1}\right) $ relate to the isometry and
isotropy Lie algebras as follows:
\begin{equation}
\mathfrak{Lie}\left( \mathbb{T}P^{1}\right) \simeq \mathfrak{isom}\left(
\mathbb{T}P^{1}\right) \ominus \mathfrak{isot}\left( \mathbb{T}P^{1}\right) ,
\end{equation}%
and whose tangent space $T\left( \mathbb{T}P^{1}\right) $ carries a $%
\mathfrak{isot}\left( \mathbb{T}P^{1}\right) $-covariant realization of $%
\mathbb{T}$ itself.

\updatered{

\begin{description}
\item[$\mathbf{I}$] By iterated branchings of $\mathfrak{isom}\left( \mathbb{%
T}P_{I}^{1}\right) $ given by (\ref{isom1}), one obtains%
\begin{eqnarray}
\mathfrak{isom}\left( \mathbb{T}P_{I}^{1}\right) &\simeq &\mathfrak{su}%
_{2}\oplus \mathfrak{so}_{8}\oplus \left( \mathbf{1}\oplus 2\cdot \mathbf{3},%
\mathbf{8}_{v}+\mathbf{1}\right) \oplus \left( \mathbf{1},\mathbf{8}%
_{v}\right)  \notag \\
&=&\mathfrak{su}_{2}\oplus \mathfrak{so}_{7}\oplus \left( 2\cdot \mathbf{3},%
\mathbf{7+}2\cdot \mathbf{1}\right) \oplus 3\cdot \left( \mathbf{1},\mathbf{%
7+1}\right)  \notag \\
&=&\mathfrak{su}_{2}\oplus \mathfrak{g}_{2}\oplus \left( 2\cdot \mathbf{3},%
\mathbf{7+}2\cdot \mathbf{1}\right) \oplus \left( \mathbf{1},4\cdot \mathbf{%
7+}3\cdot \mathbf{1}\right)  \notag \\
&=&\mathfrak{su}_{2}\oplus \mathfrak{g}_{2}\oplus \left( 2\cdot \mathbf{3},%
\mathbf{7+1}\right) \oplus \left( 2\cdot \mathbf{3},\mathbf{1}\right) \oplus
\left( 2\cdot \mathbf{1},\mathbf{7+1}\right) \oplus \left( \mathbf{1},2\cdot
\mathbf{7+1}\right) \\
&=&:\mathfrak{isot}\left( \mathbb{T}P_{I}^{1}\right) \oplus \mathfrak{c}%
\left( \mathbb{T}P_{I}^{1}\right) ,  \label{res1}
\end{eqnarray}%
thus implying that%
\begin{eqnarray}
\mathfrak{isot}\left( \mathbb{T}P_{I}^{1}\right) &:=&\mathfrak{su}_{2}\oplus
\mathfrak{g}_{2}\oplus \left( \mathbf{1}+2\cdot \mathbf{3},\mathbf{1}\right)
\oplus 2\cdot \left( \mathbf{1},\mathbf{7}\right)  \notag \\
&=&\mathfrak{su}_{2}\oplus \mathfrak{so}_{7}\oplus \left( \mathbf{1}+2\cdot
\mathbf{3},\mathbf{1}\right) \oplus \left( \mathbf{1},\mathbf{7}\right)
\notag \\
&=&\mathfrak{su}_{2}\oplus \mathfrak{so}_{8}\oplus \left( \mathbf{1}+2\cdot
\mathbf{3},\mathbf{1}\right) ,  \label{isot1} \\
\mathfrak{c}\left( \mathbb{T}P_{I}^{1}\right) &\simeq &T\left( \mathbb{T}%
P_{I}^{1}\right) \overset{\text{(\ref{TI})-(\ref{TI-2})}}{\simeq }2\cdot
\left( \mathbf{1}+\mathbf{3},\mathbf{8}_{v}\right) ~\text{of~}\mathfrak{su}%
_{2}\oplus \mathfrak{so}_{8}.
\end{eqnarray}%
Therefore, one obtains the following (non-symmetric) presentation of the
Dixon projective line $\mathbb{T}P_{I}^{1}$ as a homogeneous space:
\begin{equation}
\mathbb{T}P_{I}^{1}\simeq \frac{SO_{9}\times SU_{2}\ltimes \left( \mathbf{9},%
\mathbf{1}+2\cdot \mathbf{3}\right) }{SO_{8}\times SU_{2}\ltimes \left(
2\cdot \left( \mathbf{1},\mathbf{3}\right) +\left( \mathbf{1},\mathbf{1}%
\right) \right) },  \label{DR1}
\end{equation}%
with%
\begin{equation}
\text{dim}\left( \mathbb{T}P_{I}^{1}\right) =64=\text{dim}\mathbb{T}.
\end{equation}%
The coset (\ref{DR1}) is not symmetric, because it can be checked that%
\begin{equation}
\left[ \mathfrak{c}\left( \mathbb{T}P_{I}^{1}\right) ,\mathfrak{c}\left(
\mathbb{T}P_{I}^{1}\right) \right] \simeq 4\cdot \left( \mathbf{1}+\mathbf{3}%
,\mathbf{8}_{v}\right) \otimes _{a}\left( \mathbf{1}+\mathbf{3},\mathbf{8}%
_{v}\right) \nsubseteq \mathfrak{isot}\left( \mathbb{T}P_{I}^{1}\right) ,
\end{equation}%
where subscript \textquotedblleft $a$\textquotedblright\ denotes
anti-symmetrization of the tensor product throughout.

\item[$\mathbf{II}$] From (\ref{a_II-2}) and (\ref{a_II-enh}), the
\textquotedblleft minimally\textquotedblright\ enhanced $\mathfrak{isom}%
\left( \mathbb{T}P_{II}^{1}\right) _{\text{enh}.}$ reads%
\begin{equation}
\mathfrak{isom}\left( \mathbb{T}P_{II}^{1}\right) _{\text{enh}.}=:\mathfrak{%
isot}\left( \mathbb{T}P_{II}^{1}\right) \oplus \mathfrak{c}\left( \mathbb{T}%
P_{II}^{1}\right) ,
\end{equation}%
with%
\begin{eqnarray}
\mathfrak{isot}\left( \mathbb{T}P_{II}^{1}\right):= &&\mathfrak{so}%
_{7}\oplus \mathfrak{su}_{2}\oplus \left( \mathbf{7},\mathbf{3}\right)
\oplus \left( \mathbf{7},\mathbf{1}\right) \oplus \left( \mathbf{1},\mathbf{5%
}\right) ,  \label{isot2} \\
\mathfrak{c}\left( \mathbb{T}P_{II}^{1}\right)  &\simeq &T\left( \mathbb{T}%
P_{II}^{1}\right) \overset{\text{(\ref{pre-TI-2})}}{\simeq }T\left( \mathbb{T%
}P_{I}^{1}\right) .  \label{jj}
\end{eqnarray}%
Thence, one obtains the following (non-symmetric) presentation of the Dixon
projective line $\mathbb{T}P_{II}^{1}$ as a homogeneous space:%
\begin{equation}
\mathbb{T}P_{II}^{1}\simeq \frac{SO_{7}\times SO_{6}\ltimes \left( 3\cdot
\left( \mathbf{7},\mathbf{4}\right) \oplus \left( \mathbf{1},\mathbf{1}%
\right) \right) }{SO_{7}\times SU_{2}\ltimes \left( \left( \mathbf{7},%
\mathbf{3}+\mathbf{1}\right) \oplus \left( \mathbf{1},\mathbf{5}\right)
\right) },  \label{DR2}
\end{equation}%
once again with%
\begin{equation}
\text{dim}\left( \mathbb{T}P_{II}^{1}\right) =64=\text{dim}\mathbb{T}.
\end{equation}%
The coset (\ref{DR2}) is not symmetric, because it can be checked that%
\begin{equation}
\left[ \mathfrak{c}\left( \mathbb{T}P_{II}^{1}\right) ,\mathfrak{c}\left(
\mathbb{T}P_{II}^{1}\right) \right] \simeq 4\cdot \left( \mathbf{1}+\mathbf{3%
},\mathbf{7}+\mathbf{1}\right) \otimes _{a}\left( \mathbf{1}+\mathbf{3},%
\mathbf{7}+\mathbf{1}\right) \nsubseteq \mathfrak{isot}\left( \mathbb{T}%
P_{II}^{1}\right) .
\end{equation}

\item[$\mathbf{III}$] By iterated branchings of $\mathfrak{isom}\left(
\mathbb{T}P_{III}^{1}\right) $, given by (\ref{isom3}), one obtains%
\begin{eqnarray}
\mathfrak{isom}\left( \mathbb{T}P_{III}^{1}\right)  &=&\mathfrak{g}%
_{2}\oplus \mathfrak{su}_{2}\oplus \mathfrak{su}_{2}\oplus \left( \mathbf{1},%
\mathbf{2,2}\right) \oplus \left( 2\cdot \mathbf{7}+\mathbf{1},\mathbf{2,2}%
\right) \oplus \left( 2\cdot \mathbf{7}+\mathbf{1},\mathbf{1,1}\right)
\notag \\
&=&\mathfrak{g}_{2}\oplus \mathfrak{su}_{2,d}\oplus 2\cdot \left( \mathbf{1},%
\mathbf{3}\right) \oplus \left( \mathbf{1},\mathbf{1}\right) \oplus \left(
2\cdot \mathbf{7}+\mathbf{1},\mathbf{3}\right) \oplus 2\cdot \left( 2\cdot
\mathbf{7}+\mathbf{1},\mathbf{1}\right)   \notag \\
&\simeq &:\mathfrak{isot}\left( \mathbb{T}P_{III}^{1}\right) \oplus
\mathfrak{c}\left( \mathbb{T}P_{III}^{1}\right) ,  \label{jjj}
\end{eqnarray}%
thus implying that%
\begin{eqnarray}
\mathfrak{isot}\left( \mathbb{T}P_{III}^{1}\right):\simeq  &&\mathfrak{g}%
_{2}\oplus \mathfrak{su}_{2}\oplus \left( 2\cdot \mathbf{7}+\mathbf{1},%
\mathbf{1}\right) \oplus \left( \mathbf{1},\mathbf{3}\right) ,  \label{isot3}
\\
\mathfrak{c}\left( \mathbb{T}P_{III}^{1}\right)  &\simeq &T\left( \mathbb{T}%
P_{III}^{1}\right) \overset{\text{(\ref{TI}), (\ref{jj})}}{\simeq }T\left(
\mathbb{T}P_{I}^{1}\right) ,  \label{jjjj}
\end{eqnarray}%
and therefore leading to the following (non-symmetric) presentation of the
Dixon projective line $\mathbb{T}P_{III}^{1}$ as a homogeneous space:
\begin{equation}
\mathbb{T}P_{III}^{1}\simeq \frac{G_{2}\times SO_{5}\ltimes \left( 2\cdot
\mathbf{7}+\mathbf{1},\mathbf{5}\right) }{G_{2}\times SU_{2}\ltimes \left(
\left( 2\cdot \mathbf{7}+\mathbf{1},\mathbf{1}\right) +\left( \mathbf{1},%
\mathbf{3}\right) \right) },  \label{DR3}
\end{equation}%
once again with%
\begin{equation}
\text{dim}\left( \mathbb{T}P_{III}^{1}\right) =64=\text{dim}\mathbb{T}.
\end{equation}%
The coset (\ref{DR3}) is not symmetric, because it can be checked that%
\begin{equation}
\left[ \mathfrak{c}\left( \mathbb{T}P_{III}^{1}\right) ,\mathfrak{c}\left(
\mathbb{T}P_{III}^{1}\right) \right] \simeq 4\cdot \left( \mathbf{7}+\mathbf{%
1},\mathbf{3}+\mathbf{1}\right) \otimes _{a}\left( \mathbf{7}+\mathbf{1},%
\mathbf{3}+\mathbf{1}\right) \nsubseteq \mathfrak{isot}\left( \mathbb{T}%
P_{III}^{1}\right) .
\end{equation}

\item[\textbf{Remark}] The above analysis yields the following isometry
algebras:%
\begin{eqnarray}
\underset{\text{dim}_{\mathbb{R}}=102}{\mathfrak{isom}\left( \mathbb{T}%
P_{I}^{1}\right) } &\simeq &\mathfrak{g}_{2}\oplus \mathfrak{su}_{2}\oplus
2\cdot \left( \mathbf{7},\mathbf{3}\right) \oplus 4\cdot \left( \mathbf{1},%
\mathbf{3}\right) \oplus 4\cdot \left( \mathbf{7},\mathbf{1}\right) \oplus
3\cdot \left( \mathbf{1},\mathbf{1}\right) , \\
\underset{\text{dim}_{\mathbb{R}}=121}{\mathfrak{isom}\left( \mathbb{T}%
P_{II}^{1}\right) _{\text{enh}.}} &\simeq &\mathfrak{g}_{2}\oplus \mathfrak{%
su}_{2}\oplus 3\cdot \left( \mathbf{7},\mathbf{3}\right) \oplus 2\cdot
\left( \mathbf{1},\mathbf{3}\right) \oplus 4\cdot \left( \mathbf{7},\mathbf{1%
}\right) \oplus 2\cdot \left( \mathbf{1},\mathbf{1}\right) \oplus \left(
\mathbf{1},\mathbf{5}\right) , \\
\underset{\text{dim}_{\mathbb{R}}=99}{\mathfrak{isom}\left( \mathbb{T}%
P_{III}^{1}\right) } &=&\mathfrak{g}_{2}\oplus \mathfrak{su}_{2}\oplus
2\cdot \left( \mathbf{7},\mathbf{3}\right) \oplus 3\cdot \left( \mathbf{1},%
\mathbf{3}\right) \oplus 4\cdot \left( \mathbf{7},\mathbf{1}\right) \oplus
3\cdot \left( \mathbf{1},\mathbf{1}\right) ,
\end{eqnarray}%
as well as the following isotropy algebras:%
\begin{eqnarray}
\underset{\text{dim}_{\mathbb{R}}=38}{\mathfrak{isot}\left( \mathbb{T}%
P_{I}^{1}\right) } &\simeq &\mathfrak{g}_{2}\oplus \mathfrak{su}_{2}\oplus
2\cdot \left( \mathbf{1,3}\right) \oplus 2\cdot \left( \mathbf{7},\mathbf{1}%
\right) \oplus \left( \mathbf{1,1}\right) , \\
\underset{\text{dim}_{\mathbb{R}}=57}{\mathfrak{isot}\left( \mathbb{T}%
P_{II}^{1}\right) } &\simeq &\mathfrak{g}_{2}\oplus \mathfrak{su}_{2}\oplus
\left( \mathbf{7},\mathbf{3}\right) \oplus 2\cdot \left( \mathbf{7},\mathbf{1%
}\right) \oplus \left( \mathbf{1},\mathbf{5}\right) , \\
\underset{\text{dim}_{\mathbb{R}}=35}{\mathfrak{isot}\left( \mathbb{T}%
P_{III}^{1}\right) } &=&\mathfrak{g}_{2}\oplus \mathfrak{su}_{2}\oplus
\left( \mathbf{1},\mathbf{3}\right) \oplus 2\cdot \left( \mathbf{7},\mathbf{1%
}\right) \oplus \left( \mathbf{1},\mathbf{1}\right) ,
\end{eqnarray}%
which all imply the same coset Lie algebra locally on the tangent space,
providing a manifestly $\left( \mathfrak{g}_{2}\oplus \mathfrak{su}%
_{2}\right) $-covariant (or, equivalently, $\left( \mathfrak{so}_{7}\oplus
\mathfrak{su}_{2}\right) $-covariant) realization of the Dixon algebra $%
\mathbb{T}$, as given by (\ref{TI}) and (\ref{pre-TI-2}):%
\begin{gather}
\mathfrak{c}\left( \mathbb{T}P_{I}^{1}\right) \simeq \mathfrak{c}\left(
\mathbb{T}P_{I}^{1}\right) \simeq \mathfrak{c}\left( \mathbb{T}%
P_{I}^{1}\right) \simeq \left( \mathbf{7}+\mathbf{1},2\cdot \mathbf{3}%
+2\cdot \mathbf{1}\right) , \\
\Updownarrow   \notag \\
T\left( \mathbb{T}P_{I}^{1}\right) \simeq T\left( \mathbb{T}P_{I}^{1}\right)
\simeq T\left( \mathbb{T}P_{I}^{1}\right) \simeq \left( \mathbf{7}+\mathbf{1}%
,2\cdot \mathbf{3}+2\cdot \mathbf{1}\right) .
\end{gather}%
Thus, the three Dixon-Rosenfeld projective lines $\mathbb{T}P_{I}^{1}$, $%
\mathbb{T}P_{II}^{1}$ and $\mathbb{T}P_{III}^{1}$ have slightly different
isometry and isotropy Lie algebras; from the formul\ae\ above, it follows
that%
\begin{eqnarray}
\mathfrak{isom}\left( \mathbb{T}P_{I}^{1}\right)  &\simeq &\mathfrak{isom}%
\left( \mathbb{T}P_{III}^{1}\right) \oplus \left( \mathbf{1},\mathbf{3}%
\right) ; \\
\mathfrak{isom}\left( \mathbb{T}P_{II}^{1}\right) _{\text{enh}.} &\nsubseteq
&\nsupseteq \mathfrak{isom}\left( \mathbb{T}P_{III}^{1}\right) ; \\
\mathfrak{isom}\left( \mathbb{T}P_{II}^{1}\right) _{\text{enh}.}\cap
\mathfrak{isom}\left( \mathbb{T}P_{III}^{1}\right)  &\simeq &\mathfrak{g}%
_{2}\oplus \mathfrak{su}_{2}\oplus 2\cdot \left( \mathbf{7},\mathbf{3}%
\right) \oplus 2\cdot \left( \mathbf{1},\mathbf{3}\right) \nonumber \\
&& \oplus 4\cdot
\left( \mathbf{7},\mathbf{1}\right) \oplus 2\cdot \left( \mathbf{1},\mathbf{1%
}\right) ,
\end{eqnarray}%
and%
\begin{eqnarray}
\mathfrak{isot}\left( \mathbb{T}P_{I}^{1}\right)  &\simeq &\mathfrak{isot}%
\left( \mathbb{T}P_{III}^{1}\right) \oplus \left( \mathbf{1},\mathbf{3}%
\right) ; \\
\mathfrak{isot}\left( \mathbb{T}P_{II}^{1}\right) _{\text{enh}.} &\nsubseteq
&\nsupseteq \mathfrak{isot}\left( \mathbb{T}P_{III}^{1}\right) ; \\
\mathfrak{isot}\left( \mathbb{T}P_{II}^{1}\right) _{\text{enh}.}\cap
\mathfrak{isot}\left( \mathbb{T}P_{III}^{1}\right)  &\simeq &\mathfrak{g}%
_{2}\oplus \mathfrak{su}_{2}\oplus 2\cdot \left( \mathbf{7},\mathbf{1}%
\right) .
\end{eqnarray}%
However, the set of generators of the isometry Lie group whose non-linear
realization gives rise to the Dixon-Rosenfeld projective line is the same
for $\mathbb{T}P_{I}^{1}$, $\mathbb{T}P_{II}^{1}$ and $\mathbb{T}P_{III}^{1}$%
; since such a set of generators also provide a local realization of the
tangent space, one can conclude that $\mathbb{T}P_{I}^{1}$, $\mathbb{T}%
P_{II}^{1}$ and $\mathbb{T}P_{III}^{1}$ are \textit{locally} \textit{%
isomorphic} as homogeneous (non-symmetric) spaces.
\end{description}

}

\section{Relationship with octonionic Rosenfeld lines}

\label{OctoRosenfeld} It is interesting to point out the relationship
between the Dixon-Rosenfeld lines and the other octonionic Rosenfeld lines,
whose definition can be found in from an historical point of view in \cite%
{Rosenfeld Group2-1,Rosenf98} and in a more rigorous definition in \cite%
{MCCAI Magic Rosenfeld}. Let us just recall here the homogeneous space
realization of Rosenfeld lines over $\mathbb{A}\otimes\mathbb{O}$, with
$\mathbb{A}=\mathbb{R},\mathbb{C},\mathbb{H},\mathbb{O}$ (see \cite%
{Rosenf98,Rosenfeld Group2-1} and \cite{MCCAI Magic Rosenfeld,Corr Notes
Octo,RealF}), i.e.~for the \emph{octonionic projective line} $\left(\mathbb{R%
}\otimes\mathbb{O}\right)P^{1}$, the \emph{bioctonionic Rosenfeld line} $%
\left(\mathbb{C}\otimes\mathbb{O}\right)P^{1}$, the \emph{quateroctonionic
Rosenfeld line}$\left(\mathbb{H}\otimes\mathbb{O}\right)P^{1}$ and, finally,
for the \emph{octooctonionic Rosenfeld line} $\left(\mathbb{O}\otimes\mathbb{%
O}\right)P^{1}$:
\begin{align*}
\left(\mathbb{R}\otimes\mathbb{O}\right)P^{1} & =\frac{SO_{9}}{SO_{8}}\simeq
S^{8}, \\
\left(\mathbb{C}\otimes\mathbb{O}\right)P^{1} & =\frac{SO_{10}\times U_{1}}{%
SO_{8}\times U_{1}\times U_{1}}\simeq\frac{SO_{10}}{SO_{8}\times U_{1}}, \\
\left(\mathbb{H}\otimes\mathbb{O}\right)P^{1} & =\frac{SO_{12}\times Sp_{2}}{%
SO_{8}\times SU_{2}\times SU_{2}\times Sp_{2}}\simeq\frac{SO_{12}}{%
SO_{8}\times SU_{2}\times SU_{2}}, \\
\left(\mathbb{O}\otimes\mathbb{O}\right)P^{1} & =\frac{SO_{16}}{SO_{8}\times
SO_{8}},
\end{align*}
from which it consistently follows that
\begin{equation}
\begin{array}{ccc}
T\left(\mathbb{O}P^{1}\right) & \simeq & \mathbf{8}_{v}~\text{of~}\mathfrak{%
so}_{8} \\
& \simeq & \mathbf{7}+\mathbf{1}~\text{of~}\mathfrak{so}_{7} \\
& \simeq & \mathbf{7}+\mathbf{1}~\text{of~}\mathfrak{g}_{2},%
\end{array}%
\end{equation}
\begin{equation}
\begin{array}{ccc}
T\left(\left(\mathbb{C}\otimes\mathbb{O}\right)P^{1}\right) & \simeq &
\mathbf{8}_{v,+}\oplus\mathbf{8}_{v,-}~\text{of~}\mathfrak{so}_{8}\oplus%
\mathfrak{u}_{1} \\
& \simeq & 2\cdot\left(\mathbf{7}+\mathbf{1}\right)~\text{of~}\mathfrak{so}%
_{7} \\
& \simeq & 2\cdot\left(\mathbf{7}+\mathbf{1}\right)~\text{of~}\mathfrak{g}%
_{2},%
\end{array}%
\end{equation}
\begin{equation}
\begin{array}{ccc}
T\left(\left(\mathbb{H}\otimes\mathbb{O}\right)P^{1}\right) & \simeq & \left(%
\mathbf{8}_{v},\mathbf{2},\mathbf{2}\right)~\text{of~}\mathfrak{so}_{8}\oplus%
\mathfrak{su}_{2}\oplus\mathfrak{su}_{2} \\
& \simeq & \left(\mathbf{8}_{v},\mathbf{3}+\mathbf{1}\right)~\text{of~}%
\mathfrak{so}_{8}\oplus\mathfrak{su}_{2,d} \\
& \simeq & \left(\mathbf{7+1},\mathbf{3}+\mathbf{1}\right)~\text{of~}%
\mathfrak{g}_{2}\oplus\mathfrak{su}_{2},%
\end{array}%
\end{equation}
\begin{equation}
\begin{array}{ccc}
T\left(\left(\mathbb{O}\otimes\mathbb{O}\right)P^{1}\right) & \simeq & \left(%
\mathbf{8}_{v},\mathbf{8}_{v}\right)~\text{of~}\mathfrak{so}_{8}\oplus%
\mathfrak{so}_{8} \\
& \simeq & \left(\mathbf{7+1},\mathbf{7+1}\right)~\text{of~}\mathfrak{so}%
_{7}\oplus\mathfrak{so}_{7} \\
& \simeq & \left(\mathbf{7+1},\mathbf{7+1}\right)~\text{of~}\mathfrak{g}%
_{2}\oplus\mathfrak{g}_{2}.%
\end{array}%
\end{equation}
which illustrates how the tangent spaces of octonionic projective lines
generally carry an enhancement of the symmetry with respect to the Lie
algebra $\mathfrak{der}\left(\mathbb{A}\otimes\mathbb{O}\right)\simeq%
\mathfrak{der}\left(\mathbb{A}\right)\oplus\mathfrak{g}_{2}$.

Geometrically, the octonionic projective lines $\left(\mathbb{A}\otimes%
\mathbb{O}\right)P^{1}$ can be regarded as $\mathbb{A}\otimes\mathbb{O}$
together with a point at infinity, and thus as a $8$dim$_{\mathbb{R}}\mathbb{%
A}$-sphere, namely as a maximal totally geodesic sphere in the corresponding
octonionic Rosenfeld projective plane $\left(\mathbb{A}\otimes\mathbb{O}%
\right)P^{2}$ \cite{Rosenf98}. In the case $\mathbb{A}=\mathbb{R}$, such a
``spherical characterization''\ of octonionic projective lines is well
known, whereas for the other cases (the ``genuinely Rosenfeld'' ones) it is
less trivial (see e.g.~\cite{Nagano}).

We can now study the relations among the Dixon-Rosenfeld lines discussed
above and the octonionic Rosenfeld lines.
\updatered{
Of course,
\begin{equation}
\text{dim}\left( \left( \mathbb{O}\otimes \mathbb{O}\right) P^{1}\right) =%
\text{dim}\left( \mathbb{T}P_{I}^{1}\right) =\text{dim}\left( \mathbb{T}%
P_{II}^{1}\right) =\text{dim}\left( \mathbb{T}P_{III}^{1}\right) =64.
\end{equation}%
By recalling (\ref{TI}) and considering the $\left( \mathfrak{g}_{2}\oplus
\mathfrak{g}_{2}\right) $-covariant representation of the tensor algebra%
\begin{equation}
\mathbb{O}\otimes \mathbb{O}\simeq \left( \mathbf{7}+\mathbf{1},\mathbf{7}+%
\mathbf{1}\right) ~\text{of~}\underset{\mathfrak{der}\left( \mathbb{O}%
\right) \oplus \mathfrak{der}\left( \mathbb{O}\right) }{\mathfrak{g}%
_{2}\oplus \mathfrak{g}_{2}},  \label{OO}
\end{equation}%
one observes that, when restricting the first (or, equivalently, the second)
$\mathfrak{g}_{2}$ to a $\mathfrak{su}_{2}$ subalgebra defined by (\ref{jh})
(or, equivalently, by (\ref{jhh})), the irrepr. $\mathbf{7}$ of $\mathfrak{g}%
_{2}$ breaks\ into $2\cdot \mathbf{3}+\mathbf{1}$ of $\mathfrak{su}_{2}$,
and therefore it holds that%
\begin{equation}
\mathbb{O}\otimes \mathbb{O}\simeq \underset{\mathfrak{g}_{2}\oplus
\mathfrak{g}_{2}\simeq \mathfrak{der}\left( \mathbb{O}\right) \oplus
\mathfrak{der}\left( \mathbb{O}\right) }{\left( \mathbf{7}+\mathbf{1},%
\mathbf{7}+\mathbf{1}\right) }\overset{\mathfrak{g}_{2}\rightarrow \mathfrak{%
su}_{2}}{\simeq }\underset{\mathfrak{su}_{2}\oplus \mathfrak{g}_{2}\simeq
\mathfrak{der}\left( \mathbb{C}\otimes \mathbb{H}\right) \oplus \mathfrak{der%
}\left( \mathbb{O}\right) }{2\cdot \left( \mathbf{3}+\mathbf{1},\mathbf{7}+%
\mathbf{1}\right) }\overset{\text{(\ref{TI})}}{\simeq }\mathbb{T}.
\label{still}
\end{equation}%
In other words, as resulting from the treatment below, $\mathbb{O}\otimes
\mathbb{O}$ and $\mathbb{T}$ are isomorphic as vector spaces (but not as
algebras), with $\mathfrak{der}\left( \mathbb{O}\otimes \mathbb{O}\right)
\supsetneq \mathfrak{der}\left( \mathbb{T}\right) $: thus, octo-octonions
have a larger derivation algebra than the Dixon algebra, with an
enhancement/restriction expressed by (\ref{jh}) or, equivalently, by (\ref%
{jhh}).

\begin{description}
\item[$\mathbf{I}$] From (\ref{isom1}) and (\ref{isot1}), one respectively
obtains%
\begin{eqnarray}
\mathfrak{isom}\left( \left( \mathbb{O}\otimes \mathbb{O}\right)
P^{1}\right)  &\simeq &\mathfrak{so}_{16}=\mathfrak{so}_{9}\oplus \mathfrak{%
so}_{7}\oplus \left( \mathbf{9},\mathbf{7}\right) =\mathfrak{so}_{9}\oplus
\mathfrak{g}_{2}\oplus \left( \mathbf{9},\mathbf{7}\right) \oplus \left(
\mathbf{1},\mathbf{7}\right)   \notag \\
&=&\overset{\text{(\ref{jh})~or~(\ref{jhh})}}{...}=\mathfrak{so}_{9}\oplus
\mathfrak{su}_{2}\oplus \left( \mathbf{9,1}+2\cdot \mathbf{3}\right) \oplus
\left( \mathbf{1,}2\cdot \mathbf{3}+\mathbf{1}\right) \oplus 2\cdot \left(
\mathbf{1},\mathbf{3}\right) \oplus \left( \mathbf{1},\mathbf{5}\right)
\notag \\
&=&\mathfrak{so}_{8}\oplus \mathfrak{su}_{2}\oplus \left( \mathbf{8}_{v}+%
\mathbf{1,1}+2\cdot \mathbf{3}\right) \oplus \left( \mathbf{1,}2\cdot
\mathbf{3}+\mathbf{1}\right) \oplus 2\cdot \left( \mathbf{1},\mathbf{3}%
\right) \oplus \left( \mathbf{1},\mathbf{5}\right) \oplus \left( \mathbf{8}%
_{v},\mathbf{1}\right)   \notag \\
&=&\mathfrak{so}_{7}\oplus \mathfrak{su}_{2}\oplus \left( \mathbf{7}+2\cdot
\mathbf{1,1}+2\cdot \mathbf{3}\right) \oplus \left( \mathbf{1,}2\cdot
\mathbf{3}+\mathbf{1}\right)   \notag \\
&&\oplus 2\cdot \left( \mathbf{1},\mathbf{3}\right) \oplus \left( \mathbf{1},%
\mathbf{5}\right) \oplus \left( 2\cdot \mathbf{7+1},\mathbf{1}\right)
\notag \\
&=&\mathfrak{so}_{7}\oplus \mathfrak{su}_{2}\oplus 2\cdot \left( \mathbf{7,3}%
\right) \oplus 8\cdot \left( \mathbf{1,3}\right) \oplus 3\cdot \left(
\mathbf{7,1}\right) \oplus 4\cdot \left( \mathbf{1,1}\right) \oplus \left(
\mathbf{1},\mathbf{5}\right)   \notag \\
&\simeq &\mathfrak{isom}\left( \mathbb{T}P_{I}^{1}\right) \oplus 4\cdot
\left( \mathbf{1,3}\right) \oplus \left( \mathbf{1,1}\right) \oplus \left(
\mathbf{1},\mathbf{5}\right) ,
\end{eqnarray}%
and%
\begin{eqnarray}
\mathfrak{isot}\left( \left( \mathbb{O}\otimes \mathbb{O}\right)
P^{1}\right)  &\simeq &\mathfrak{so}_{8}\oplus \mathfrak{so}_{8}=\mathfrak{so%
}_{8}\oplus \mathfrak{so}_{7}\oplus \left( \mathbf{1},\mathbf{7}\right) =%
\mathfrak{so}_{8}\oplus \mathfrak{g}_{2}\oplus 2\cdot \left( \mathbf{1},%
\mathbf{7}\right)   \notag \\
&=&\overset{\text{(\ref{jh})~or~(\ref{jhh})}}{...}=\mathfrak{so}_{8}\oplus
\mathfrak{su}_{2}\oplus 2\cdot \left( \mathbf{1},\mathbf{1}+2\cdot \mathbf{3}%
\right) \oplus 2\cdot \left( \mathbf{1},\mathbf{3}\right) \oplus \left(
\mathbf{1},\mathbf{5}\right)   \notag \\
&=&\mathfrak{so}_{7}\oplus \mathfrak{su}_{2}\oplus 2\cdot \left( \mathbf{1},%
\mathbf{1}\right) \oplus 6\cdot \left( \mathbf{1},\mathbf{3}\right) \oplus
\left( \mathbf{1},\mathbf{5}\right) \oplus \left( \mathbf{7},\mathbf{1}%
\right)   \notag \\
&\simeq &\mathfrak{isot}\left( \mathbb{T}P_{I}^{1}\right) \oplus 4\cdot
\left( \mathbf{1,3}\right) \oplus \left( \mathbf{1,1}\right) \oplus \left(
\mathbf{1},\mathbf{5}\right) .
\end{eqnarray}%
Thus, it holds that%
\begin{eqnarray}
\mathfrak{isom}\left( \mathbb{T}P_{I}^{1}\right)  &\subsetneq &\mathfrak{isom%
}\left( \left( \mathbb{O}\otimes \mathbb{O}\right) P^{1}\right) , \\
\mathfrak{isot}\left( \mathbb{T}P_{I}^{1}\right)  &\subsetneq &\mathfrak{isot%
}\left( \left( \mathbb{O}\otimes \mathbb{O}\right) P^{1}\right) ,
\end{eqnarray}%
and%
\begin{eqnarray}
\mathfrak{c}\left( \left( \mathbb{O}\otimes \mathbb{O}\right) P^{1}\right)
:\simeq  &&\mathfrak{isom}\left( \left( \mathbb{O}\otimes \mathbb{O}\right)
P^{1}\right) \ominus \mathfrak{isot}\left( \left( \mathbb{O}\otimes \mathbb{O%
}\right) P^{1}\right)   \notag \\
&=&\left( \mathfrak{isom}\left( \mathbb{T}P_{I}^{1}\right) \oplus \left(
\mathbf{5}+4\cdot \mathbf{3}+\mathbf{1,1}\right) \right) \nonumber \\
&& \ominus \left(
\mathfrak{isot}\left( \mathbb{T}P_{I}^{1}\right) \oplus \left( \mathbf{5}%
+4\cdot \mathbf{3}+\mathbf{1,1}\right) \right)   \notag \\
&\simeq &\mathfrak{isom}\left( \mathbb{T}P_{I}^{1}\right) \ominus \mathfrak{%
isot}\left( \mathbb{T}P_{I}^{1}\right) \simeq:\mathfrak{c}\left( \mathbb{T}%
P_{I}^{1}\right) .
\end{eqnarray}

\item[$\mathbf{II}$] Analogously, from (\ref{TTitsEnh}) and (\ref{isot2}), one
respectively obtains%
\begin{eqnarray}
\mathfrak{isom}\left( \left( \mathbb{O}\otimes \mathbb{O}\right)
P^{1}\right)  &\nsubseteq &\nsupseteq \mathfrak{isom}\left( \mathbb{T}%
P_{II}^{1}\right) _{\text{enh.}}; \\
\mathfrak{isom}\left( \left( \mathbb{O}\otimes \mathbb{O}\right)
P^{1}\right) \cap \mathfrak{isom}\left( \mathbb{T}P_{II}^{1}\right) _{\text{%
enh.}} &\simeq &\mathfrak{so}_{7}\oplus \mathfrak{su}_{2}\oplus 2\cdot
\left( \mathbf{7},\mathbf{3}\right) \oplus 2\cdot \left( \mathbf{1},\mathbf{3%
}\right) \nonumber \\
&& \oplus 3\cdot \left( \mathbf{7},\mathbf{1}\right) \oplus 2\cdot
\left( \mathbf{1},\mathbf{1}\right) \oplus \left( \mathbf{1},\mathbf{5}%
\right) ,
\end{eqnarray}%
and%
\begin{eqnarray}
\mathfrak{isot}\left( \left( \mathbb{O}\otimes \mathbb{O}\right)
P^{1}\right)  &\nsubseteq &\nsupseteq \mathfrak{isot}\left( \mathbb{T}%
P_{II}^{1}\right) ; \\
\mathfrak{isot}\left( \left( \mathbb{O}\otimes \mathbb{O}\right)
P^{1}\right) \cap \mathfrak{isot}\left( \mathbb{T}P_{II}^{1}\right)  &\simeq
&\mathfrak{so}_{7}\oplus \mathfrak{su}_{2}\oplus \left( \mathbf{7},\mathbf{1}%
\right) \oplus \left( \mathbf{1},\mathbf{5}\right) .
\end{eqnarray}%
However,%
\begin{eqnarray}
\mathfrak{c}\left( \left( \mathbb{O}\otimes \mathbb{O}\right) P^{1}\right)
:\simeq  &&\mathfrak{isom}\left( \left( \mathbb{O}\otimes \mathbb{O}\right)
P^{1}\right) \ominus \mathfrak{isot}\left( \left( \mathbb{O}\otimes \mathbb{O%
}\right) P^{1}\right)   \notag \\
&\simeq &\mathfrak{isom}\left( \mathbb{T}P_{II}^{1}\right) _{\text{enh.}%
}\ominus \mathfrak{isot}\left( \mathbb{T}P_{II}^{1}\right) \simeq:\mathfrak{%
c}\left( \mathbb{T}P_{II}^{1}\right) .
\end{eqnarray}

\item[$\mathbf{III}$] Again, from (\ref{isom3}) and (\ref{isot3}), one
respectively obtains%
\begin{equation}
\mathfrak{isom}\left( \left( \mathbb{O}\otimes \mathbb{O}\right)
P^{1}\right) \simeq \mathfrak{isom}\left( \mathbb{T}P_{III}^{1}\right)
\oplus 5\cdot \left( \mathbf{1,3}\right) \oplus 2\cdot \left( \mathbf{7,1}%
\right) \oplus \left( \mathbf{1,1}\right) \oplus \left( \mathbf{1},\mathbf{5}%
\right) ,
\end{equation}%
and%
\begin{eqnarray}
\mathfrak{isot}\left( \left( \mathbb{O}\otimes \mathbb{O}\right)
P^{1}\right)  &=&\mathfrak{so}_{7}\oplus \mathfrak{su}_{2}\oplus 2\cdot
\left( \mathbf{1},\mathbf{1}\right) \oplus 6\cdot \left( \mathbf{1},\mathbf{3%
}\right) \oplus \left( \mathbf{1},\mathbf{5}\right) \oplus \left( \mathbf{7},%
\mathbf{1}\right)   \notag \\
&\simeq &\mathfrak{isot}\left( \mathbb{T}P_{III}^{1}\right) \oplus 5\cdot
\left( \mathbf{1,3}\right) \oplus 2\cdot \left( \mathbf{7,1}\right) \oplus
\left( \mathbf{1,1}\right) \oplus \left( \mathbf{1},\mathbf{5}\right) .
\end{eqnarray}%
Thus, it holds that%
\begin{eqnarray}
\mathfrak{isom}\left( \mathbb{T}P_{III}^{1}\right)  &\subsetneq &\mathfrak{%
isom}\left( \left( \mathbb{O}\otimes \mathbb{O}\right) P^{1}\right) , \\
\mathfrak{isot}\left( \mathbb{T}P_{III}^{1}\right)  &\subsetneq &\mathfrak{%
isot}\left( \left( \mathbb{O}\otimes \mathbb{O}\right) P^{1}\right) ,
\end{eqnarray}%
and%
\begin{eqnarray}
\mathfrak{c}\left( \left( \mathbb{O}\otimes \mathbb{O}\right) P^{1}\right)
:\simeq  &&\mathfrak{isom}\left( \left( \mathbb{O}\otimes \mathbb{O}\right)
P^{1}\right) \ominus \mathfrak{isot}\left( \left( \mathbb{O}\otimes \mathbb{O%
}\right) P^{1}\right)   \notag \\
&=&\left( \mathfrak{isom}\left( \mathbb{T}P_{III}^{1}\right) \oplus 5\cdot
\left( \mathbf{1,3}\right) \oplus 2\cdot \left( \mathbf{7,1}\right) \oplus
\left( \mathbf{1,1}\right) \oplus \left( \mathbf{1},\mathbf{5}\right)
\right)   \notag \\
&&\ominus \left( \mathfrak{isot}\left( \mathbb{T}P_{I}^{1}\right) \oplus
5\cdot \left( \mathbf{1,3}\right) \oplus 2\cdot \left( \mathbf{7,1}\right)
\oplus \left( \mathbf{1,1}\right) \oplus \left( \mathbf{1},\mathbf{5}\right)
\right)   \notag \\
&\simeq &\mathfrak{isom}\left( \mathbb{T}P_{III}^{1}\right) \ominus
\mathfrak{isot}\left( \mathbb{T}P_{III}^{1}\right) \simeq:\mathfrak{c}%
\left( \mathbb{T}P_{III}^{1}\right) .
\end{eqnarray}
\end{description}

In other words, the Dixon-Rosenfeld projective lines $\mathbb{T}P_{I}^{1}$
and $\mathbb{T}P_{III}^{1}$ have the isometry resp. isotropy Lie algebra
strictly contained in the isometry resp. isotropy Lie algebra of the
octo-octonionic projective line $\left( \mathbb{O}\otimes \mathbb{O}\right)
P^{1}$, whereas the Dixon-Rosenfeld projective line $\mathbb{T}P_{II}^{1}$
does not contain nor is contained into $\left( \mathbb{O}\otimes \mathbb{O}%
\right) P^{1}$. Nonetheless, as pointed out above, the set of generators of
the isometry Lie group whose non-linear realization gives rise to the
Dixon-Rosenfeld projective line is the same for $\mathbb{T}P_{I}^{1}$, $%
\mathbb{T}P_{II}^{1}$ and $\mathbb{T}P_{III}^{1}$; thus, one can conclude
that all such spaces are \textit{locally} \textit{isomorphic} as homogeneous
spaces:%
\begin{equation}
T\left( \mathbb{T}P_{I}^{1}\right) \simeq T\left( \mathbb{T}%
P_{II}^{1}\right) \simeq T\left( \mathbb{T}P_{III}^{1}\right) \simeq T\left(
\left( \mathbb{O}\otimes \mathbb{O}\right) P^{1}\right) .  \label{isom2}
\end{equation}%
It is interesting to remark that this holds notwithstanding the fact that,
while the three Dixon-Rosenfeld projective lines have \textit{non-symmetric}
presentations, the octo-octonionic Rosenfeld projective line is a \textit{%
symmetric} space.
}

\section{Projective lines over $\mathbb{C}\otimes\mathbb{H}$ via $\mathbb{C}\otimes J_{2}(\mathbb{H})$}

\label{CHphysics}

\subsection{Generalized minimal left ideals of $\mathbb{C}\otimes\mathbb{H}$}


In pursuing the Standard Model physics of $\mathbb{C}\otimes\mathbb{H}\otimes\mathbb{O}$,
Furey started by considering generalized minimal left ideals of $\mathbb{C}\otimes\mathbb{H}$
and demonstrated how scalar, chiral spinors, vector, and 2-form representations
of the Lorentz spacetime group may be identified \cite{Furey2012}.
Given some algebra $\mathfrak{g}$, a (generalized) minimal ideal
$\mathfrak{i}\subset\mathfrak{g}$ is a subalgebra where $m(a,v)\in\mathfrak{i}$
for all $a\in\mathfrak{g}$ and $v\in\mathfrak{i}$ with $m$ as a
(generalized) multiplication. The generalized minimal left ideal that
Furey considered for spinors from $\mathfrak{g}=\mathbb{C}\otimes\mathbb{H}$
is
\begin{equation}
m_{1}(a,v)=v^{\prime}=avP+a^{*}vP^{*}\label{mult1}
\end{equation}
where $P=(1+Ik)/2$ such that $P^{*}=(1-Ik)/2$ are projectors satisfying
$P^{2}=P,P^{*2}=P^{*}$, and $PP^{*}=P^{*}P=0$. The 4-vectors (1-forms)
were found as generalized minimal ideals via the the following generalized
multiplication,
\begin{equation}
m_{2}(a,v)=v^{\prime}=ava^{\dagger},\label{mult2}
\end{equation}
where $a^{\dagger}=\hat{a}^{*}$ is used just for this subsection
when $a\in\mathbb{C}\otimes\mathbb{H}$, with $\hat{}$ and $^{*}$ denoting the quaternionic and complex
conjugate, respectively. The symbol $\dagger$
is used throughout as a Hermitian conjugate of the algebra, but the explicit
mathematical operation will differ depending on the algebra under
consideration. The scalars and field strength
(2-forms) were found as generalized minimal ideals via the generalized
multiplication below,
\begin{equation}
m_{3}(a,v)=v^{\prime}=av\hat{a}.\label{mult3}
\end{equation}

Focusing on the spinors, a Dirac spinor $\psi_{D}$ as an element
of $\mathbb{C}\otimes\mathbb{H}$ is decomposed into left- and right-chiral
(Weyl) spinors $\psi_{L}$ and $\psi_{R}$ as minimal left ideals
with respect to Eq.~\eqref{mult1},
\begin{eqnarray}
\psi_{L} & = & v_{1}=\left(c_{1}+c_{3}j\right)P \nonumber \\
&=&\frac{1}{2}\left(\left(c_{1,1}+c_{1,2}I\right)-\left(c_{3,2}-c_{3,1}I\right)i+\left(c_{3,1}+c_{3,2}I\right)j-\left(c_{1,2}-c_{1,1}I\right)k\right),\nonumber \\
\psi_{R} & = & v_{2}=\left(c_{2}-c_{4}j\right)P^{*}\nonumber \\
&=&\frac{1}{2}\left(\left(c_{2,1}+c_{2,2}I\right)-\left(c_{4,2}-c_{4,1}I\right)i-\left(c_{4,1}+c_{4,2}I\right)j+\left(c_{2,2}-c_{2,1}I\right)k\right),\label{complexcoeff}
\end{eqnarray}
where $c_{i}$ for $i=1,\ldots4$ are complex coefficients $c_{i}=c_{i,1}+c_{i,2}I$.
Since $\mathbb{C}\otimes\mathbb{H}$ is associative, it is straightforward
to verify that $\psi_{L}P=\psi_{L},\psi_{R}P^{*}=\psi_{R}$, and $\psi_{L}P^{*}=\psi_{R}P=0$.
Additionally, the Lorentz transformations can be found as the exponentiation
of linear combinations of vectors and bivectors of $Cl(3)$.

The basis of minimal ideals is less clear with $\mathbb{C}\otimes\mathbb{H}$
and improved with reference to another basis spanned by $\left\{ P,P^{*},jP,\hat{\jmath}P^{*},IP,IP^{*},IjP,I\hat{\jmath}P^{*}\right\} $.
To provide a dictionary of various representations used by Furey for
the spinor minimal ideal bases \cite{Furey2012,Furey2014,Furey2015},
consider
\begin{eqnarray}
P & = & [\uparrow L]=|\uparrow\rangle\langle\uparrow|=\epsilon_{\uparrow\uparrow}=\frac{1+Ik}{2},\nonumber \\
P^{*} & = & [\downarrow R]=|\downarrow\rangle\langle\downarrow|=\epsilon_{\downarrow\downarrow}=\frac{1-Ik}{2},\nonumber \\
jP & = & [\downarrow L]=|\downarrow\rangle\langle\uparrow|=\epsilon_{\downarrow\uparrow}=\frac{j+Ii}{2}=\alpha,\\
\hat{\jmath}P^{*}=-jP^{*} & = & [\uparrow R]=|\uparrow\rangle\langle\downarrow|=\epsilon_{\uparrow\downarrow}=\frac{-j+Ii}{2}=\alpha^{\dagger}.\nonumber
\end{eqnarray}
We found it convenient to confirm that $\psi_{L}$ and $\psi_{R}$
are minimal left ideals in Mathematica when converting to the basis
above (along with the four elements multiplied by $I$ ). The following
anti-commutation relations can be found,
\begin{eqnarray}
\left\{ \alpha,\alpha^{\dagger}\right\}  & = & 1,\nonumber \\
\{\alpha,\alpha\} & = & 0,\\
\left\{ \alpha^{\dagger},\alpha^{\dagger}\right\}  & = & 0.\nonumber
\end{eqnarray}
Note that $Ii$ and $Ij$ act as bases of $\mathbb{C}l(2)$.

\subsection{Generalized minimal left ideals of $\mathbb{C}\otimes J_{2}(\mathbb{H})$}

To build up to projective lines of $\mathbb{C}\otimes\mathbb{H}\otimes\mathbb{O}$,
the physics of spinors for $\mathbb{C}\otimes\mathbb{H}$ are uplifted
to $\mathbb{C}\otimes J_{2}(\mathbb{H})$. The $\mathbb{C}\otimes\mathbb{H}$
spinors are also embedded into $\mathbb{C}\otimes J_{2}(\mathbb{H})$
by placing $\psi_{D}$ in the upper-right component and adding by
its quaternionic Hermitian conjugate to obtain an element of $\mathbb{C}\otimes J_{2}(\mathbb{H})$,
\begin{equation}
\psi_{D}\rightarrow J\left(\psi_{D}\right)\equiv\left(\begin{array}{cc}
0 & \psi_{D}\\
0 & 0
\end{array}\right)+\left(\begin{array}{cc}
0 & \psi_{D}\\
0 & 0
\end{array}\right)^{\dagger}=\left(\begin{array}{cc}
0 & \psi_{D}\\
\hat{\psi}_{D} & 0
\end{array}\right).
\end{equation}
Note that here $\dagger $ denotes matrix transpose and quaternionic
conjugation.

This brings in a complication for generalizing $P$, as $2\times2$
matrices admit two projectors as idempotents, yet $\mathbb{C}\otimes J_{2}(\mathbb{H})$
does not contain $P=(1+Ik)/2$ on any diagonal elements. The action
of $\mathbb{C}\otimes\mathbb{H}$ must occur on the off-diagonals.
Despite not giving projectors, the bases are embedded as follows
\begin{eqnarray}
P & \rightarrow & J_{P}\equiv J(P)=\left(\begin{array}{cc}
0 & P\\
\hat{P} & 0
\end{array}\right)=\frac{1}{2}\left(\begin{array}{cc}
0 & 1+Ik\\
1-Ik & 0
\end{array}\right),\nonumber \\
P^{*} & \rightarrow & J_{P^{*}}\equiv J\left(P^{*}\right)=\left(\begin{array}{cc}
0 & P^{*}\\
P^{\dagger} & 0
\end{array}\right)=\frac{1}{2}\left(\begin{array}{cc}
0 & 1-Ik\\
1+Ik & 0
\end{array}\right),\nonumber \\
jP & \rightarrow & J_{jP}\equiv J(jP)=\left(\begin{array}{cc}
0 & jP\\
-jP & 0
\end{array}\right)=\frac{1}{2}\left(\begin{array}{cc}
0 & j+Ii\\
-j-Ii & 0
\end{array}\right),\label{CHbases}\\
\hat{\jmath}P^{*} & \rightarrow & J_{\hat{\jmath}P^{*}}\equiv J\left(\hat{\jmath}P^{*}\right)=\left(\begin{array}{cc}
0 & \hat{\jmath}P^{*}\\
jP^{*} & 0
\end{array}\right)=\frac{1}{2}\left(\begin{array}{cc}
0 & -j+Ii\\
j-Ii & 0
\end{array}\right).\nonumber
\end{eqnarray}

A new generalized multiplication was identified for spinors as elements
of $\mathbb{C}\otimes J_{2}(\mathbb{H})$ by taking the Jordan product
with two matrices from the right to replace $P$ and $P^{*}$ in Eq.~\eqref{mult1},
\begin{eqnarray}
m_{4}(a,v)&=&2\left[\left((A\circ v)\circ J_{P^{*}}\right)\circ J_{P}-\left((A\circ v)\circ J_{\hat{\jmath}P^{*}}\right)\circ J_{jP}\right. \nonumber \\
&& \left.
+\left((A\circ v)\circ J_{P}\right)\circ J_{P^{*}}-\left((A\circ v)\circ J_{jP}\right)\circ J_{\hat{\jmath}^{*}}\right].\label{CHspinor}
\end{eqnarray}
where $a\in\mathbb{C}\otimes J_{2}(\mathbb{H})$ and $a\circ b=(ab+ba)/2$
is the Jordan product. We verified in Mathematica that $m_{4}(a,v)$
gives spinorial ideals for arbitrary $a\in\mathbb{C}\otimes J_{2}(\mathbb{H})$.
Since $\mathbb{C}\otimes J_{2}(\mathbb{H})$ is larger than the piece
of $\mathbb{C}\otimes\mathbb{H}$ embedded in $\mathbb{C}\otimes J_{2}(\mathbb{H})$,
the existence of such a generalized ideal may hold for the entire
algebra constructed from the Dixon-Rosenfeld line via the Freudenthal-Tits
construction.

For Hermitian and anti-Hermitian vectors, the following generalized
multiplication rule is found,
\begin{equation}
m_{5}(a,v)=(a\circ v)\circ\hat{a}^{*}+a\circ\left(v\circ\hat{a}^{*}\right),\label{CHvector}
\end{equation}
where $m_{5}$ is identified as a Jordan anti-associator. If $a$
is chosen to be a purely off-diagonal element of $\mathbb{C}\otimes J_{2}(\mathbb{H})$,
then $m_{5}$ leads to an element of $\mathfrak{i}$ for $v$ as a
Hermitian or anti-Hermitian vector. If $a$ is chosen as an arbitrary
element of $\mathbb{C}\otimes J_{2}(\mathbb{H})$, then the Hermitian
vector uplifted to $\mathbb{C}\otimes J_{2}(\mathbb{H})$ develops
a purely real diagonal term, while the antiHermitian vector uplifted
develops a purely imaginary diagonal term. It is also anticipated
that diagonals of $\mathbb{C}\otimes J_{2}(\mathbb{H})$ not found
in $\mathbb{C}\otimes\mathbb{H}$ should be purely bosonic, which
motivates a higher-dimensional Hermitian and anti-Hermitian vector
to be found as ideals of $\mathbb{C}\otimes J_{2}(\mathbb{H})$.

For scalars and two-forms, the following generalized multiplication
rule is found with a Jordan anti-associator and slightly different
conjugation,
\begin{equation}
m_{6}(a,v)=(a\circ v)\circ\hat{a}+a\circ(v\circ\hat{a}).\label{CH2form}
\end{equation}
It turns out that the 2-form uplifted to $\mathbb{C}\otimes J_{2}(\mathbb{H})$
is a minimal ideal, while the scalar uplifted must be generalized
to include a complex diagonal.

For concreteness, the left- and right-chiral spinors embedded in $\mathbb{C}\otimes J_{2}(\mathbb{H})$
as minimal ideals of $m_{4}$ in Eq.~\eqref{CHspinor} are
\begin{eqnarray}
J_{\psi_{L}} & = & \left(\begin{array}{cc}
0 & \left(c_{1}+c_{3}j\right)P\\
c_{1}P^{*}-c_{3}P & 0
\end{array}\right)\\
J_{\psi_{R}} & = & \left(\begin{array}{cc}
0 & \left(c_{2}-c_{4}j\right)P^{*}\\
c_{2}P+c_{4}P^{*} & 0
\end{array}\right).\nonumber
\end{eqnarray}
The vectors \updatered{$h^\mu$ and pseudo-vectors $g^\mu$ for $\mu = 0,1,2,3$ represented as elements of $\mathbb{C}\otimes\mathbb{H}$
to be used with Eq.~\eqref{CHvector} are generalized to the following
minimal ideals of $\mathbb{C}\otimes J_{2}(\mathbb{H})$} with diagonal components
\begin{eqnarray}
J_{h} & = & \left(\begin{array}{cc}
h_{4} & h_{0}I+h_{1}i+h_{2}j+h_{3}k\\
h_{0}I-h_{1}i-h_{2}j-h_{3}k & h_{5}
\end{array}\right),\\
J_{g} & = & \left(\begin{array}{cc}
g_{4}I & g_{0}+g_{1}iI+g_{2}jI+g_{3}kI\\
g_{0}-g_{1}iI-g_{2}jI-g_{3}kI & g_5 I
\end{array}\right),\nonumber
\end{eqnarray}
\updatered{where $h_4$, $h_5$, $g_4$, and $g_5$ are scalar degrees of freedom found on the diagonals of the minimal ideals that extend the 4-vector and 4-pseudo-vector.}
The scalars $\phi$ and 2-forms $F$ embedded in $\mathbb{C}\otimes J_{2}(\mathbb{H})$
with Eq.~\eqref{CH2form} are found as minimal ideals when a complex
diagonal is added to the scalars
\begin{eqnarray}
J_{\phi} & = & \left(\begin{array}{ll}
\phi_{3}+\phi_{4}I & \phi_{1}+\phi_{2}I\\
\phi_{1}-\phi_{2}I & \phi_{5}+\phi_{6}I
\end{array}\right), \nonumber \\
J_{F} & = & \left(\begin{array}{cc}
0 & \updatered{J_{F,12}}\\
\updatered{J_{F,21}} & 0
\end{array}\right), \\
\updatered{J_{F,12}} &=& \updatered{F^{32}i+F^{13}j+F^{21}k+F^{01}iI+F^{02}jI+F^{03}kI
}, \nonumber \\
\updatered{J_{F,21}} &=& \updatered{-F^{32}i-F^{13}j-F^{21}k-F^{01}iI-F^{02}jI-F^{03}kI}.\nonumber
\end{eqnarray}

One may anticipate that the vector, spinor, and conjugate spinor representations
can be embedded in the three independent off-diagonal components of
$\mathbb{C}\otimes J_{3}(\mathbb{H})$, but this is left for future
work.

\section{Projective lines over $\mathbb{C}\otimes\mathbb{O}$ via $\mathbb{C}\otimes J_{2}(\mathbb{O})$}

\label{COphysics}

\subsection{Minimal left ideals of $\mathbb{C}l(6)$ via chain algebra $\mathbb{C}\otimes\protect\overleftarrow{\mathbb{O}}$}

To establish our conventions for octonions, we review the complexification
of the octonionic chain algebra applied to raising and lowering operators
for $SU(3)_{c}\times U(1)_{em}$ fermionic charge states \cite{Furey2012,Furey2015}.
For $\mathbb{C}\otimes\mathbb{O}$, we use $I$ and $e_{i}$ for $i=1,\ldots,7$
as the imaginary units. To convert from Furey's octonionic basis to
ours, take $\left\{ e_{1},e_{2},e_{3},e_{4},e_{5},e_{6},e_{7}\right\} \rightarrow$
$\left\{ e_{2},e_{3},e_{6},e_{1},e_{5},e_{7},-e_{4}\right\} $. A
system of ladder operators was constructed from the complexification
of the octonionic chain algebra $\mathbb{C}\otimes\overleftarrow{\mathbb{O}}\cong\mathbb{C}l(6)$,
which allows contact with $SU(3)_{c}\times U(1)_{em}$ \cite{Furey2015B}.
Due to the nonassociative nature of the octonions, the following association
of multiplication is always assumed, where an arbitrary element $f\in\mathbb{C}\otimes\mathbb{O}$
must be considered,
\begin{equation}
\left\{ \alpha_{i},\alpha_{j}\right\} :=\left\{ \alpha_{i},\alpha_{j}\right\} f=\alpha_{i}\left(\alpha_{j}f\right)+\alpha_{j}\left(\alpha_{i}f\right).
\end{equation}
If $a^{*}$ refers to complex conjugation and $\tilde{a}$ refers
to octonionic conjugation, denote $a^{\dagger}=\tilde{a}^{*}$ as
the Hermitian conjugate only when acting on $a\in\mathbb{C}\otimes\mathbb{O}$.

Our basis of raising and lowering operators is chosen as
\begin{eqnarray}
\alpha_{1}=q_{1}=\frac{1}{2}\left(-e_{5}+Ie_{1}\right), & \qquad & \alpha_{1}^{\dagger}=-q_{1}^{*}=\frac{1}{2}\left(e_{5}+Ie_{1}\right),\nonumber \\
\alpha_{2}=q_{2}=\frac{1}{2}\left(-e_{6}+Ie_{2}\right), & \qquad & \alpha_{2}^{\dagger}=-q_{2}^{*}=\frac{1}{2}\left(e_{6}+Ie_{2}\right),\\
\alpha_{3}=q_{3}=\frac{1}{2}\left(-e_{7}+Ie_{3}\right), & \qquad & \alpha_{3}^{\dagger}=-q_{3}^{*}=\frac{1}{2}\left(e_{7}+Ie_{3}\right).\nonumber
\end{eqnarray}
With this basis, we explicitly confirmed in Mathematica that the following
relations hold,
\begin{eqnarray}
\left\{ \alpha_{i},\alpha_{j}^{\dagger}\right\} f & = & \delta_{ij}f,\nonumber \\
\left\{ \alpha_{i},\alpha_{j}\right\} f & = & 0,\\
\left\{ \alpha_{i}^{\dagger},\alpha_{j}^{\dagger}\right\} f & = & 0\nonumber
\end{eqnarray}
It was also confirmed that $\left\{ \alpha_{i}^{*},\tilde{\alpha}_{j}\right\} =\delta_{ij}$.
For later convenience, a leptonic sector of operators is also introduced
as
\begin{equation}
\alpha_{0}=Il^{*}=\frac{1}{2}\left(-e_{4}+I\right),\qquad\tilde{\alpha}_{0}=Il=\frac{1}{2}\left(e_{4}+I\right).
\end{equation}

Due to the non-associativity of octonions, acting from the left once
does not span all of the possible transformations, which motivates
nested multiplication. This naturally motivates $\mathbb{C}\otimes\overleftarrow{\mathbb{O}}$
as the octonionic chain algebra corresponding to $\mathbb{C}l(6)$.
This chooses $-e_{4}$ as a pseudoscalar, such that the $k$-vector
decomposition of $\mathbb{C}l(6)$ is spanned by 1-vectors $\left\{ Ie_{2},Ie_{3},Ie_{6},Ie_{1},Ie_{5},Ie_{7}\right\} $.

Next, a nilpotent object $\omega=\alpha_{1}\alpha_{2}\alpha_{3}$
is introduced, where the parentheses of the chain algebra mentioned
above is assumed below. The Hermitian conjugate is $\omega^{\dagger}=\alpha_{3}^{\dagger}\alpha_{2}^{\dagger}\alpha_{1}^{\dagger}$.
The state $v_{c}=\omega\omega^{\dagger}$ is considered roughly as
a vacuum state (perhaps renormalized with weak isospin up), since
$\alpha_{i}\omega\omega^{\dagger}=0$. Fermionic charge states of
isospin up are identified as minimal left ideals via
\begin{equation}
\begin{array}{cccc}
S^{u}\equiv &  & \nu\omega\omega^{\dagger}\\
 & +\bar{d}^{r}\alpha_{1}^{\dagger}\omega\omega^{\dagger}\quad & +\bar{d}^{g}\alpha_{2}^{\dagger}\omega\omega^{\dagger} & +\bar{d}^{b}\alpha_{3}^{\dagger}\omega\omega^{\dagger}\\
 & +u^{r}\alpha_{3}^{\dagger}\alpha_{2}^{\dagger}\omega\omega^{\dagger} & +u^{g}\alpha_{1}^{\dagger}\alpha_{3}^{\dagger}\omega\omega^{\dagger} & +u^{b}\alpha_{2}^{\dagger}\alpha_{1}^{\dagger}\omega\omega^{\dagger}\\
 &  & +\bar{e}\alpha_{3}^{\dagger}\alpha_{2}^{\dagger}\alpha_{1}^{\dagger}\omega\omega^{\dagger}
\end{array},
\end{equation}
where $\nu,\bar{d}^{i},u^{i}$, and $\bar{e}$ are complex coefficients.
The weak isospin down states are found by building off of $v_{c}^{*}=\omega^{\dagger}\omega$,
giving
\begin{equation}
\begin{array}{cccc}
S^{d}\equiv &  & \bar{\nu}\omega^{\dagger}\omega\\
 & -d^{r}\alpha_{1}\omega^{\dagger}\omega & -d^{g}\alpha_{2}\omega^{\dagger}\omega & -d^{b}\alpha_{3}\omega^{\dagger}\omega\\
 & +\bar{u}^{r}\alpha_{3}\alpha_{2}\omega^{\dagger}\omega & +\bar{u}^{g}\alpha_{1}\alpha_{3}\omega^{\dagger}\omega & +\bar{u}^{b}\alpha_{2}\alpha_{1}\omega^{\dagger}\omega\\
 &  & +e\alpha_{1}\alpha_{2}\alpha_{3}\omega^{\dagger}\omega
\end{array}.
\end{equation}
These algebraic operators represent charge states associated with
one generation of the Standard Model with reference to $SU(3)_{c}\times U(1)_{em}$.

A notion of Pauli's exclusion principle is found, since the following
relations hold,
\begin{eqnarray}
\omega\omega^{\dagger}\omega\omega^{\dagger} & = & \omega\omega^{\dagger},\nonumber \\
\alpha_{i}^{\dagger}\omega\omega^{\dagger}\omega\omega^{\dagger} & = & \alpha_{i}^{\dagger}\omega\omega^{\dagger}\\
\alpha_{i}^{\dagger}\omega\omega^{\dagger}\alpha_{i}^{\dagger}\omega\omega^{\dagger} & = & \alpha_{i}^{\dagger}\alpha_{j}^{\dagger}\omega\omega^{\dagger}\alpha_{i}^{\dagger}\alpha_{j}^{\dagger}\omega\omega^{\dagger}=\alpha_{3}^{\dagger}\alpha_{2}^{\dagger}\alpha_{1}^{\dagger}\omega\omega^{\dagger}\alpha_{3}^{\dagger}\alpha_{2}^{\dagger}\alpha_{1}^{\dagger}\omega\omega^{\dagger}=0.\nonumber
\end{eqnarray}
The above equations imply that it is impossible to create two identical
fermionic states.

As implied, the three raising/lowering operators are associated with
three color charges. Furey also demonstrated that the electric charge
is associated with the mean of the number operators $N_{i}=\alpha_{i}^{\dagger}\alpha_{i}$
\cite{Furey2015B}. To obtain spinors associated with these charge
configurations, Furey advocates for $(\mathbb{C}\otimes\mathbb{H})\otimes_{\mathbb{C}}(\mathbb{C}\otimes\mathbb{O})=\mathbb{C}\otimes\mathbb{H}\otimes\mathbb{O}$.
Before reviewing this procedure, we first generalize the results of
$\mathbb{C}\otimes\mathbb{O}$ to $\mathbb{C}\otimes J_{2}(\mathbb{O})$.

\subsection{Uplift of $\mathbb{C}l(6)$ in $\mathbb{C}\otimes\protect\overleftarrow{J_{2}(\mathbb{O})}$}

Next, the analogous raising and lowering operators associated with
one generation of the Standard Model are constructed with elements
of $\mathbb{C}\otimes\overleftarrow{J_{2}(\mathbb{O})}$. Our guiding
principle is to take elements of $\mathbb{C}\otimes\mathbb{O}$, place
them on the upper off-diagonal component of $\mathbb{C}\otimes\overleftarrow{J_{2}(\mathbb{O})}$,
and add the Hermitian octonionic conjugate. We seek a new generalized
multiplication that implements the same particle dynamics as $\mathbb{C}\otimes\overleftarrow{\mathbb{O}}$.
For concreteness, consider $J_{f}$ as an arbitrary element of $\mathbb{C}\otimes J_{2}(\mathbb{O})$,
\begin{equation}
J_{f}=\left(\begin{array}{cc}
f_{8} & f\\
\tilde{f} & f_{9}
\end{array}\right)=\left(\begin{array}{cc}
f_{8} & f_{0}+\sum_{i=1}^{7}e_{i}f_{i}\\
f_{0}-\sum_{i=1}^{7}e_{i}f_{i} & f_{9}
\end{array}\right),
\end{equation}
where $f_{i}=f_{i,0}+If_{i,1}$ for $i=0,1,\ldots,9$.

The Jordan product is utilized to restore elements of $\mathbb{C}\otimes J_{2}(\mathbb{O})$.
However, this conflicts with left multiplication utilized in the chain
algebra $\mathbb{C}\otimes\overleftarrow{\mathbb{O}}$. The natural
multiplication for $\mathbb{C}\otimes\overleftarrow{J_{2}(\mathbb{O})}$
used throughout uses a nested commutator of Jordan products, 
\begin{equation}
m_{7}\left(J_{1},J_{2},J_{f}\right)\equiv J_{1}\circ\left(J_{2}\circ J_{f}\right)-J_{2}\circ\left(J_{1}\circ J_{f}\right),\label{nested}
\end{equation}
where $J_{1},J_{2}\in \mathbb{C}\otimes J_{2}(\mathbb{O})$ as arbitrary elements. Rather than having a single element of $\mathbb{C}\otimes J_{2}(\mathbb{O})$
to implement $\alpha_{i}$ and $\alpha_{j}^{\dagger}$, the multiplication
above is utilized. The following $\mathbb{C}\otimes\mathbb{O}$ variables
are first uplifted to elements of $\mathbb{C}\otimes J_{2}(\mathbb{O})$,
\begin{eqnarray}
J_{\alpha_{0}} & \equiv & \left(\begin{array}{cc}
0 & \alpha_{0}\\
\tilde{\alpha}_{0} & 0
\end{array}\right)=\frac{1}{2}\left(\begin{array}{cc}
0 & -e_{4}+I\\
e_{4}+I & 0
\end{array}\right),\nonumber \\
J_{\alpha_{1}} & \equiv & \left(\begin{array}{cc}
0 & \alpha_{1}\\
\tilde{\alpha}_{1} & 0
\end{array}\right)=\frac{1}{2}\left(\begin{array}{cc}
0 & -e_{5}+e_{1}I\\
e_{5}-e_{1}I & 0
\end{array}\right),\nonumber \\
J_{\alpha_{2}} & \equiv & \left(\begin{array}{cc}
0 & \alpha_{2}\\
\tilde{\alpha}_{2} & 0
\end{array}\right)=\frac{1}{2}\left(\begin{array}{cc}
0 & -e_{6}+e_{2}I\\
e_{6}-e_{2}I & 0
\end{array}\right),\nonumber \\
J_{\alpha_{3}} & \equiv & \left(\begin{array}{cc}
0 & \alpha_{3}\\
\tilde{\alpha}_{3} & 0
\end{array}\right)=\frac{1}{2}\left(\begin{array}{cc}
0 & -e_{7}+e_{3}I\\
e_{7}-e_{3}I & 0
\end{array}\right),\nonumber \\
J_{\tilde{\alpha}_{0}} & \equiv & \left(\begin{array}{cc}
0 & \tilde{\alpha_{0}}\\
\alpha_{0} & 0
\end{array}\right)=\frac{1}{2}\left(\begin{array}{cc}
0 & e_{4}+I\\
-e_{4}+I & 0
\end{array}\right),\\
J_{\alpha_{1}^{\dagger}} & \equiv & \left(\begin{array}{cc}
0 & \alpha_{1}^{\dagger}\\
\tilde{\alpha}_{1}^{\dagger} & 0
\end{array}\right)=\frac{1}{2}\left(\begin{array}{cc}
0 & e_{5}+e_{1}I\\
-e_{5}-e_{1}I & 0
\end{array}\right),\nonumber \\
J_{\alpha_{2}^{\dagger}} & \equiv & \left(\begin{array}{cc}
0 & \alpha_{2}^{\dagger}\\
\tilde{\alpha}_{2}^{\dagger} & 0
\end{array}\right)=\frac{1}{2}\left(\begin{array}{cc}
0 & e_{6}+e_{2}I\\
-e_{6}-e_{2}I & 0
\end{array}\right),\nonumber \\
J_{\alpha_{3}^{\dagger}} & \equiv & \left(\begin{array}{cc}
0 & \alpha_{3}^{\dagger}\\
\tilde{\alpha}_{3}^{\dagger} & 0
\end{array}\right)=\frac{1}{2}\left(\begin{array}{cc}
0 & e_{7}+e_{3}I\\
-e_{7}-e_{3}I & 0
\end{array}\right),\nonumber
\end{eqnarray}
where $\alpha_{0}=-e_{4}+I$ was introduced for later convenience.
We also introduce $J_{I\alpha_{i}}=IJ_{\alpha_{i}}$ as a shorthand.

These matrices allow for the following nested multiplications to mimic
the action of $\alpha_{i}$ and $\alpha_{j}^{\dagger}$,
\begin{eqnarray}
m_{\alpha_{1}}\left(J_{f}\right) & = & 2\left(m_{7}\left(J_{I\tilde{\alpha}_{0}},J_{\alpha_{1}},J_{f}\right)+m_{7}\left(J_{I\alpha_{2}^{\dagger}},J_{\alpha_{3}^{\dagger}},J_{f}\right)\right),\nonumber \\
m_{\alpha_{2}}\left(J_{f}\right) & = & 2\left(m_{7}\left(J_{I\tilde{\alpha}_{0}},J_{\alpha_{2}},J_{f}\right)+m_{7}\left(J_{I\alpha_{3}^{\dagger}},J_{\alpha_{1}^{\dagger}},J_{f}\right)\right),\nonumber \\
m_{\alpha_{3}}\left(J_{f}\right) & = & 2\left(m_{7}\left(J_{I\tilde{\alpha}_{0}},J_{\alpha_{3}},J_{f}\right)+m_{7}\left(J_{I\alpha_{1}^{\dagger}},J_{\alpha_{2}^{\dagger}},J_{f}\right)\right),\nonumber \\
m_{\alpha_{1}^{\dagger}}\left(J_{f}\right) & = & 2\left(m_{7}\left(J_{I\alpha_{0}},J_{\alpha_{1}^{\dagger}},J_{f}\right)+m_{7}\left(J_{I\alpha_{2}},J_{\alpha_{3}},J_{f}\right)\right),\\
m_{\alpha_{2}^{\dagger}}\left(J_{f}\right) & = & 2\left(m_{7}\left(J_{I\alpha_{0}},J_{\alpha_{2}^{\dagger}},J_{f}\right)+m_{7}\left(J_{I\alpha_{3}},J_{\alpha_{1}},J_{f}\right)\right),\nonumber \\
m_{\alpha_{3}^{\dagger}}\left(J_{f}\right) & = & 2\left(m_{7}\left(J_{I\alpha_{0}},J_{\alpha_{3}^{\dagger}},J_{f}\right)+m_{7}\left(J_{I\alpha_{1}},J_{\alpha_{2}},J_{f}\right)\right).\nonumber
\end{eqnarray}
The following anticommutation relations were explicitly verified,
\begin{eqnarray}
\left\{ m_{\alpha_{i}},m_{\alpha_{j}^{\dagger}}\right\} J_{f} & \equiv & m_{\alpha_{i}}\left(m_{\alpha_{j}^{\dagger}}\left(J_{f}\right)\right)+m_{\alpha_{j}^{\dagger}}\left(m_{\alpha_{i}}\left(J_{f}\right)\right)=\delta_{ij}J_{f}^{\mathrm{off}},\nonumber \\
\left\{ m_{\alpha_{i}},m_{\alpha_{j}}\right\} J_{f} & \equiv & m_{\alpha_{i}}\left(m_{\alpha_{j}}\left(J_{f}\right)\right)+m_{\alpha_{j}}\left(m_{\alpha_{i}}\left(J_{f}\right)\right)=0,\\
\left\{ m_{\alpha_{i}^{\dagger}},m_{\alpha_{j}^{\dagger}}\right\} J_{f} & \equiv & m_{\alpha_{i}^{\dagger}}\left(m_{\alpha_{j}^{\dagger}}\left(J_{f}\right)\right)+m_{\alpha_{j}^{\dagger}}\left(m_{\alpha_{i}^{\dagger}}\left(J_{f}\right)\right)=0,\nonumber
\end{eqnarray}
where $J_{f}^{\text{off }}$ contains only the off-diagonal components
of $J_{f}$. This suffices to generalize the fermionic degrees of
freedom from $\mathbb{C}\otimes\mathbb{O}$ since they are uplifted
to the off-diagonals of $\mathbb{C}\otimes J_{2}(\mathbb{O})$.

As an abuse of notation, $m_{\alpha_{i}}m_{\alpha_{j}}$ is shorthand
for $m_{\alpha_{i}}\left(m_{\alpha_{j}}\left(J_{f}\right)\right)$.
The nilpotent operator of $\mathbb{C}\otimes\overleftarrow{J_{2}(\mathbb{O})}$
is given by $m_{\omega}$,
\begin{equation}
m_{\omega}=m_{\alpha_{1}}m_{\alpha_{2}}m_{\alpha_{3}},\quad m_{\omega^{\dagger}}=m_{\alpha_{3}^{\dagger}}m_{\alpha_{2}^{\dagger}}m_{\alpha_{1}^{\dagger}}
\end{equation}
One may verify that $m_{\omega}m_{\omega}=m_{\omega^{\dagger}}m_{\omega^{\dagger}}=0$,
while $m_{\omega}m_{\omega^{\dagger}}$ acts on $J_{f}$ to give a
generalized minimal ideal of $\mathbb{C}\otimes\overleftarrow{J_{2}(\mathbb{O})}$,
\begin{eqnarray}
m_{\omega}m_{\omega^{\dagger}}J_{f}&=&\left(\begin{array}{cc}
0 & \omega\omega^{\dagger}f\\
\left(\omega\omega^{\dagger}f\right)^{*\dagger} & 0
\end{array}\right) \\
&=&\frac{1}{2}\left(\begin{array}{cc}
0 & f_{0}\left(1-e_{4}I\right)+f_{4}\left(e_{4}+I\right)\\
f_{0}\left(1+e_{4}I\right)+f_{4}\left(-e_{4}+I\right) & 0
\end{array}\right),\nonumber
\end{eqnarray}
where $f$ is the upper-right component of $J_{f}$ and $\left(\omega\omega^{\dagger}f\right)^{*\dagger}$
is a shorthand for the octonionic conjugate. This allows for the assignment
of a neutrino ``vacuum'' state, which allows for the following assignments
of particles,
\begin{eqnarray}
 & m_{\nu}=m_{\omega}m_{\omega^{\dagger}},\nonumber \\
m_{\bar{d}^{r}}=m_{\alpha_{1}^{\dagger}}m_{\omega}m_{\omega^{\dagger}}, & m_{\bar{d}^{g}}=m_{\alpha_{2}^{\dagger}}m_{\omega}m_{\omega^{\dagger}}, & m_{\bar{d}^{b}}=m_{\alpha_{3}^{\dagger}}m_{\omega}m_{\omega^{\dagger}}\nonumber \\
m_{u^{r}}=m_{\alpha_{3}^{\dagger}}m_{\alpha_{2}^{\dagger}}m_{\omega}m_{\omega^{\dagger}}, & m_{u^{g}}=m_{\alpha_{1}^{\dagger}}m_{\alpha_{3}^{\dagger}}m_{\omega}m_{\omega^{\dagger}}, & m_{u^{b}}=m_{\alpha_{2}^{\dagger}}m_{\alpha_{1}^{\dagger}}m_{\omega}m_{\omega^{\dagger}},\\
 & m_{\bar{e}}=m_{\alpha_{3}^{\dagger}}m_{\alpha_{2}^{\dagger}}m_{\alpha_{1}^{\dagger}}m_{\omega}m_{\omega^{\dagger}},\nonumber
\end{eqnarray}
and
\begin{eqnarray}
 & m_{\bar{\nu}}=m_{\omega^{\dagger}}m_{\omega},\nonumber \\
m_{d^{r}}=-m_{\alpha_{1}}m_{\omega^{\dagger}}m_{\omega}, & m_{d^{g}}=-m_{\alpha_{2}}m_{\omega^{\dagger}}m_{\omega}, & m_{d^{b}}=-m_{\alpha_{3}}m_{\omega^{\dagger}}m_{\omega}\nonumber \\
m_{\bar{u}^{r}}=m_{\alpha_{3}}m_{\alpha_{2}}m_{\omega^{\dagger}}m_{\omega}, & m_{\bar{u}^{g}}=m_{\alpha_{1}}m_{\alpha_{3}}m_{\omega^{\dagger}}m_{\omega}, & m_{\bar{u}^{b}}=m_{\alpha_{2}}m_{\alpha_{1}}m_{\omega^{\dagger}}m_{\omega},\\
 & m_{e}=m_{\alpha_{3}}m_{\alpha_{2}}m_{\alpha_{1}}m_{\omega^{\dagger}}m_{\omega}.\nonumber
\end{eqnarray}
In summary, the collection of weak-isospin up and down states are
\begin{eqnarray}
m^{u}\left(\nu,\bar{d}^{r},\bar{d}^{g},\bar{d}^{b},u^{r},u^{g},u^{b},\bar{e}\right) & = & \nu m_{\nu}+\bar{d}^{r}m_{\bar{d}^{r}}+\bar{d}^{g}m_{\bar{d}^{g}}+\bar{d}^{b}m_{\bar{d}^{b}}\nonumber \\
&&+u^{r}m_{u^{r}}+u^{g}m_{u^{g}}+u^{b}m_{u^{b}}+\bar{e}m_{\bar{e}},\nonumber \\
m^{d}\left(\bar{\nu},d^{r},d^{g},d^{b},\bar{u}^{r},\bar{u}^{g},\bar{u}^{b},e\right) & = & \bar{\nu}m_{\bar{\nu}}+d^{r}m_{d^{r}}+d^{g}m_{d^{g}}+d^{b}m_{d^{b}} \nonumber \\
&&+\bar{u}^{r}m_{\bar{u}^{r}}+\bar{u}^{g}m_{\bar{u}^{g}}+\bar{u}^{b}m_{\bar{u}^{b}}+em_{e},
\end{eqnarray}
where $\nu,\bar{d}^{r}$, etc.~are complex coefficients.


\section{Projective lines over $\mathbb{C}\otimes\mathbb{H}\otimes\mathbb{O}$}

\label{CHOphysics}


\subsection{One generation from $\mathbb{C}\otimes\mathbb{H}\otimes\mathbb{O}$}

Furey provided a formulation of the electroweak sector \cite{Furey2018A},
which led to the Standard Model embedded in $SU(5)$ and allows for
$U(1)_{B-L}$ symmetry \cite{Furey2018C,GresnigtB}. The construction
relies on identifying $\mathbb{C}l(10)=\mathbb{C}l(6)\otimes_{\mathbb{C}}\mathbb{C}l(4)$,
which can be found from a double-sided chain algebra over $\mathbb{C}\otimes\mathbb{H}\otimes\mathbb{O}$.
For instance, the left- and right-chiral spinors can be brought together
via $\psi_{D}=\psi_{R}+\psi_{L}$ with the gamma matrices implemented
as
\begin{equation}
\gamma^{0}=1\left|Ii,\quad\gamma^{1}=Ii\right|j,\quad\gamma^{2}=Ij\left|j,\quad\gamma^{3}=Ik\right|j,
\end{equation}
where $a|b$ acting on $z$ is $azb$, which is well-defined when
$(az)b=a(zb)$. This allows for left and right action of $\mathbb{C}\otimes\mathbb{H}$
to give $\mathbb{C}l(4)=\mathbb{C}l(2)\otimes_{\mathbb{C}}\mathbb{C}l(2)$.
This idea can be taken further to give $\mathbb{C}l(10)$ to identify
$Spin(10)$ and make contact with $SU(3)\times SU(2)\times U(1)$
for the Standard Model. In this manner, $\mathbb{C}\otimes\mathbb{H}\otimes\mathbb{O}$
allows for $Spin(10)$ to act from the left. While the full $\mathbb{C}l(4)$
spacetime algebra cannot be found, the remaining right action remarkably
picks out $SL(2,\mathbb{C})$ as $SU(2)_{\mathbb{C}}$.

A collection of left-chiral Weyl spinors in the $({\bf 2},{\bf 1},{\bf 16})$
representation of $SL(2,\mathbb{C})\times Spin(10)$ also contains
degrees of freedom for right-chiral antiparticles with opposite charges
via $({\bf 1},{\bf 2},\overline{{\bf 16}})$, which leads to a physicist's
convention to ignore writing down the conjugate representation. Each
of the 16 Weyl spinors is an element of $\mathbb{C}^{2}$. When working
with $\mathbb{C}\otimes\mathbb{H}\otimes\overleftarrow{\mathbb{O}}$,
there are no two-component vectors, so it is necessary to find two
copies of ${\bf 16}$. When Furey explored $\mathbb{C}l(10)$ from
$\mathbb{C}\otimes\mathbb{H}\otimes\overleftarrow{\mathbb{O}}$, a
${\bf 16}$ with its conjugate representation was found, instead of
two ${\bf 16}$'s to give $({\bf 2},{\bf 1},{\bf 16})$ for a single
generation of Standard Model fermions. This led to the so-called fermion
doubling problem.

Recent work by Furey and Hughes introduced fermions in the non-associative
$\mathbb{C}\otimes\mathbb{H}\otimes\mathbb{O}$ algebra to solve this
fermion doubling problem, which can be resolved by taking a slightly
different route to $Spin(10)$, rather than taking bivectors of $\mathbb{C}l(10)$
\cite{FureyHughes}. Instead, consider the following generalization
of Pauli matrices,
\begin{eqnarray}
\sigma_{i}=-e_{i}j|1,\quad\sigma_{8}=-Ii|1,\quad\sigma_{9}=-Ik|1,\quad\sigma_{10}=-I|1,
\end{eqnarray}
where $i=1,\dots,7$ and $\{\sigma_{i},\sigma_{8},\sigma_{9}\}$ allow
for a basis of $\mathbb{C}l(9)$. The ten ``generators'' $\sigma_{I}$
for $I=1,\dots,10$ lead to transformations on $f$ via
\begin{equation}
\frac{1}{2}\sigma_{[I}\bar{\sigma}_{J]}\psi=\frac{1}{4}\left(\sigma_{I}\left(\bar{\sigma}_{J}f\right)-\sigma_{J}\left(\bar{\sigma}_{I}f\right)\right),
\end{equation}
where $\bar{\sigma}_{a}=-\sigma_{a}$ for $a=1,\dots,9$ and $\bar{\sigma}_{10}=\sigma_{10}$.
This allows for $Spin(10)$ to act on a Weyl spinor in the ${\bf 16}$
representation instead of two 1-component objects of ${\bf 16}\oplus\overline{{\bf 16}}$
to resolve the fermion doubling problem.


With $\alpha_{\mu}=(Il^{*},q_{1},q_{2},q_{3})$ and $\alpha_{\mu}^{*}=(-Il,q_{1}^{*},q_{2}^{*},q_{3}^{*})$
for $\mu=0,1,2,3$ as an electrostrong sector and $\epsilon_{\alpha\beta}$
with $\alpha=\uparrow,\downarrow$ as an electroweak sector, the non-associative
algebra $\mathbb{C}\otimes\mathbb{H}\otimes\mathbb{O}$ can be used
to implement particle states for a single generation of the Standard Model fermions. 
We specify the particle states by using the notation and assignments recently introduced by Furey and Hughes in their solution to the fermion doubling problem \cite{FureyHughes}, namely
\begin{eqnarray}
\psi & = & \left(\mathcal{V}_{L}^{\uparrow}\euu+\mathcal{V}_{L}^{\downarrow}\eud+\mathcal{E}_{L}^{\uparrow}\edu+\mathcal{E}_{L}^{\downarrow}\edd\right)l \nonumber \\
&& +\left(\mathcal{E}_{R}^{\downarrow*}\euu-\mathcal{E}_{R}^{\uparrow*}\eud-\mathcal{V}_{R}^{\downarrow*}\edu+\mathcal{V}_{R}^{\uparrow*}\edd\right)l^{*}\\
 &  & -I\left(\mathcal{U}_{L}^{a\uparrow}\euu+\mathcal{U}_{L}^{a\downarrow}\eud+\mathcal{D}_{L}^{a\uparrow}\edu+\mathcal{D}_{L}^{a\downarrow}\edd\right)q_{a} \nonumber \\
 &&+I\left(\mathcal{D}_{R}^{a\downarrow*}\euu-\mathcal{D}_{R}^{a\uparrow*}\eud-\mathcal{U}_{R}^{a\downarrow*}\edu+\mathcal{U}_{R}^{a\uparrow*}\edd\right)q_{a}^{*}.\nonumber
\end{eqnarray}
The coefficients such as $\mathcal{V}_{L}^{\uparrow}$ are complex.

In our conventions, the $SU(3)$ Gell-Mann matrices are represented
as elements of $\mathbb{C}\otimes\overleftarrow{\mathbb{O}}$ given
by
\begin{eqnarray}
\Lambda_{1} & = & -\frac{I}{2}(e_{61}-e_{25}),\qquad\Lambda_{2}=-\frac{I}{2}(e_{21}+e_{65}),\nonumber \\
\Lambda_{3} & = & -\frac{I}{2}(e_{26}-e_{15}),\qquad\Lambda_{4}=\frac{I}{2}(e_{35}-e_{17}),\label{GellMann}\\
\Lambda_{5} & = & -\frac{I}{2}(e_{31}-e_{57}),\qquad\Lambda_{6}=\frac{I}{2}(e_{27}+e_{36}),\nonumber \\
\Lambda_{7} & = & \frac{I}{2}(e_{23}+e_{67}),\qquad\Lambda_{8}=\frac{I}{2\sqrt{3}}(e_{26}+e_{15}-e_{37}),\nonumber
\end{eqnarray}
where $e_{ij}f$ stands for $e_{i}(e_{j}f)$. For the electroweak
sector with $SU(2)\times U(1)$ symmetry, the $SU(2)$ generators
are represented in terms of imaginary quaternions and a weak isospin
projector $s=(1-Ie_{4})/2$,
\begin{equation}
\tau_{9}=\frac{I}{2}si,\qquad\tau_{10}=\frac{I}{2}sj,\qquad\tau_{11}=\frac{I}{2}sk.
\end{equation}
The weak hypercharge is given by
\begin{equation}
Y=-\frac{I}{2}\left(\frac{1}{3}\left(e_{15}+e_{26}+e_{37}\right)-s^{*}k\right).
\end{equation}
Note that all operators from $SU(3)\times SU(2)\times U(1)$ are elements
of $\mathbb{C}\otimes\mathbb{H}\otimes\overleftarrow{\mathbb{O}}$
and act from the left. The electric charge operator $Q$ is
\begin{equation}
Q=\tau_{11}+Y=-\frac{I}{2}\left(\frac{1}{3}\left(e_{15}+e_{26}+e_{37}\right)-k\right).
\end{equation}


By separating $\psi$ into $\psi_{l}+\psi_{q}+\psi_{\nu}^{c}+\psi_{e}^{c}+\psi_{u}^{c}+\psi_{d}^{c}$,
the following fields are found to correspond to the appropriate representations
of the Standard Model,
\begin{eqnarray}
({\bf 1},{\bf 2})_{-1/2}:\psi_{l} & = & \left(\mathcal{V}_{L}^{\uparrow}\euu+\mathcal{V}_{L}^{\downarrow}\eud+\mathcal{E}_{L}^{\uparrow}\edu+\mathcal{E}_{L}^{\downarrow}\edd\right)l,\nonumber \\
({\bf 3},{\bf 2})_{1/6}:\psi_{q} & = & -I\left(\mathcal{U}_{L}^{a\uparrow}\euu+\mathcal{U}_{L}^{a\downarrow}\eud+\mathcal{D}_{L}^{a\uparrow}\edu+\mathcal{D}_{L}^{a\downarrow}\edd\right)q_{a},\nonumber \\
({\bf 1},{\bf 1})_{0}:\psi_{\nu}^{c} & = & \left(-\mathcal{V}_{R}^{\downarrow*}\edu+\mathcal{V}_{R}^{\uparrow*}\edd\right)l^{*},\\
({\bf 1},{\bf 1})_{1}:\psi_{e}^{c} & = & \left(\mathcal{E}_{R}^{\downarrow*}\euu-\mathcal{E}_{R}^{\uparrow*}\eud\right)l^{*}\nonumber \\
(\overline{{\bf 3}},{\bf 1})_{-2/3}:\psi_{u}^{c} & = & I\left(-\mathcal{U}_{R}^{a\downarrow*}\edu+\mathcal{U}_{R}^{a\uparrow*}\edd\right)q_{a}^{*},\nonumber \\
(\overline{{\bf 3}},{\bf 1})_{1/3}:\psi_{d}^{c} & = & I\left(\mathcal{D}_{R}^{a\downarrow*}\euu-\mathcal{D}_{R}^{a\uparrow*}\eud\right)q_{a}^{*},\nonumber
\end{eqnarray}
where we confirmed that the above states have the appropriate weak
hypercharge values as well as weak isospin and electric charges. 
Note that complex conjugation leads to the appropriate conjugate states,
which turns left(right)-chiral particles into right(left)-chiral anti-particles.
Finally, the largest algebra commuting with $\mathfrak{so}_{10}$
derived from $\mathbb{C}\otimes\mathbb{H}\otimes\overleftrightarrow{\mathbb{O}}$
when considering action from the left and right is given by $\mathfrak{sl}_{2,\mathbb{C}}$,
which are generated by $\{1|i,1|j,1|k,1|Ii,1|Ij,1|Ik\}$.

\subsection{Uplift to 
$\mathbb{C}\otimes\mathbb{H}\otimes J_{2}(\mathbb{O})$}

To uplift the physics of $\mathbb{C}\otimes\mathbb{H}\otimes\mathbb{O}$
to $\mathbb{C}\otimes\mathbb{H}\otimes J_{2}(\mathbb{O})$, we start
by considering $f\in\mathbb{C}\otimes\mathbb{H}\otimes\mathbb{O}$
uplifted to an off-diagonal matrix $J_{f}^{\textrm{off}}\in\mathbb{C}\otimes\mathbb{H}\otimes J_{2}(\mathbb{O})$.
Our first goal is to understand how to implement left multiplication
of $\mathbb{C}\otimes\mathbb{H}\otimes\mathbb{O}$ basis elements
on $f$ by the analogous construction in $\mathbb{C}\otimes\mathbb{H}\otimes J_{2}(\mathbb{O})$
acting on $J_{f}^{\textrm{off}}$, where
\begin{equation}
J_{f}^{\textrm{off}}=\left(\begin{array}{cc}
0 & f\\
\tilde{f} & 0
\end{array}\right).
\end{equation}
For $\mathbb{C}\otimes\mathbb{H}$ bases, these can be implemented
by mapping the basis elements to the same elements times the identity
matrix. The same cannot be done for $\mathbb{O}$, as the elements
$e_{i}$ must map to $J_{2}(\mathbb{O})$ via the eight off-diagonal
octonionic Pauli matrices $J_{e_{i}}$,
\begin{equation}
J_{e_{i}}=\left(\begin{array}{cc}
0 & e_{i}\\
\tilde{e_{i}} & 0
\end{array}\right).
\end{equation}

To understand how to multiply $f$ from the left by $e_{i}$ generalized
to $\mathbb{C}\otimes\mathbb{H}\otimes J_{2}(\mathbb{O})$, the Fano
plane is crucial. A single octonionic unit can always be implemented
by multiplying by two units in four different ways. For instance,
$e_{1}=e_{1}1=e_{2}e_{3}=e_{4}e_{5}=e_{7}e_{6}$. If $e_{1}f$ is
uplifted to $J_{e_{1}f}^{\textrm{off}}$, by recalling the definition \eqref{nested} of nested commutator of Jordan products, a generalized multiplication rule can be found to give $J_{e_{1}f}^{\textrm{off}}$ from $J_{f}^{\textrm{off}}$,
\begin{eqnarray}
J_{e_{1}f}^{\textrm{off}}=m_{e_{1}}(J_{f}^{\textrm{off}}) & \equiv & \jcomm{J_{e_{1}}}{J_{1}}{J_{f}^{\textrm{off}}}+\jcomm{J_{e_{3}}}{J_{e_{2}}}{J_{f}^{\textrm{off}}}\nonumber \\
&& +\jcomm{J_{e_{5}}}{J_{e_{4}}}{J_{f}^{\textrm{off}}}+\jcomm{J_{e_{6}}}{J_{e_{7}}}{J_{f}^{\textrm{off}}},\nonumber \\
J_{e_{2}f}^{\textrm{off}}=m_{e_{2}}(J_{f}^{\textrm{off}}) & \equiv & \jcomm{J_{e_{2}}}{J_{1}}{J_{f}^{\textrm{off}}}+\jcomm{J_{e_{1}}}{J_{e_{3}}}{J_{f}^{\textrm{off}}} \nonumber \\
&& +\jcomm{J_{e_{6}}}{J_{e_{4}}}{J_{f}^{\textrm{off}}}+\jcomm{J_{e_{7}}}{J_{e_{5}}}{J_{f}^{\textrm{off}}},\nonumber \\
J_{e_{3}f}^{\textrm{off}}=m_{e_{3}}(J_{f}^{\textrm{off}}) & \equiv & \jcomm{J_{e_{3}}}{J_{1}}{J_{f}^{\textrm{off}}}+\jcomm{J_{e_{2}}}{J_{e_{1}}}{J_{f}^{\textrm{off}}} \nonumber \\
&& +\jcomm{J_{e_{7}}}{J_{e_{4}}}{J_{f}^{\textrm{off}}}+\jcomm{J_{e_{5}}}{J_{e_{6}}}{J_{f}^{\textrm{off}}},\nonumber \\
J_{e_{4}f}^{\textrm{off}}=m_{e_{4}}(J_{f}^{\textrm{off}}) & \equiv & \jcomm{J_{e_{4}}}{J_{1}}{J_{f}^{\textrm{off}}}+\jcomm{J_{e_{1}}}{J_{e_{5}}}{J_{f}^{\textrm{off}}} \nonumber \\
&& +\jcomm{J_{e_{2}}}{J_{e_{6}}}{J_{f}^{\textrm{off}}}+\jcomm{J_{e_{3}}}{J_{e_{7}}}{J_{f}^{\textrm{off}}},\\
J_{e_{5}f}^{\textrm{off}}=m_{e_{5}}(J_{f}^{\textrm{off}}) & \equiv & \jcomm{J_{e_{5}}}{J_{1}}{J_{f}^{\textrm{off}}}+\jcomm{J_{e_{4}}}{J_{e_{1}}}{J_{f}^{\textrm{off}}}\nonumber \\
&& +\jcomm{J_{e_{2}}}{J_{e_{7}}}{J_{f}^{\textrm{off}}}+\jcomm{J_{e_{3}}}{J_{e_{6}}}{J_{f}^{\textrm{off}}},\nonumber \\
J_{e_{6}f}^{\textrm{off}}=m_{e_{6}}(J_{f}^{\textrm{off}}) & \equiv & \jcomm{J_{e_{6}}}{J_{1}}{J_{f}^{\textrm{off}}}+\jcomm{J_{e_{4}}}{J_{e_{2}}}{J_{f}^{\textrm{off}}}\nonumber \\
&& +\jcomm{J_{e_{7}}}{J_{e_{1}}}{J_{f}^{\textrm{off}}}+\jcomm{J_{e_{3}}}{J_{e_{5}}}{J_{f}^{\textrm{off}}},\nonumber \\
J_{e_{7}f}^{\textrm{off}}=m_{e_{7}}(J_{f}^{\textrm{off}}) & \equiv & \jcomm{J_{e_{7}}}{J_{1}}{J_{f}^{\textrm{off}}}+\jcomm{J_{e_{4}}}{J_{e_{3}}}{J_{f}^{\textrm{off}}}\nonumber \\
&& +\jcomm{J_{e_{1}}}{J_{e_{6}}}{J_{f}^{\textrm{off}}}+\jcomm{J_{e_{5}}}{J_{e_{2}}}{J_{f}^{\textrm{off}}}.\nonumber
\end{eqnarray}
Above, $J_1$ represents the uplift of $1$ to the real traceless symmetric $2\times 2$ matrix, not an arbitrary element.
Even though we are implementing octonionic multiplication, the above
relations hold for $f\in\mathbb{C}\otimes\mathbb{H}\otimes\mathbb{O}$.
This allows for a representation of the Gell-Mann matrices in terms
of elements of $\mathbb{C}\otimes\mathbb{H}\otimes\overleftarrow{J_{2}(\mathbb{O})}$,
\begin{eqnarray}
m_{\Lambda_{1}} & = & -\frac{I}{2}(m_{e_{6}}m_{e_{1}}-m_{e_{2}}m_{e_{5}}),\qquad m_{\Lambda_{2}}=-\frac{I}{2}(m_{e_{2}}m_{e_{1}}+m_{e_{6}}m_{e_{5}}),\nonumber \\
m_{\Lambda_{3}} & = & -\frac{I}{2}(m_{e_{2}}m_{e_{6}}-m_{e_{1}}m_{e_{5}}),\qquad m_{\Lambda_{4}}=\frac{I}{2}(m_{e_{3}}m_{e_{5}}-m_{e_{1}}m_{e_{7}}),\\
m_{\Lambda_{5}} & = & -\frac{I}{2}(m_{e_{3}}m_{e_{1}}-m_{e_{5}}m_{e_{7}}),\qquad m_{\Lambda_{6}}=\frac{I}{2}(m_{e_{2}}m_{e_{7}}+m_{e_{3}}m_{e_{6}}),\nonumber \\
m_{\Lambda_{7}} & = & \frac{I}{2}(m_{e_{2}}m_{e_{3}}+m_{e_{6}}m_{e_{7}}),\qquad m_{\Lambda_{8}}=\frac{I}{2\sqrt{3}}(m_{e_{2}}m_{e_{6}}+m_{e_{1}}m_{e_{5}}-2m_{e_{3}}m_{e_{7}}),\nonumber
\end{eqnarray}
where $m_{\Lambda_{1}}(J_{f}^{\textrm{off}})=-\frac{I}{2}(m_{e_{6}}(m_{e_{1}}(J_{f}^{\textrm{off}}))-m_{e_{2}}(m_{e_{5}}(J_{f}^{\textrm{off}})))$
more precisely. From here, particle states associated with elements
of $\mathbb{C}\otimes\mathbb{H}\otimes\mathbb{O}$ can be uplifted
to $\mathbb{C}\otimes\mathbb{H}\otimes J_{2}(\mathbb{O})$. It was
confirmed that the $SU(3)$ generators above annihilate leptons and
apply color rotations to the quarks in the appropriate manner.

The same relations found in $\mathbb{C}\otimes\mathbb{H}\otimes\mathbb{O}$
for $SU(2)\times U(1)$ generators are also found by the appropriate
uplift to $\mathbb{C}\otimes\mathbb{H}\otimes\overleftarrow{J_{2}(\mathbb{O})}$.
The appropriate left action of $g\in\mathbb{C}\otimes\mathbb{H}$
on $f$ uplifted to $J_{f}^{\textrm{off}}$ can be found simply by
taking $gJ_{f}^{\textrm{off}}$, since the diagonal elements of $\mathbb{C}\otimes\mathbb{H}\otimes J_{2}(\mathbb{O})$
can contain $\mathbb{C}\otimes\mathbb{H}$. Uplifting the generators
of $SU(2)\times U(1)$ therefore gives
\begin{eqnarray}
m_{\tau_{9}} & = & \frac{i}{4}\left(I+m_{e_{4}}\right),\qquad m_{\tau_{10}}=\frac{j}{4}\left(I+m_{e_{4}}\right),\qquad m_{\tau_{11}}=\frac{k}{4}\left(I+m_{e_{4}}\right),\\
m_{Y} & = & -\frac{1}{2}\left(\frac{I}{3}\left(m_{e_{1}}m_{e_{5}}+m_{e_{2}}m_{e_{6}}+m_{e_{3}}m_{e_{7}}\right)-\frac{k}{2}\left(I-m_{e_{4}}\right)\right),\nonumber
\end{eqnarray}
where all multiplication is assumed to act from the left. Similarly,
the electric charge operator becomes
\begin{equation}
m_{Q}=m_{\tau_{11}}+m_{Y}=-\frac{I}{2}\left(\frac{1}{3}\left(m_{e_{1}}m_{e_{5}}+m_{e_{2}}m_{e_{6}}+m_{e_{3}}m_{e_{7}}\right)-k\right).
\end{equation}

The fermionic states in the $\mathbb{C}\otimes\mathbb{H}\otimes J_{2}(\mathbb{O})$
are identified as
\begin{eqnarray}
({\bf 1},{\bf 2})_{-1/2}:J_{\psi_{l}} & = & \left(\mathcal{V}_{L}^{\uparrow}\euu+\mathcal{V}_{L}^{\downarrow}\eud+\mathcal{E}_{L}^{\uparrow}\edu+\mathcal{E}_{L}^{\downarrow}\edd\right)J_{l},\nonumber \\
({\bf 3},{\bf 2})_{1/6}:J_{\psi_{q}} & = & -I\left(\mathcal{U}_{L}^{a\uparrow}\euu+\mathcal{U}_{L}^{a\downarrow}\eud+\mathcal{D}_{L}^{a\uparrow}\edu+\mathcal{D}_{L}^{a\downarrow}\edd\right)J_{q_{a}},\nonumber \\
({\bf 1},{\bf 1})_{0}:J_{\psi_{\nu}^{c}} & = & \left(-\mathcal{V}_{R}^{\downarrow*}\edu+\mathcal{V}_{R}^{\uparrow*}\edd\right)J_{l^{*}},\\
({\bf 1},{\bf 1})_{1}:J_{\psi_{e}^{c}} & = & \left(\mathcal{E}_{R}^{\downarrow*}\euu-\mathcal{E}_{R}^{\uparrow*}\eud\right)J_{l^{*}}\nonumber \\
(\overline{{\bf 3}},{\bf 1})_{-2/3}:J_{\psi_{u}^{c}} & = & I\left(-\mathcal{U}_{R}^{a\downarrow*}\edu+\mathcal{U}_{R}^{a\uparrow*}\edd\right)J_{q_{a}^{*}},\nonumber \\
(\overline{{\bf 3}},{\bf 1})_{1/3}:J_{\psi_{d}^{c}} & = & I\left(\mathcal{D}_{R}^{a\downarrow*}\euu-\mathcal{D}_{R}^{a\uparrow*}\eud\right)J_{q_{a}^{*}},\nonumber
\end{eqnarray}
where in our conventions, the $\mathbb{C}\otimes\mathbb{O}$ quantities
such as $l$ and $q_{a}$ are uplifted explicitly to give
\begin{eqnarray}
J_{l}=\frac{1}{2}\left(\begin{array}{cc}
0 & 1-e_{4}I\\
1+e_{4}I & 0
\end{array}\right), & \qquad & J_{l^{*}}=\frac{1}{2}\left(\begin{array}{cc}
0 & 1+e_{4}I\\
1-e_{4}I & 0
\end{array}\right),\nonumber \\
J_{q_{1}}=\frac{1}{2}\left(\begin{array}{cc}
0 & -e_{5}+e_{1}I\\
e_{5}-e_{1}I & 0
\end{array}\right), & \qquad & J_{q_{1}^{*}}=\frac{1}{2}\left(\begin{array}{cc}
0 & -e_{5}-e_{1}I\\
e_{5}+e_{1}I & 0
\end{array}\right),\\
J_{q_{2}}=\frac{1}{2}\left(\begin{array}{cc}
0 & -e_{6}+e_{2}I\\
e_{6}-e_{2}I & 0
\end{array}\right), & \qquad & J_{q_{2}^{*}}=\frac{1}{2}\left(\begin{array}{cc}
0 & -e_{6}-e_{2}I\\
e_{6}+e_{2}I & 0
\end{array}\right),\nonumber \\
J_{q_{3}}=\frac{1}{2}\left(\begin{array}{cc}
0 & -e_{7}+e_{3}I\\
e_{7}-e_{3}I & 0
\end{array}\right), & \qquad & J_{q_{3}^{*}}=\frac{1}{2}\left(\begin{array}{cc}
0 & -e_{7}-e_{3}I\\
e_{7}+e_{3}I & 0
\end{array}\right).\nonumber
\end{eqnarray}
It was confirmed that $m_{\tau_{11}}$, $m_{Y}$, and $m_{Q}$ give
the appropriate eigenvalues for these states.

\subsection{Uplift to 
$\mathbb{O}\otimes J_{2}(\mathbb{C}\otimes\mathbb{H})$}

Next, we seek to obtain the physics of $\mathbb{C}\otimes\mathbb{H}\otimes\mathbb{O}$
by uplifting to $\mathbb{O}\otimes J_{2}(\mathbb{C}\otimes\mathbb{H})$.
The Hermitian conjugate of $\mathbb{O}\otimes J_{2}(\mathbb{C}\otimes\mathbb{H})$
takes conjugation with respect to both $\mathbb{C}$ and $\mathbb{H}$.
Uplifting an element $f\in\mathbb{C}\otimes\mathbb{H}\otimes\mathbb{O}$
to $J_{f}^{\textrm{off}}\in\mathbb{O}\otimes J_{2}(\mathbb{C}\otimes\mathbb{H})$
is given by
\begin{equation}
J_{f}^{\textrm{off}}=\left(\begin{array}{cc}
0 & f\\
\hat{f}^{*} & 0
\end{array}\right),
\end{equation}
where $f^{*}$ is the complex conjugate and $\hat{f}$ is the quaternionic
conjugate. Finding the corresponding left action of $\mathbb{C}\otimes\mathbb{H}\otimes\mathbb{O}$
within $\mathbb{O}\otimes J_{2}(\mathbb{C}\otimes\mathbb{H})$ is
straightforward for $\mathbb{O}$, yet requires care with $\mathbb{C}\otimes\mathbb{H}$.


Left multiplication of $I$ on $f$ uplifted to $J_{If}^{\textrm{off}}$
must be implemented with the nested Jordan commutator product \eqref{nested},
\begin{equation}
J_{If}^{\textrm{off}}=m_{I}(J_{f}^{\textrm{off}})\equiv\{J_{I},J_{1},J_{f}^{\textrm{off}}\}_{\circ}.
\end{equation}
This holds for arbitrary elements $f\in\mathbb{C}\otimes\mathbb{H}\otimes\mathbb{O}$.
The analogous relationship for imaginary quaternionic units are
\begin{eqnarray}
J_{if}^{\textrm{off}} & = & m_{i}(J_{f}^{\textrm{off}})\equiv\{J_{i},J_{1},J_{f}^{\textrm{off}}\}_{\circ}+\{J_{k},J_{j},J_{f}^{\textrm{off}}\}_{\circ},\nonumber \\
J_{jf}^{\textrm{off}} & = & m_{j}(J_{f}^{\textrm{off}})\equiv\{J_{j},J_{1},J_{f}^{\textrm{off}}\}_{\circ}+\{J_{i},J_{k},J_{f}^{\textrm{off}}\}_{\circ},\label{CxHuplift}\\
J_{kf}^{\textrm{off}} & = & m_{k}(J_{f}^{\textrm{off}})\equiv\{J_{k},J_{1},J_{f}^{\textrm{off}}\}_{\circ}+\{J_{j},J_{i},J_{f}^{\textrm{off}}\}_{\circ}.\nonumber
\end{eqnarray}
The corresponding uplift of left multiplication by imaginary octonions
is given by left multiplication, such that $J_{e_{i}f}^{\textrm{off}}=e_{i}J_{f}^{\textrm{off}}$.

From here, the uplift of the fermionic states and the action of bosonic
operators on the fermions is similar to the previous discussion on
$\mathbb{C}\otimes\mathbb{H}\otimes J_{2}(\mathbb{O})$. To highlight
this uplift with more detail and for a specific example, consider
$\psi_{e}^{c}$ as a left-chiral positron and weak isospin singlet,
\begin{equation}
\psi_{\nu}^{c}=\left(\mathcal{E}_{R}^{\downarrow*}\euu-\mathcal{E}_{R}^{\uparrow*}\eud\right)l^{*}=\frac{1}{4}\left(\mathcal{E}_{R}^{\downarrow*}\left(1+Ik+e_{4}I-e_{4}k\right)+\mathcal{E}_{R}^{\uparrow*}\left(-j+Ii-e_{4}i-e_{4}Ij\right)\right).
\end{equation}
Uplifting to $\mathbb{O}\otimes J_{2}(\mathbb{C}\otimes\mathbb{H})$
explicitly gives
\begin{eqnarray}
J_{\psi_{\nu}^{c}} & = & \left(\begin{array}{cc}
0 & \left(\mathcal{E}_{R}^{\downarrow*}\euu-\mathcal{E}_{R}^{\uparrow*}\eud\right)l^{*}\\
\left(\mathcal{E}_{R}^{\downarrow}\euu-\mathcal{E}_{R}^{\uparrow}\edu\right)l & 0
\end{array}\right)\\
\updatered{\left(\mathcal{E}_{R}^{\downarrow*}\euu-\mathcal{E}_{R}^{\uparrow*}\eud\right)l^{*}} &=& \updatered{\frac{1}{4}\left(\mathcal{E}_{R}^{\downarrow*}\left(1+Ik+e_{4}I-e_{4}k\right)+\mathcal{E}_{R}^{\uparrow*}\left(-j+Ii-e_{4}i-e_{4}Ij\right)\right)} \nonumber \\
\updatered{\left(\mathcal{E}_{R}^{\downarrow}\euu-\mathcal{E}_{R}^{\uparrow}\edu\right)l} &=& \updatered{\frac{1}{4} \left(\mathcal{E}_{R}^{\downarrow}\left(1+Ik-e_{4}I+e_{4}k\right)+\mathcal{E}_{R}^{\uparrow}\left(j+Ii+e_{4}i-e_{4}Ij\right)\right)} .\nonumber
\end{eqnarray}

The action of the Gell-Mann generators uplifted to $\mathbb{O}\otimes J_{2}(\mathbb{C}\otimes\mathbb{H})$
is
\begin{eqnarray}
m_{\Lambda_{1}} & = & -\frac{m_{I}}{2}(e_{61}-e_{25}),\qquad m_{\Lambda_{2}}=-\frac{m_{I}}{2}(e_{21}+e_{65}),\nonumber \\
m_{\Lambda_{3}} & = & -\frac{m_{I}}{2}(e_{26}-e_{15}),\qquad m_{\Lambda_{4}}=\frac{m_{I}}{2}(e_{35}-e_{17}),\\
m_{\Lambda_{5}} & = & -\frac{m_{I}}{2}(e_{31}-e_{57}),\qquad m_{\Lambda_{6}}=\frac{m_{I}}{2}(e_{27}+e_{36}),\nonumber \\
m_{\Lambda_{7}} & = & \frac{m_{I}}{2}(e_{23}+e_{67}),\qquad m_{\Lambda_{8}}=\frac{m_{I}}{2\sqrt{3}}(e_{26}+e_{15}-e_{37}).\nonumber
\end{eqnarray}
The electroweak generators are given by
\begin{eqnarray}
m_{\tau_{9}} & = & \frac{m_{I}}{2}m_{s}m_{i},\qquad m_{\tau_{10}}=\frac{m_{I}}{2}m_{s}m_{j},\qquad m_{\tau_{11}}=\frac{m_{I}}{2}m_{s}m_{k},\nonumber \\
m_{Y} & = & -\frac{m_{I}}{2}\left(\frac{1}{3}\left(e_{15}+e_{26}+e_{37}\right)-m_{s^{*}}m_{k}\right),
\end{eqnarray}
where
\begin{equation}
m_{s}(J_{f})=\frac{1}{2}\left(1-e_{4}m_{I}\right)J_{f},\qquad m_{s^{*}}(J_{f})=\frac{1}{2}\left(1+e_{4}m_{I}\right)J_{f}.
\end{equation}
The electric charge operator is given by
\begin{equation}
m_{Q}=m_{\tau_{11}}+m_{Y}=-\frac{m_{I}}{2}\left(\frac{1}{3}\left(e_{15}+e_{26}+e_{37}\right)-m_{k}\right).
\end{equation}

The action of these generators leads to the expected results when
acting on $J_{\psi_{\nu}^{c}}$. For instance, all of the $SU(3)$
generators vanish and $J_{\psi_{\nu}^{c}}$ is an eigenstate of $m_{\tau_{11}}$
and $m_{Y}$,
\begin{eqnarray}
m_{\Lambda_{i}}(J_{\psi_{\nu}^{c}}) & = & 0,\nonumber \\
m_{\tau_{11}}(J_{\psi_{\nu}^{c}}) & = & 0,\nonumber \\
m_{Y}(J_{\psi_{\nu}^{c}}) & = & 1J_{\psi_{\nu}^{c}},\\
m_{Q}(J_{\psi_{\nu}^{c}}) & = & 1J_{\psi_{\nu}^{c}},\nonumber
\end{eqnarray}
where $1$ is found as an eigenvalue for electric charge and weak
hypercharge with the left-chiral positron.

\subsection{Uplift to 
$\mathbb{C}\otimes\mathbb{O}\otimes J_{2}(\mathbb{H})$}

Finally, the physics of $\mathbb{C}\otimes\mathbb{H}\otimes\mathbb{O}$
is uplifted to $\mathbb{C}\otimes\mathbb{O}\otimes J_{2}(\mathbb{H})$.
Uplifting an element $f\in\mathbb{C}\otimes\mathbb{H}\otimes\mathbb{O}$
to $J_{f}^{\textrm{off}}\in\mathbb{C}\otimes\mathbb{O}\otimes J_{2}(\mathbb{H})$
is given by
\begin{equation}
J_{f}^{\textrm{off}}=\left(\begin{array}{cc}
0 & f\\
\hat{f} & 0
\end{array}\right),\label{Huplift}
\end{equation}
where, as above, $\hat{f}$ denotes the quaternionic conjugation of $f$. From here, it is clear that the uplift of left multiplication by imaginary
quaternionic units is identical to Eq.~\eqref{CxHuplift}. Less care
is needed with the complex numbers and octonions, as they are on the
diagonals of $\mathbb{C}\otimes\mathbb{O}\otimes J_{2}(\mathbb{H})$.

The action of the Gell-Mann generators uplifted to $\mathbb{C}\otimes\mathbb{O}\otimes J_{2}(\mathbb{H})$
is identical to Eq.~\eqref{GellMann}. The electroweak generators
are given by
\begin{eqnarray}
m_{\tau_{9}} & = & \frac{I}{2}sm_{i},\qquad m_{\tau_{10}}=\frac{I}{2}sm_{j},\qquad m_{\tau_{11}}=\frac{I}{2}sm_{k},\nonumber \\
m_{Y} & = & -\frac{I}{2}\left(\frac{1}{3}\left(e_{15}+e_{26}+e_{37}\right)-{s^{*}}m_{k}\right).
\end{eqnarray}
The electric charge operator is given by
\begin{equation}
m_{Q}=m_{\tau_{11}}+m_{Y}=-\frac{I}{2}\left(\frac{1}{3}\left(e_{15}+e_{26}+e_{37}\right)-m_{k}\right).
\end{equation}
The fermions of $\mathbb{C}\otimes\mathbb{H}\otimes\mathbb{O}$ can
be uplifted to $\mathbb{C}\otimes\mathbb{O}\otimes J_{2}(\mathbb{H})$
via Eq.~\eqref{Huplift} and the generators shown above can be found
to act appropriately on the fermionic states.

\section{Conclusions}

\label{Conclusions}

In this work, we showed how to construct three homogeneous
spaces that, following Rosenfeld's interpretation of the Magic Square, correspond to his ``generalized'' projective lines over \updatered{the Dixon algebra,} $\mathbb{C}\otimes\mathbb{H}\otimes\mathbb{O}$.
Such spaces are obtained from three non-simple Lie algebras
obtained from Tits' construction for the Freudenthal Magic Square.
The quotient space of these isometry groups modded out by derivations
lead to $\mathbb{C}\otimes\mathbb{H}\otimes J_{2}\left(\mathbb{O}\right)$,
$\mathbb{O}\otimes J_{2}\left(\mathbb{C}\otimes\mathbb{H}\right)$,
and $\mathbb{C}\otimes\mathbb{O}\otimes J_{2}\left(\mathbb{H}\right)$,
which contains the three newly found Dixon-Rosenfeld projective lines.
The physics of $\mathbb{C}\otimes\mathbb{H}\otimes\mathbb{O}$ can
be uplifted to each of these extended Jordan algebras and the generators
of $SU(3)\times SU(2)\times U(1)$ for the Standard Model can be uplifted
into a (nested) chain algebra over $\overleftarrow{\mathbb{C}\otimes\mathbb{H}\otimes J_{2}\left(\mathbb{O}\right)}$,
$\overleftarrow{\mathbb{O}\otimes J_{2}\left(\mathbb{C}\otimes\mathbb{H}\right)}$,
and $\overleftarrow{\mathbb{C}\otimes\mathbb{O}\otimes J_{2}\left(\mathbb{H}\right)}$. We provided explicit states for one generation of fermions in the standard model within these projective lines, including operators for gauge boson interactions and identification of charges.

While non-simple Lie algebras were found from the Dixon-Rosenfeld
projective lines and one generation of the Standard Model fermions
were embedded into these projective lines, further work is needed
to see if the appropriate representations of the Standard Model are
contained within the corresponding isometry groups. For instance,
while the bosonic interactions with fermions were demonstrated to
be in the chain algebras over division algebras tensored with Jordan
algebras and various $SU(3)\times SU(2)\times U(1)$ groups can be
found in the derivation groups, the representations with respect to
these groups do not isolate the Standard Model fermionic representations
and charges. This is similar to how $Spin(9)$, $SU(3)\times SU(3)$,
and $F_{4}$ are not GUT groups, but the octonions and $F_{4}$ have
been used to encode Standard Model fermions \cite{Todorov:2018yvi,Todorov:2019hlc,Krasnov:2019auj}.

\updatered{
It appears that the Freudenthal-Tits formula should work for $\mathbb{A}=\mathbb{O}$ and $\mathbb{B} = \mathbb{C}\otimes \mathbb{H}$ to give a Lie algebra $\mathfrak{a}_{II}$. However, there is not a single formula for the $2\times 2$ case, as setting $\mathbb{A}=\mathbb{O}$ already leads to a difference. Here, we articulated the structure of $J_2(\mathbb{C}\otimes\mathbb{H})$ and found $\mathfrak{der}(J_2(\mathbb{C}\otimes\mathbb{H}))$. However, applying the $2\times 2$ analogue of the Freudenthal-Tits construction did not lead to the anticipated representations of $\mathbb{T}$ with respect to $\mathfrak{der}(\mathbb{T})$. To further complicate matters, it is known that $\mathbb{C}\otimes \mathbb{H}$ can lead to multiple representations. For now, we merely claim that some non-simple Lie algebra $\mathfrak{a}_{II}$ exists that contains at least 120 dimensions. By exploring the $3\times 3$ case in future work, we hope to gain a further understanding of the true definition of $\mathfrak{a}_{II}$.
}

Additional work is needed to see if other subalgebras of these non-simple
Lie algebras exist that can isolate the appropriate representation
theory for the Standard Model. Otherwise, chain algebras such as $\overleftarrow{\mathbb{A}\otimes J_{2}(\mathbb{B})}$
may lead to Clifford algebras that would be large enough to contain
the Standard Model gauge group, just as $\mathbb{C}\otimes\mathbb{H}\otimes\overleftarrow{\mathbb{O}}$
can lead to $Cl(10)$. In future work, we seek to investigate the
notion of Dixon-Rosenfeld projective planes to see if this may provide
applications for three generations of the Standard Model fermions
with $\mathbb{C}\otimes\mathbb{H}\otimes\mathbb{O}$. Interactions with the Higgs boson would also be worth exploring, which has been discussed recently \cite{FureyHughesB}.

\section{Acknowledgments}

Thanks to Cohl Furey for countless helpful discussions related to
$\mathbb{C}\otimes\mathbb{H}\otimes\mathbb{O}$ and Mia Hughes for
additional support. Thanks to Richard Clawson, Dugan Hammock,
and Garrett Lisi for discussions on Clifford algebras and minimal
ideals. The work of D.~Corradetti is supported by a grant of the Quantum
Gravity Research Institute. The work of AM is supported by a ``Maria
Zambrano'' distinguished researcher fellowship, financed by the European
Union within the NextGenerationEU program.

\appendix
\section{$J_2(\mathbb{C}\otimes\mathbb{H})$ as $4\times 4$ complex matrices}

\updatered{
The Jordan algebra $J_2(\mathbb{C}\otimes\mathbb{H})$ is 16-dimensional and can be expressed in terms of a set of matrices in $M_4(\mathbb{C})$. We review this isomorphism and determine the action of the double conjugation with respect to $\mathbb{C}$ and $\mathbb{H}$ in the language of $4\times 4$ complex matrices. 
Before introducing $J_2(\mathbb{C}\otimes \mathbb{H})$, we first clarify how the double conjugation of $\mathbb{C}\otimes \mathbb{H}$ with respect to $\mathbb{C}$ and $\mathbb{H}$ leads to a 4-dimensional element and specify how this maps into the isomorphism with $M_2(\mathbb{C})$. 

First, $f \in \mathbb{C}\otimes \mathbb{H}$ is recast in $\mathcal{M}(f) \in M_2(\mathbb{C})$ by the following isomorphism, 
using our notation of \eqref{res2-bis} and \eqref{complexcoeff}: 
\begin{eqnarray}
 f &=& \left(c_{1,1}+c_{1,2}I\right)+\left(c_{2,1}+c_{2,2}I\right)i+\left(c_{3,1}+c_{3,2}I\right)j+\left(c_{4,1}+c_{4,2}I\right)k  \nonumber\\
 &=& \left\{ \underset{%
\mathbf{1}\oplus \mathbf{1}}{\underbrace{1,I}},\underset{\mathbf{3}}{%
\underbrace{i,j,k}},\underset{\mathbf{3}}{\underbrace{Ii,Ij,Ik}}\right\}. 
\left\{c_{1,1}, c_{1,2}, c_{2,1}, c_{2,2}, c_{3,1}, c_{3,2}, c_{4,1}, c_{4,2}\right\}^{T}  \nonumber\\
\simeq \mathcal{M}(f) &=&
\left(
\begin{array}{cc}
 c_{1,1}-c_{2,2}+i \left(c_{1,2}+c_{2,1}\right) & c_{3,1}-c_{4,2}+i \left(c_{3,2}+c_{4,1}\right) \\
 -c_{3,1}-c_{4,2}+i \left(c_{4,1}-c_{3,2}\right) & c_{1,1}+c_{2,2}+i \left(c_{1,2}-c_{2,1}\right) \\
\end{array}
\right).
\end{eqnarray}
Next, we clarify how double conjugation of $\mathbb{C}\otimes \mathbb{H}$ maps into $M_2(\mathbb{C})$, 
\begin{eqnarray}
    \hat{f}^*&=& \left(c_{1,1}-c_{1,2}I\right)+\left(-c_{2,1}+c_{2,2}I\right)i+\left(-c_{3,1}+c_{3,2}I\right)j+\left(-c_{4,1}+c_{4,2}I\right)k   \nonumber \\
    \simeq \overline{\mathcal{M}(f)}^\top &=&
\left(
\begin{array}{cc}
 c_{1,1}-c_{2,2}-i \left(c_{1,2}+c_{2,1}\right) & -c_{3,1}-c_{4,2}-i \left(c_{4,1}-c_{3,2}\right) \\
  c_{3,1}-c_{4,2}-i \left(c_{3,2}+c_{4,1}\right) & c_{1,1}+c_{2,2}-i \left(c_{1,2}-c_{2,1}\right) \\
\end{array}
\right).
\end{eqnarray}
As shown above, $\hat{f}^* \cong \overline{\mathcal{M}(f)}^\top$. 

By using the conventions of conjugation as shown in Eqs.~\eqref{mult1}-\eqref{mult2}, we find a ``real'' element of 4 dimensions by
\begin{eqnarray}
    r &=& \frac{1}{2}\left(f + \hat{f}^*\right) = c_{1,1}+c_{2,2}Ii+c_{3,2}Ij+c_{4,2}Ik, \nonumber \\
    \mathcal{M}(r)&=& \mathcal{M}\left( \frac{1}{2}\left(f + \hat{f}^*\right) \right) = \frac{1}{2}\left(\mathcal{M}(f) + \overline{\mathcal{M}(f)}^\top \right).
\end{eqnarray}

A 16-dim representation of $X \in J_{2}\left( \mathbb{C}\otimes \mathbb{H}\right)$ is built as a Hermitian block-matrix in $\mathcal{M}(X) \in M_4(\mathbb{C})$, 
\begin{eqnarray}
\label{eq:M(X)}
    X &=& \left(\begin{array}{cc} r & f \\ \hat{f}^*& s\end{array} \right), \\
    \cong \mathcal{M}(X) &=& \left(\begin{array}{cc} \mathcal{M}(r) & \mathcal{M}(f) \\ \overline{\mathcal{M}(f)}^\top& \mathcal{M}(s)\end{array} \right). \nonumber 
\end{eqnarray}
where $r$ and $s$ real with respect to the double conjugation on  $\mathbb{C}\otimes \mathbb{H}$, leading to $r$ and $s$ as 4-dimensional elements spanning $\{1, Ii, Ij, Ik\}$ with $\mathcal{M}(r) = \left(\begin{array}{cc} r_{1,1}-r_{2,2} & -r_{4,2}+i r_{3,2}\\
 -r_{4,2}-i r_{3,2} & r_{1,1}+r_{2,2} \end{array}\right)$ 
 $ \in H_2(\mathbb{C})$ 
 , while $\mathcal{M}(f) \in M_2(\mathbb{C})$. As such, we refer to $X$ as $X(r,s,f)$.


}


\section{Demonstration of $\mathfrak{der}\left( J_{2}\left( \mathbb{C}\otimes \mathbb{H}\right)
\right)\cong \mathfrak{su(4)}$}
\label{App-der(J2(biquat))}

\updatered{
The derivation of an alternative algebra is defined (\cite{schafer_introduction_1966} page 77) as a Bracket algebra satisfying the Leibniz rule, and a theorem shows it is of the form
\begin{equation}
    D_{X,Y} = [L_X,L_Y] + [L_X,R_Y] + [R_X, R_Y].
\end{equation}
When the Jordan algebra is not only alternative but is commutative, $L_X = R_X$ and the inner derivations are (\cite{schafer_introduction_1966} page 92)
\begin{equation}
    D_{X,Y} = [L_X,L_Y]. \label{eq:LeftJordanDerivation}
\end{equation}
A derivation parameterized by $X$ and $Y$ applied to an element $Z \in J_{2}\left( \mathbb{C}\otimes \mathbb{H}\right)$ gives
\begin{eqnarray}
    D_{X,Y}(Z) &=& [L_X,L_Y](Z) = L_X(L_Y(Z))-L_Y(L_X(Z)) \nonumber \\
    &=& X . (Y . Z) - Y . ( X . Z) = X . ( Z . Y) - (X . Z ) . Y = -[ X, Z, Y ], \label{eq:JordanAssociatorDerivation}
\end{eqnarray}
where $[X, Z, Y]$ is the Jordan associator, sandwiching $Z$ between $X$ and $Y$.

A pair $X$,$Y \in J_{2}\left( \mathbb{C}\otimes \mathbb{H}\right)$ elements: $X(r,s,f)\cong \mathcal{M}(\left(\begin{array}{cc}
r & f\\
 \hat{f}^* & s \\
\end{array}\right))$
and $Y(R,S,F) \cong \mathcal{M}(\left(\begin{array}{cc}
R & F\\
 \hat{F}^* & S \\
\end{array}\right))$ is bracketed by a commutator to give an anti-hermitian matrix in $M_4(\mathbb{C})$.
Though, \eqref{eq:JordanAssociatorDerivation} is rewritten from note (2) page 7 and introducing $i$ two times, the derivation becomes a bracket commutator between Hermitian matrices, and  $\mathfrak{der}\left( J_{2}\left( \mathbb{C}\otimes \mathbb{H}\right)\right)$  is represented by the Hermitian matrix $\delta$,  
\begin{eqnarray}
\delta=i[X,Y] &=& \delta(\rho,\sigma,\phi) \cong \left(\begin{array}{cc}
\mathcal{M}(\rho) & \mathcal{M}(\phi)\\
 \overline{\mathcal{M}(\phi)}^\top & \mathcal{M}(\sigma) \\
\end{array}\right), \nonumber \\ 
    D_{X,Y}(Z) &=& -[ X, Z, Y ] = i[ i[ X, Y], Z ] = i[ \delta, Z ]. \label{eq:HermitianDerivation}
\end{eqnarray}

Next, consider the embedding of $J_2(\mathbb{C}\otimes\mathbb{H})$ in $M_4(\mathbb{C})$ to find the commutator of two elements,
defined below as block matrices using \eqref{eq:M(X)}:
\begin{eqnarray}
   [ \mathcal{M}(X),  \mathcal{M}(Y) ] &=& [
   \left(\begin{array}{cc} \mathcal{M}(r) & \mathcal{M}(f) \\ \overline{\mathcal{M}(f)}^\top& \mathcal{M}(s)\end{array} \right),
   \left(\begin{array}{cc} \mathcal{M}(R) & \mathcal{M}(F) \\ \overline{\mathcal{M}(F)}^\top& \mathcal{M}(S)\end{array} \right) ]
\label{M(XY)}
\end{eqnarray}
The commutator of these two matrices leads to 
\begin{equation}
[\mathcal{M}(X), \mathcal{M}(Y)] = -i \left(\begin{array}{cc}
\mathcal{M}(\rho) & \mathcal{M}(\phi)\\
 \overline{\mathcal{M}(\phi)}^\top & \mathcal{M}(\sigma) \\
\end{array}\right). \label{[M(X),M(Y)]}
\end{equation}
The block $2\times 2$ matrices are given by
\begin{eqnarray}
\mathcal{M}(\rho) &=& \left(\begin{array}{cc} \rho_{1,1}-\rho_{2,2} & -\rho_{4,2}+i \rho_{3,2}\\
 -\rho_{4,2}-i \rho_{3,2} & \rho_{1,1}+\rho_{2,2} \end{array}\right), \nonumber \\ 
\mathcal{M}(\sigma) &=& \left(\begin{array}{cc} \sigma_{1,1}-\sigma_{2,2} & -\sigma_{4,2}+i \sigma_{3,2}\\
 -\sigma_{4,2}-i \sigma_{3,2} & \sigma_{1,1}+\sigma_{2,2} \end{array}\right), \\ 
\mathcal{M}(\phi) &=& \left(\begin{array}{cc}\phi_{1,1}-\phi_{2,2}+i \left(\phi_{1,2}+\phi_{2,1}\right) & \phi_{3,1}-\phi_{4,2}+i \left(\phi_{3,2}+\phi_{4,1}\right) \\
-\phi_{3,1}-\phi_{4,2}+i \left(\phi_{4,1}-\phi_{3,2}\right) & \phi_{1,1}+\phi_{2,2}+i \left(\phi_{1,2}-\phi_{2,1}\right)\end{array}\right). \nonumber
\end{eqnarray}

The solution for the matrix components above are found by plugging Eq.~\eqref{M(XY)} into Eq.~\eqref{[M(X),M(Y)]} to give
\begin{eqnarray}
    \rho_{1,1} &=& -f_{1,2} F_{1,1}+f_{1,1} F_{1,2}-f_{2,2} F_{2,1}+f_{2,1} F_{2,2} \nonumber \\
    && -f_{3,2} F_{3,1}+f_{3,1} F_{3,2}-f_{4,2} F_{4,1}+f_{4,1} F_{4,2}, \nonumber \\ 
    \rho_{2,2} &=& f_{2,1} F_{1,1}+f_{2,2} F_{1,2}-f_{1,1} F_{2,1}-f_{1,2} F_{2,2} \nonumber \\ 
    && +f_{4,1} F_{3,1}+f_{4,2} F_{3,2}-f_{3,1} F_{4,1}-f_{3,2} F_{4,2} + r_{4,2} R_{3,2}-r_{3,2} R_{4,2}, \nonumber \\ 
    \rho_{3,2} &=& f_{3,1} F_{1,1}+f_{3,2} F_{1,2}-f_{4,1} F_{2,1}-f_{4,2} F_{2,2} \nonumber \\
    && -f_{1,1} F_{3,1}-f_{1,2} F_{3,2}+f_{2,1} F_{4,1}+f_{2,2} F_{4,2} -r_{4,2} R_{2,2}+r_{2,2} R_{4,2}, \nonumber \\ 
    \rho_{4,2} &=& f_{4,1} F_{1,1}+f_{4,2} F_{1,2}+f_{3,1} F_{2,1}+f_{3,2} F_{2,2} \nonumber \\
    && -f_{2,1} F_{3,1}-f_{2,2} F_{3,2}-f_{1,1} F_{4,1}-f_{1,2} F_{4,2} + r_{3,2} R_{2,2}-r_{2,2} R_{3,2}, \nonumber \\ 
    \sigma_{1,1} &=& f_{1,2} F_{1,1}-f_{1,1} F_{1,2}+f_{2,2} F_{2,1}-f_{2,1} F_{2,2} \\
    && +f_{3,2} F_{3,1}-f_{3,1} F_{3,2}+f_{4,2} F_{4,1}-f_{4,1} F_{4,2}, \nonumber \\ 
    \sigma_{2,2} &=& -f_{2,1} F_{1,1}-f_{2,2} F_{1,2}+f_{1,1} F_{2,1}+f_{1,2} F_{2,2} \nonumber \\
    && +f_{4,1} F_{3,1}+f_{4,2} F_{3,2}-f_{3,1} F_{4,1}-f_{3,2} F_{4,2} + s_{4,2} S_{3,2}-s_{3,2} S_{4,2}, \nonumber \\ 
    \sigma_{3,2} &=& -f_{3,1} F_{1,1}-f_{3,2} F_{1,2}-f_{4,1} F_{2,1}-f_{4,2} F_{2,2} \nonumber \\
    && +f_{1,1} F_{3,1}+f_{1,2} F_{3,2}+f_{2,1} F_{4,1}+f_{2,2} F_{4,2} -s_{4,2} S_{2,2}+s_{2,2} S_{4,2}, \nonumber \\ 
    \sigma_{4,2} &=& -f_{4,1} F_{1,1}-f_{4,2} F_{1,2}+f_{3,1} F_{2,1}+f_{3,2} F_{2,2} \nonumber \\
    && -f_{2,1} F_{3,1}-f_{2,2} F_{3,2}+f_{1,1} F_{4,1}+f_{1,2} F_{4,2} + s_{3,2} S_{2,2}-s_{2,2} S_{3,2}, \nonumber \\
    \phi_{1,1} &=& -\frac{1}{2}\left(r_{1,1} F_{1,2}-r_{2,2} F_{2,1}-r_{3,2} F_{3,1}-r_{4,2} F_{4,1}-R_{1,1} f_{1,2}+R_{2,2} f_{2,1}+R_{3,2} f_{3,1}+R_{4,2} f_{4,1} \right. \nonumber \\
    && \left.-s_{1,1} F_{1,2}+s_{2,2} F_{2,1}+s_{3,2} F_{3,1}+s_{4,2} F_{4,1}+S_{1,1} f_{1,2}-S_{2,2} f_{2,1}-S_{3,2} f_{3,1}-S_{4,2} f_{4,1}\right), \nonumber \\ 
    \phi_{1,2} &=& \frac{1}{2}\left(r_{1,1} F_{1,1}+r_{2,2} F_{2,2}+r_{3,2} F_{3,2}+r_{4,2} F_{4,2}-R_{1,1} f_{1,1}-R_{2,2} f_{2,2}-R_{3,2} f_{3,2}-R_{4,2} f_{4,2}\right. \nonumber \\ 
    && \left. -s_{1,1} F_{1,1}-s_{2,2} F_{2,2}-s_{3,2} F_{3,2}-s_{4,2} F_{4,2}+S_{1,1} f_{1,1}+S_{2,2} f_{2,2}+S_{3,2} f_{3,2}+S_{4,2} f_{4,2}\right), \nonumber
\end{eqnarray}
where the solutions to the other six $\phi_{i,j}$ can be found similarly.

Next, consider the trace of this $4\times 4$ matrix $\mathcal{M}(\delta)$ by considering the traces of $\mathcal{M}(\rho)$ and $\mathcal{M}(\sigma)$,
\begin{equation}
    \mbox{Tr}(\mathcal{M}(\rho)) = - \mbox{Tr}(\mathcal{M}(\sigma)) = 2\rho_{1,1} .
\end{equation}
Since $\mbox{Tr}(\mathcal{M}(\delta)) = \mbox{Tr}(\mathcal{M}(\rho))+ \mbox{Tr}(\mathcal{M}(\sigma)) = 0$, it is derived that the $4\times 4$ matrices are traceless. While $\rho_{i,j}$, $\sigma_{i,j}$, and $\tau_{i,j}$ lead to 16 degrees of freedom, since $\rho_{1,1} = -\sigma_{1,1}$, there are 15 linearly independent elements, which form a basis for the Hermitian traceless generators of $\mathfrak{su}_4$,
\begin{equation}
\begin{array}{llll}
 L_{1}= \left(
\begin{smallmatrix}
 1 & 0 & 0 & 0 \\
 0 & 1 & 0 & 0 \\
 0 & 0 & -1 & 0 \\
 0 & 0 & 0 & -1 \\
\end{smallmatrix}
\right), & L_{2}= \left(
\begin{smallmatrix}
 1 & 0 & 0 & 0 \\
 0 & -1 & 0 & 0 \\
 0 & 0 & 1 & 0 \\
 0 & 0 & 0 & -1 \\
\end{smallmatrix}
\right), & L_{3}= \left(
\begin{smallmatrix}
 1 & 0 & 0 & 0 \\
 0 & -1 & 0 & 0 \\
 0 & 0 & -1 & 0 \\
 0 & 0 & 0 & 1 \\
\end{smallmatrix}
\right), & L_{4}= \left(
\begin{smallmatrix}
 0 & 0 & 0 & 1 \\
 0 & 0 & -1 & 0 \\
 0 & -1 & 0 & 0 \\
 1 & 0 & 0 & 0 \\
\end{smallmatrix}
\right), \\
 L_{5}= \left(
\begin{smallmatrix}
 0 & 0 & 0 & -1 \\
 0 & 0 & -1 & 0 \\
 0 & -1 & 0 & 0 \\
 -1 & 0 & 0 & 0 \\
\end{smallmatrix}
\right), & L_{6}= \left(
\begin{smallmatrix}
 0 & 0 & 0 & i \\
 0 & 0 & i & 0 \\
 0 & -i & 0 & 0 \\
 -i & 0 & 0 & 0 \\
\end{smallmatrix}
\right), & L_{7}= \left(
\begin{smallmatrix}
 0 & 0 & 0 & i \\
 0 & 0 & -i & 0 \\
 0 & i & 0 & 0 \\
 -i & 0 & 0 & 0 \\
\end{smallmatrix}
\right), & L_{8}= \left(
\begin{smallmatrix}
 0 & 0 & -1 & 0 \\
 0 & 0 & 0 & 1 \\
 -1 & 0 & 0 & 0 \\
 0 & 1 & 0 & 0 \\
\end{smallmatrix}
\right), \\
 L_{9}= \left(
\begin{smallmatrix}
 0 & 0 & 1 & 0 \\
 0 & 0 & 0 & 1 \\
 1 & 0 & 0 & 0 \\
 0 & 1 & 0 & 0 \\
\end{smallmatrix}
\right), & L_{10}= \left(
\begin{smallmatrix}
 0 & 0 & -i & 0 \\
 0 & 0 & 0 & -i \\
 i & 0 & 0 & 0 \\
 0 & i & 0 & 0 \\
\end{smallmatrix}
\right), & L_{11}= \left(
\begin{smallmatrix}
 0 & 0 & i & 0 \\
 0 & 0 & 0 & -i \\
 -i & 0 & 0 & 0 \\
 0 & i & 0 & 0 \\
\end{smallmatrix}
\right), & L_{12}= \left(
\begin{smallmatrix}
 0 & 1 & 0 & 0 \\
 1 & 0 & 0 & 0 \\
 0 & 0 & 0 & -1 \\
 0 & 0 & -1 & 0 \\
\end{smallmatrix}
\right), \\
 L_{13}= \left(
\begin{smallmatrix}
 0 & -1 & 0 & 0 \\
 -1 & 0 & 0 & 0 \\
 0 & 0 & 0 & -1 \\
 0 & 0 & -1 & 0 \\
\end{smallmatrix}
\right), & L_{14}= \left(
\begin{smallmatrix}
 0 & i & 0 & 0 \\
 -i & 0 & 0 & 0 \\
 0 & 0 & 0 & i \\
 0 & 0 & -i & 0 \\
\end{smallmatrix}
\right), & L_{15}= \left(
\begin{smallmatrix}
 0 & i & 0 & 0 \\
 -i & 0 & 0 & 0 \\
 0 & 0 & 0 & -i \\
 0 & 0 & i & 0 \\
\end{smallmatrix}
\right). & \text{} \\
\end{array}
\end{equation}

Next, we demonstrate that a collection of matrices $X_a$ and $Y_b$ lead to $L_{A} \rightarrow L_{a,b} =\frac{i}{2}[X_a, Y_b]$. While there is not a unique set of matrices, we found a small collection of $X_a$ for $a=1,\dots,4$ and $Y_b$ for $b=1,\dots 12$ that lead to the 15 generators $L_A$. $X_a$ are given by 
\begin{eqnarray}
X_1 &=& X(0,0,1) = X(0,0,f_{1,1}=1), \nonumber \\ 
X_2 &=& X(0,0,I) = X(0,0,f_{1,2}=1), \\
X_3 &=& X(j I,-j I,0) = X(r_{3,2}=-1,s_{3,2}=1,0), \nonumber \\  
X_4 &=& X(-k I,k I,0) =X(r_{4,2}=-1,s_{4,2}=1,0). \nonumber 
\end{eqnarray}
12 elements $Y_1$ to $Y_{12}$ in $J_{2}\left( \mathbb{C}\otimes \mathbb{H}\right)$ are given by
\begin{eqnarray}
   Y_1 &=& Y(0,0,1) = Y(0,0,F_{1,1}=1), \nonumber \\
   Y_2 &=& Y(0,0,I) =Y(0,0,F_{1,2}=1), \nonumber \\
 Y_3 &=& Y(j I,-j I,0) = Y(R_{3,2}=-S_{3,2}=1), \nonumber \\
   Y_4 &=& Y(-k I,k I,0) = Y(-R_{4,2}=S_{4,2}=1), \nonumber \\
   Y_5 &=& Y(0,0,i) = Y(0,0,F_{2,1}=1), \nonumber \\
   Y_6 &=& Y(0,0,j) = Y(0,0,F_{3,1}=-1), \\
 Y_7 &=& Y(0,0,k) = Y(0,0,F_{4,1}=1), \nonumber \\
   Y_8 &=& Y(i I,-i I,0) = Y(R_{2,2}=-S_{2,2}=1),\nonumber \\
 Y_9 &=& Y(2,0,0) =Y(R_{1,1}=2,0,0), \nonumber \\
   Y_{10} &=& Y(2 i I,0,0) =Y(R_{2,2}=2,0,0), \nonumber \\
   Y_{11} &=& Y(2 j I,0,0) =Y(R_{3,2}=2,0,0), \nonumber \\
   Y_{12} &=& Y(2 k I,0,0) =Y(R_{4,2}=2,0,0). 
\end{eqnarray}
These $Y_b$ can be combined with $X_a$ to give the generators of $\mathfrak{su}_4$.
we build from them the 15 pairs $L_{a,b}= \frac{i}{2}[X_a, Y_b]$ from the following index pairs $\{a,b\}$:
\begin{eqnarray}
    L_1 = L_{1,2}, \quad L_2 = L_{4,3}, \quad L_3 = L_{1,5},
    \quad L_4 = L_{1,11}, \quad L_5 = L_{2,12}, \nonumber \\
    L_6 = L_{1,12}, \quad L_7 = L_{2,11}, \quad L_8 = L_{2,10}, \quad L_9 = L_{2,9}, \quad L_{10} = L_{1,9}, \\
    L_{11} = L_{1,10}, \quad L_{12} = L_{1,7}, \quad L_{13} = L_{3,8} \quad L_{14} = L_{4,8} \quad L_{15} = L_{1,6}. \nonumber
\end{eqnarray}
The three Cartan generators of $\mathfrak{su}_4$ are the three first expressed above, diagonal and commuting.
By construction as commutators of Hermitian matrices scaled by the imaginary factor $i$, the matrices $L_{a,b}$ are all Hermitian, and span at most the 16-dimensional space $\mathfrak{u}_4$, but from the property that $\rho_{1,1}+\sigma_{1,1}=0$ and that $\mathcal{M}(\delta)$ is traceless,
the non-traceless generator of $\mathfrak{u_4}$ can not be obtained as a derivation $L_{a,b}$, and therefore the derivation of $J_{2}\left( \mathbb{C}\otimes \mathbb{H}\right)$ is  $\mathfrak{su}_4$.
}

\section{Demonstration that the algebra $\mathfrak{a}_{II}$ given by Tits' formula
does not contain the Dixon algebra $\mathbb{T}$ with a ${\bf3}$ representation of $\mathfrak{der}(\mathbb{H})$}
\label{aIInotFT}

\updatered{
\textbf{Theorem:} The algebra $\mathfrak{a}_{II}$ given by Tits' formula
does not contain the Dixon algebra $\mathbb{T}= \mathbb{C}\otimes \mathbb{H}\otimes \mathbb{O}$ when $\mathbb{H} \subset \mathbb{T}$ corresponds to the representations ${\bf3}\oplus {\bf1}$ with respect to $\mathfrak{der}(\mathbb{H}) = \mathfrak{su}_2$.

\textbf{Proof:} By applying Tits' formula (with Barton-Sudbery's
modification), one obtains%
\begin{eqnarray}
\mathfrak{a}_{II} &=&\mathcal{L}_{2}\left( \mathbb{O},\mathbb{C}\otimes
\mathbb{H}\right) =\mathfrak{isom}\left( \mathbb{T}P_{II}^{1}\right)
\nonumber \\
&:=&\mathfrak{so}\left( \mathbb{O}^{\prime }\right) \oplus \mathfrak{der}%
\left( J_{2}(\mathbb{C}\otimes \mathbb{H})\right) \oplus \mathbb{O}^{\prime
}\otimes J_{2}^{\prime }(\mathbb{C}\otimes \mathbb{H})  \nonumber \\
&=&\mathfrak{so}_{7}\oplus \mathfrak{so}_{6}\oplus 3\cdot \left( \mathbf{7},%
\mathbf{4}\right)   \nonumber \\
&=&\mathfrak{g}_{2}\oplus \mathfrak{so}_{6}\oplus 3\cdot \left( \mathbf{7},%
\mathbf{4}\right) \oplus \left( \mathbf{7},\mathbf{1}\right) .\label{tthis2}
\end{eqnarray}%
We also recall that%
\begin{equation}
\mathbb{T}:=\mathbb{R}\otimes \mathbb{C}\otimes \mathbb{H}\otimes \mathbb{O}%
\simeq 2\cdot \left( \mathbf{7}+\mathbf{1},\mathbf{3}+\mathbf{1}\right) ~%
\text{of~}\mathfrak{der}\left( \mathbb{T}\right) =\mathfrak{g}_{2}\oplus
\mathfrak{su}_{2}.  \label{TI2}
\end{equation}%
Let us now find all $\mathfrak{su}_{2}$ subalgebras of $\mathfrak{so}_{6}$
in (\ref{tthis2}):

\begin{enumerate}
\item
\begin{eqnarray}
\mathfrak{so}_{6} &\rightarrow &\underset{\text{symm~}I\leftrightarrow II}{%
\mathfrak{su}_{2,I}\oplus \mathfrak{su}_{2,II}}\rightarrow \left\{
\begin{array}{l}
\mathfrak{su}_{2,I\text{~or~}II}; \\
\mathfrak{su}_{2,d};%
\end{array}%
\right.  \\
\mathbf{15} &=&(\mathbf{3},\mathbf{1})+(\mathbf{1},\mathbf{3})+(\boldsymbol{3%
},\mathbf{3})=\left\{
\begin{array}{l}
4\cdot \mathbf{3}+3\cdot \mathbf{1}; \\
\mathbf{5}+3\cdot \mathbf{3}+\mathbf{1};%
\end{array}%
\right.  \\
\mathbf{4} &=&(\mathbf{2},\mathbf{2})=\left\{
\begin{array}{l}
2\cdot \mathbf{2}; \\
\mathbf{3}+\mathbf{1},%
\end{array}%
\right.
\end{eqnarray}%
such that (\ref{tthis2}) can be further branched as%
\begin{eqnarray}
\mathfrak{a}_{II} &=&\mathfrak{g}_{2}\oplus \mathfrak{su}_{2,I}\oplus
\mathfrak{su}_{2,II}\oplus (\mathbf{1},\boldsymbol{3},\mathbf{3})\oplus
3\cdot \left( \mathbf{7},\mathbf{2,2}\right) \oplus \left( \mathbf{7},%
\mathbf{1,1}\right)   \nonumber \\
&&  \nonumber \\
&=&\left\{
\begin{array}{l}
\mathfrak{g}_{2}\oplus \mathfrak{su}_{2,I\text{~or~}II}\oplus 3\cdot (%
\boldsymbol{1},\mathbf{3})\oplus 3\cdot (\boldsymbol{1},\mathbf{1})\oplus
6\cdot \left( \mathbf{7},\mathbf{2}\right) \oplus \left( \mathbf{7},\mathbf{1%
}\right) \nsupseteq \mathbb{T}; \\
\\
\mathfrak{g}_{2}\oplus \mathfrak{su}_{2,d}\oplus 2\cdot (\mathbf{1},%
\boldsymbol{3})\oplus (\mathbf{1},\boldsymbol{1})\oplus (\mathbf{1},%
\boldsymbol{5})\oplus 3\cdot \left( \mathbf{7},\mathbf{3}\right) \oplus
4\cdot \left( \mathbf{7},\mathbf{1}\right) \nsupseteq \mathbb{T}.%
\end{array}%
\right.
\end{eqnarray}

\item
\begin{eqnarray}
\mathfrak{so}_{6} &\rightarrow &\mathfrak{so}_{5}\rightarrow \left\{
\begin{array}{l}
\underset{\text{symm~}I\leftrightarrow II}{\mathfrak{su}_{2,I}\oplus
\mathfrak{su}_{2,II}}\rightarrow \left\{
\begin{array}{l}
\mathfrak{su}_{2,I\text{~or~}II}; \\
\mathfrak{su}_{2,d};%
\end{array}%
\right.  \\
\mathfrak{su}_{2}\left( \oplus u_{1}\right) ; \\
\mathfrak{su}_{2,P};%
\end{array}%
\right.  \\
\mathbf{15} &=&\mathbf{10}+\mathbf{5}=\left\{
\begin{array}{l}
(\mathbf{3},\mathbf{1})+(\mathbf{1},\mathbf{3})+2\cdot (\mathbf{2},\mathbf{2}%
)+(\mathbf{1},\mathbf{1})=\left\{
\begin{array}{l}
\mathbf{3}+4\cdot \mathbf{1}+4\cdot \mathbf{2}; \\
4\cdot \mathbf{3}+3\cdot \mathbf{1};%
\end{array}%
\right.  \\
4\cdot \mathbf{3}+3\cdot \mathbf{1}; \\
\mathbf{3}+\mathbf{5}+\mathbf{7};%
\end{array}%
\right.  \\
\mathbf{4} &=&\mathbf{4}=\left\{
\begin{array}{l}
(\mathbf{2},\mathbf{1})+(\mathbf{1},\mathbf{2})=\left\{
\begin{array}{l}
\mathbf{2}+2\cdot \mathbf{1}; \\
2\cdot \mathbf{2};%
\end{array}%
\right.  \\
2\cdot \mathbf{2}; \\
\mathbf{4},%
\end{array}%
\right.
\end{eqnarray}%
such that (\ref{tthis2}) can be further branched as%
\begin{eqnarray}
\mathfrak{a}_{II} &=&\mathfrak{g}_{2}\oplus \mathfrak{so}_{5}\oplus \left(
\mathbf{1},\mathbf{5}\right) \oplus 3\cdot \left( \mathbf{7},\mathbf{4}%
\right) \oplus \left( \mathbf{7},\mathbf{1}\right)   \nonumber \\
&&  \nonumber \\
&=&\left\{
\begin{array}{l}
\mathfrak{g}_{2}\oplus \mathfrak{su}_{2,I}\oplus \mathfrak{su}_{2,II}\oplus
2\cdot \left( \mathbf{1},\mathbf{2,2}\right) \oplus \left( \mathbf{1},%
\mathbf{1,1}\right) \oplus 3\cdot \left( \mathbf{7},\mathbf{2,1}\right)
\oplus 3\cdot \left( \mathbf{7},\mathbf{1,2}\right) \oplus \left( \mathbf{7,1%
}\right)  \\
\\
=\left\{
\begin{array}{l}
\mathfrak{g}_{2}\oplus \mathfrak{su}_{2,I\text{~or~}II}\oplus 3\cdot \left(
\mathbf{1},\mathbf{1}\right) \oplus 4\cdot \left( \mathbf{1},\mathbf{2}%
\right) \oplus \left( \mathbf{1},\mathbf{1}\right) \oplus 3\cdot \left(
\mathbf{7},\mathbf{2}\right) \oplus 7\cdot \left( \mathbf{7},\mathbf{1}%
\right) \nsupseteq \mathbb{T}; \\
\\
\mathfrak{g}_{2}\oplus \mathfrak{su}_{2,d}\oplus 3\cdot \left( \mathbf{1},%
\mathbf{3}\right) \oplus 3\cdot \left( \mathbf{1},\mathbf{1}\right) \oplus
6\cdot \left( \mathbf{7},\mathbf{2}\right) \oplus \left( \mathbf{7,1}\right)
\nsupseteq \mathbb{T};%
\end{array}%
\right.  \\
\\
\mathfrak{g}_{2}\oplus \mathfrak{su}_{2}\oplus \mathfrak{u}_{1}\oplus 3\cdot
\left( \mathbf{1},\mathbf{3}\right) \oplus 2\cdot \left( \mathbf{1},\mathbf{1%
}\right) \oplus 6\cdot \left( \mathbf{7},\mathbf{2}\right) \oplus \left(
\mathbf{7},\mathbf{1}\right) \nsupseteq \mathbb{T}; \\
\\
\mathfrak{g}_{2}\oplus \mathfrak{su}_{2,P}\oplus \left( \mathbf{1},\mathbf{5}%
\right) \oplus \left( \mathbf{1},\mathbf{7}\right) \oplus 3\cdot \left(
\mathbf{7},\mathbf{4}\right) \oplus \left( \mathbf{7},\mathbf{1}\right)
\nsupseteq \mathbb{T}.%
\end{array}%
\right.   \nonumber \\
&&
\end{eqnarray}

\item
\begin{eqnarray}
\mathfrak{so}_{6} &\rightarrow &\underset{\text{symm~}I\leftrightarrow II}{%
\mathfrak{su}_{2,I}\oplus \mathfrak{su}_{2,II}}\left( \oplus \mathfrak{u}%
_{1}\right) \rightarrow \left\{
\begin{array}{l}
\mathfrak{su}_{2,I\text{~or~}II}\left( \oplus \mathfrak{u}_{1}\right) ; \\
\mathfrak{su}_{2,d}\left( \oplus \mathfrak{u}_{1}\right) ;%
\end{array}%
\right.  \\
\mathbf{15} &=&(\mathbf{3},\mathbf{1})+(\mathbf{1},\mathbf{3})+(\boldsymbol{1%
},\mathbf{1})+2\cdot (\boldsymbol{2},\mathbf{2})=\left\{
\begin{array}{l}
\mathbf{3}+4\cdot \mathbf{1}+4\cdot \boldsymbol{2}; \\
4\cdot \mathbf{3}+3\cdot \mathbf{1};%
\end{array}%
\right.  \\
\mathbf{4} &=&(\mathbf{2},\mathbf{1})+(\mathbf{1},\mathbf{2})=\left\{
\begin{array}{l}
\mathbf{2}+2\cdot \mathbf{1}; \\
2\cdot \mathbf{2},%
\end{array}%
\right.
\end{eqnarray}%
such that (\ref{tthis2}) can be further branched as%
\begin{eqnarray}
\mathfrak{a}_{II} &=&\mathfrak{g}_{2}\oplus \mathfrak{su}_{2,I}\oplus
\mathfrak{su}_{2,II}\left( \oplus \mathfrak{u}_{1}\right) \oplus 2\cdot
\left( \mathbf{1},\mathbf{2},\mathbf{2}\right) \oplus 3\cdot \left( \mathbf{7%
},\mathbf{2,1}\right) +3\cdot \left( \mathbf{7},\mathbf{1,2}\right) \oplus
\left( \mathbf{7},\mathbf{1,1}\right)   \nonumber \\
&&  \nonumber \\
&=&\left\{
\begin{array}{l}
\mathfrak{g}_{2}\oplus \mathfrak{su}_{2,I\text{~or~}II}\left( \oplus
\mathfrak{u}_{1}\right) \oplus 3\cdot \left( \mathbf{1},\mathbf{1}\right)
\oplus 4\cdot \left( \mathbf{1},\mathbf{2}\right) \oplus 3\cdot \left(
\mathbf{7},\mathbf{2}\right) +7\cdot \left( \mathbf{7},\mathbf{1}\right)
\nsupseteq \mathbb{T}; \\
\\
\mathfrak{g}_{2}\oplus \mathfrak{su}_{2,d}\left( \oplus \mathfrak{u}%
_{1}\right) \oplus 3\cdot (\mathbf{1},\mathbf{3})\oplus 2\cdot \left(
\mathbf{1},\mathbf{1}\right) \oplus 6\cdot \left( \mathbf{7},\mathbf{2}%
\right) \oplus \left( \mathbf{7},\mathbf{1}\right) \nsupseteq \mathbb{T}.%
\end{array}%
\right.
\end{eqnarray}

\item
\begin{eqnarray}
\mathfrak{so}_{6} &\rightarrow &\mathfrak{su}_{3}\left( \oplus \mathfrak{u}%
_{1}\right) \rightarrow \left\{
\begin{array}{l}
\mathfrak{su}_{2}\left( \oplus 2\mathfrak{u}_{1}\right) ; \\
\mathfrak{su}_{2,P}\left( \oplus \mathfrak{u}_{1}\right) ;%
\end{array}%
\right.  \\
\mathbf{15} &=&\mathbf{8}+\mathbf{1}+2\cdot \mathbf{3}=\left\{
\begin{array}{l}
\mathbf{3}+4\cdot \mathbf{1+}4\cdot \mathbf{2}; \\
\mathbf{5}+3\cdot \mathbf{3}+\mathbf{1};%
\end{array}%
\right.  \\
\mathbf{4} &=&\mathbf{3}+\mathbf{1}=\left\{
\begin{array}{l}
\mathbf{2}+2\cdot \mathbf{1}; \\
\mathbf{3}+\mathbf{1},%
\end{array}%
\right.
\end{eqnarray}%
such that (\ref{tthis2}) can be further branched as%
\begin{eqnarray}
\mathfrak{a}_{II} &=&\mathfrak{g}_{2}\oplus \mathfrak{su}_{3}\left( \oplus
\mathfrak{u}_{1}\right) \oplus 2\cdot \left( \mathbf{1},\mathbf{3}\right)
\oplus 3\cdot \left( \mathbf{7},\mathbf{3}\right) \oplus 4\cdot \left(
\mathbf{7},\mathbf{1}\right)   \nonumber \\
&&  \nonumber \\
&=&\left\{
\begin{array}{l}
\mathfrak{g}_{2}\oplus \mathfrak{su}_{2}\left( \oplus 2\mathfrak{u}%
_{1}\right) \oplus 4\cdot \left( \mathbf{1},\mathbf{2}\right) \oplus 2\cdot
\left( \mathbf{1},\mathbf{1}\right) \oplus 3\cdot \left( \mathbf{7},\mathbf{2%
}\right) \oplus 7\cdot \left( \mathbf{7},\mathbf{1}\right) \nsupseteq
\mathbb{T}; \\
\\
\mathfrak{g}_{2}\oplus \mathfrak{su}_{2,P}\left( \oplus \mathfrak{u}%
_{1}\right) \oplus \left( \mathbf{1},\mathbf{5}\right) \oplus 2\cdot \left(
\mathbf{1},\mathbf{3}\right) \oplus 3\cdot \left( \mathbf{7},\mathbf{3}%
\right) \oplus 4\cdot \left( \mathbf{7},\mathbf{1}\right) \nsupseteq \mathbb{%
T}.%
\end{array}%
\right.
\end{eqnarray}
\end{enumerate}

This concludes the proof that there is no $\mathfrak{su}_{2}\simeq \mathfrak{%
der}\left( \mathbb{C}\otimes \mathbb{H}\right) $ subalgebra of $\mathfrak{so}%
_{6}\simeq \mathfrak{der}\left( J_{2}\left( \mathbb{C}\otimes \mathbb{H}%
\right) \right) $ such that $\mathfrak{a}_{II}$ given by (\ref{tthis2})
contains the Dixon algebra $\mathbb{T}$ (\ref{TI2}), presuming that $\mathbb{H} \subset \mathbb{T}$ contains a ${\bf1}\oplus {\bf3}$ representation of $\mathfrak{der}(\mathbb{H}) = \mathfrak{su}_2$. $\blacksquare $
}


\end{document}